%% file: draft_submit.tex
\documentclass[twocolumn,aip,jcp,showpacs,showkeys,preprintnumbers,amsmath,amssymb,reprint,floatfix]{revtex4-2}

\usepackage{graphicx}
\usepackage{dcolumn}
\usepackage{bm}
\usepackage{xcolor}
\usepackage{threeparttable}
\usepackage{multirow}
\usepackage{longtable}
\usepackage{booktabs}

\usepackage{placeins}
\usepackage{fancyvrb}

\newcommand\T{\rule{0pt}{2.6ex}}       %
\begin{document}

\title{%
NMR chemical shielding for solid-state systems using spin-orbit coupled ZORA GIPAW
}

\author{T.~Speelman}
\affiliation{%
Radboud University, Institute for Molecules and Materials,
Heyendaalseweg 135, NL-6525 AJ Nijmegen, The Netherlands
}

\author{M.-T.~Huebsch}
\affiliation{%
VASP Software GmbH, Berggasse 21/14, A-1090 Vienna, Austria
}

\author{R.~W.~A.~Havenith}
\affiliation{%
Stratingh Institute for Chemistry, Zernike Institute for Advanced Materials, University of Groningen, Nijenborgh~3, NL-9747~AG Groningen, The Netherlands
}
\affiliation{%
Ghent Quantum Chemistry Group, Department of Inorganic and Physical Chemistry,
Ghent University, Krijgslaan~281~(S3),
B-9000 Gent, Belgium
}

\author{M.~Marsman}
\affiliation{%
VASP Software GmbH, Berggasse 21/14, A-1090 Vienna, Austria
}

\author{G.~A.~de Wijs}
\affiliation{%
Radboud University, Institute for Molecules and Materials,
Heyendaalseweg 135, NL-6525 AJ Nijmegen, The Netherlands
}

\date{\today}

\begin{abstract}
We present an implementation of spin-orbit coupling (SOC) for the computation of nuclear magnetic resonance (NMR) chemical shielding tensors within linear response theory. Our implementation in the Vienna {\it Ab initio} Simulation Package (VASP) is tailored to solid-state systems by employing periodic boundary conditions and the gauge-including projector augmented waves (GIPAW) approach. Relativistic effects are included on the level of the zeroth-order regular approximation (ZORA). We discuss the challenges posed by the PAW partial wave basis in describing SOC regarding chemical shielding tensors. Our method is in good agreement with existing local-basis ZORA implementations for a series of Sn, Hg, and Pb molecules and cluster approximations for crystalline systems.
\end{abstract}


\maketitle

\section{\label{sec:introduction}Introduction}
Nuclear Magnetic Resonance (NMR) is a powerful tool for the structural characterization of materials. Measuring the local response of nuclei to an externally applied magnetic field reveals information about local structure and dynamics. However, the obtained spectra are sometimes difficult to resolve. First principles simulations aid the resolution of these spectra and help reveal additional insights, see, e.g., the reviews in Refs.~\onlinecite{Charpentier2011, Ashbrook2013, Bonhomme2012}. Hence, accurate first-principles calculations of the chemical shielding tensor are essential in the emerging field of NMR crystallography.\cite{Duer2012}

For molecular systems, the calculation of shielding tensors is available at the density functional theory (DFT) level and beyond \cite{Helgaker1999}, and several packages incorporate relativistic effects employing two- or even four-component theory, e.g., Refs.~\onlinecite{Wolff1998-Pauli, Wolff1999-ZORA, DiracMain}. With molecular codes, solid-state systems can be treated {\it via} a cluster approximation. However, constructing clusters is non-trivial, and solid-state systems are more efficiently and accurately treated by solid-state codes that use periodic boundary conditions and a suitable basis set.

A popular and efficient method for solid-state calculations is the Gauge-Including Projector Augmented Wave (GIPAW) method.\cite{PM2001, YPM2007} Currently, the highest level of relativistic theory available within GIPAW, for calculating chemical shielding tensors, is scalar relativistic using the zeroth-order regular approximation (ZORA).\cite{YPPM-ZORA2003, vanLenthe1996, Bouten2000} Although this level of theory is sufficient for lighter nuclei in the upper half of the periodic table, it is not adequate for heavy elements ($Z \gtrapprox 54$). That is, an accurate description of heavy elements requires spin-orbit coupling (SOC). 

To the best of our knowledge, we are the first to present the GIPAW formalism for a spin-orbit coupled ZORA relativistic treatment of the chemical shielding within linear response theory. This entails a modified Hamiltonian, including SOC, and additional contributions to the chemical shielding due to SOC. We extend the scalar relativistic ZORA GIPAW method of Yates, Pickard and Mauri (YPM).\cite{YPPM-ZORA2003} This results in a dual-approach in which orbital effects are treated through currents (as per YPM), and spin effects are captured as hyperfine interactions with an induced magnetization density. The latter is inspired by the work of \citet{dAvezac2007} on non-relativistic Knight shifts. The developed theory is implemented in the Vienna {\it Ab initio} simulation package (VASP)\cite{VASP_web, *VASP_1, *VASP_2, *VASP_3} as a continuation of our earlier work.\cite{deWijs2017}

It is worthwhile to mention at this point that there is an alternative to the direct approach of calculating the chemical shielding via linear response to an external magnetic field, namely the converse approach\cite{thonhauser2009converse}. In the converse approach, the chemical shielding is derived from the orbital magnetization induced by a magnetic dipole. In a parallel work, \citet{Zwanziger2025} included SOC in the chemical shielding following the converse approach and reported the implementation in a solid-state code.

The structure of the present paper is as follows: In Sect.~\ref{sec:method}, the chemical shielding tensor is defined and the ZORA Hamiltonian in a magnetic field is summarized in its all-electron (AE) form. Consecutively, we introduce the ZORA AE current and magnetization operators and demonstrate the validity of the dual approach. In Sect.~\ref{ssec:GIPAW}, the GIPAW Hamiltonian, orbital current, and magnetization operators are developed and applied in a linear response framework for extended and periodic systems. In Sect.~\ref{sec:results}, we benchmark our implementation on molecules with respect to accurate quantum-chemical methods. 
This includes a discussion on the limitations of the currently available PAW datasets in capturing orbital features relevant to the spin-orbit coupled description of the chemical shielding. 
We conclude by applying our method on a series of crystalline systems containing heavy elements and comparing them to results using cluster approximations.

\section{\label{sec:method}Method}
\subsection{\label{ssec:ChemSh}Chemical Shielding}
The chemical shielding tensor ${\overset \leftrightarrow {\bm{\sigma}}}\left( \bm{R} \right)$ is a $3\times3$ tensor is defined as the ratio between the induced magnetic field $\bm{B}_{\text{in}}$ and the externally applied magnetic field $\bm{B}_\text{ext}$ at the nuclear position $\bm{R}$:
\begin{equation}
    {\overset \leftrightarrow {\bm{\sigma}}}\left( \bm{R} \right) = -\frac{\bm{B}_{\text{in}}(\bm{R})}{\bm{B}_{\text{ext}}}.
\end{equation}
In the following, we represent the shielding tensor as a $3 \times 3$ matrix with Cartesian components.
The induced field is calculated as the first-order induced magnetic field, $\bm{B}^{(1)}_{\text{in}}$. It can be obtained from the Biot-Savart law:
\begin{equation}
    \bm{B}^{(1)}_{\text{in}}(\bm{R}) = \frac{1}{c} \int \bm{j}^{(1)}(\bm{r}') \bm{\times} \frac{\bm{R}-\bm{r}'}{|\bm{R}-\bm{r}'|^3} \ d^3\bm{r}',
    \label{eqn:BS}
\end{equation}
where $\bm{j}^{(1)}(\bm{r}')$ is the first-order induced current density that we compute within linear response theory.

First, we introduce the Hamiltonian with SOC for a system in a magnetic field. Subsequently, we explain the current operators and introduce an alternative magnetization-based approach for the spin-related currents.

\subsection{\label{ssec:AE-ZORA}All-Electron ZORA Theory}
The all-electron two-component ZORA Hamiltonian in an external field is given by:\cite{Ad-EPR}
\begin{equation}
    H = V + \bm{\sigma \cdot \pi} \frac{K}{2} \bm{\sigma \cdot \pi}
    \label{eqn:HamiltAE}
\end{equation}
where $\bm{\pi} = \bm{p} + \bm{A}/c$. For the vector potential $\bm{A}$, we adopt the symmetric gauge: $\bm{A} = (\bm{B \times r})/2$. The ZORA $K$-factor is
$
    K(\bm{r}) = \left( 1 - V(\bm{r})/(2c^2) \right)^{-1}
$. Here, and in the following, we use Gaussian CGS units and assume that $g_e = 2$. Solving the associated Kohn-Sham equations self-consistently yields two-component spinors $|\Psi_n\rangle$ where $n$ is the band index. The formalism and implementation of the two-component spinors \cite{hobbs2000fully} and the SOC term in the Hamiltonian\cite{Steiner2016} have been reported previously.

In a scalar relativistic calculation, the spin and orbital degrees of freedom are \textit{not} coupled. Therefore, the application of an external field $\bm{B}_{\text{ext}}$ results only in an orbital current, that consists of diamagnetic and paramagnetic contributions $\bm{\mathcal{J}}^p (\bm{r}')$ and $\bm{\mathcal{J}}^d (\bm{r}')$, viz.\ the scalar relativistic ZORA approach of \citet{YPPM-ZORA2003}:
\begin{gather}
    \bm{\mathcal{J}}^p (\bm{r}') =
        \frac{K(\bm{r}')}{2} (\bm{p} | \bm{r}' \rangle\langle \bm{r}' | + | \bm{r}' \rangle\langle \bm{r}' | \bm{p}),
    \label{eqn:Jrelp}\\
    \bm{\mathcal{J}}^d (\bm{r}') =
        -\frac{K(\bm{r}')}{c}
            \bm{A}(\bm{r}') | \bm{r}' \rangle\langle \bm{r}' |.
    \label{eqn:Jreld}
\end{gather}

However, SOC mixes spin and orbital degrees of freedom. This lifts the Kramers degeneracy and results in an additional induced magnetization $\bm{m}(\bm{r}')$ which gives rise to a corresponding spin current. Now, the induced current $\bm{j}(\bm{r}')$ consists of the induced orbital current and an induced spin current. We define the magnetization operator:
\begin{equation}
    \bm{M}(\bm{r}') = | \bm{r}' \rangle \frac{\bm{\sigma}}{2c} \langle \bm{r}' |,
    \label{eqn:M}
\end{equation}
the corresponding current operator is (see also Ref.~\onlinecite{Romaniello2006}):
\begin{equation}
    \bm{\mathcal{J}}^s(\bm{r}') = c K(\bm{r}') \bm{\nabla}' \times \bm{M}(\bm{r'}).
    \label{eqn:Jsm}
\end{equation}
Hence, the total induced current can be calculated from the spinors by:
\begin{equation}
    \bm{j}_\text{tot}(\bm{r}') =
    \sum_n^\text{occ}
        \langle \Psi_n |
            \bm{\mathcal{J}}^p (\bm{r}') + \bm{\mathcal{J}}^d (\bm{r}') + \bm{\mathcal{J}}^s(\bm{r}')
        | \Psi_n \rangle.
        \label{eqn:totcur}
\end{equation}
For paramagnetic systems - where there is no Kramers degeneracy - and metallic systems additional contributions arise which are beyond the scope of this paper.

In principle, the induced magnetic field can be calculated from the total induced current through the Biot-Savart law (Eq.~\ref{eqn:BS}) for both the scalar relativistic and spin-orbit coupled case. However, for the spin contribution, we find it more straightforward to follow an alternative route. We deviate from the YPM paradigm of currents and treat the spin contribution through the magnetization directly, inspired by \citet{dAvezac2007} We calculate the induced magnetic field of the induced magnetization $\bm{m}(\bm{r}')$ as Fermi-contact and dipolar fields. The equivalence is demonstrated in Appendix~\ref{app:equivalence}, where we obtain Eq.~\ref{eqn:curtomag} for the induced magnetic field emerging due to the induced magnetization:
\begin{equation}
    \begin{split}
        \bm{B}_{\bm{m}}(\bm{R})
        = \int &
            \frac{8\pi}{3} K(\bm{r}') \bm{m}(\bm{r}') \delta(\bm{R}-\bm{r}')
            \\&+
            \left[\bm{m}(\bm{r}')\bm{\cdot \nabla'} K(\bm{r}')\right] \frac{\bm{R}-\bm{r}'}{|\bm{R}-\bm{r}'|^3}
            \\&-
            \bm{m} (\bm{r}') \left[\frac{\bm{R}-\bm{r}'}{|\bm{R}-\bm{r}'|^3} \bm{\cdot\nabla'} K(\bm{r}')\right]
            \\&- \frac{K(\bm{r}')\bm{m}(\bm{r}')}{|\bm{R}-\bm{r}'|^3}
            \\&+
            3K(\bm{r}') \left[ \bm{m}(\bm{r}')\bm{\cdot}(\bm{R}-\bm{r}') \right]
            \frac{\bm{R}-\bm{r}'}{|\bm{R}-\bm{r}'|^5}
        \ d^3 r'.
    \label{eqn:BSM}
    \end{split}
\end{equation}
Here, we neglect the surface integral of Eq.~\ref{eqn:curtomag}. This surface integral relates to the macroscopic susceptibility, which we reflect on in section~\ref{sssec:Bmag}. The first three terms of Eq.~\ref{eqn:BSM} relate to the Fermi-contact interaction, while the latter two are the dipolar contributions.

We can compare these terms to the early relativistic generalization of the hyperfine interaction by \citet{Bluegel1987} Particularly, the second and third term of Eq.~\ref{eqn:BSM} correspond to the additional relativistic Fermi-contact terms defined in Eq.~27 of Ref.~\onlinecite{Bluegel1987}. Note that within ZORA, the ``classical'' Fermi-contact term (first term of Eq.~\ref{eqn:BSM}) vanishes because the ZORA $K$-factor becomes~0 at the nucleus\cite{Harriman1978, Wolff1999-ZORA, Autschbach2000}, however the additional relativistic Fermi-contact terms yield a finite contribution. The $\bm{\nabla'} K$ in Eq.~\ref{eqn:BSM} corresponds to the smeared out delta function in Refs.~\onlinecite{Bluegel1987, Bloechl2000}. 

In the non-relativistic limit ($K=1$) our expression for the induced magnetic field in Eq.~\ref{eqn:BSM} reduces to Eq.~5 of \citet{dAvezac2007} Note, they can treat the orbital and spin contributions separately as their Hamiltonian does not include SOC. In our case, the orbital Zeeman term of the Hamiltonian has an effect on the magnetization (see below). Therefore, we cannot treat the magnetization separately and use it within a formalism similar to the current contributions.

Summarizing, in the following the induced field is obtained as a sum of orbital and magnetization contributions:
\begin{equation}
    \bm{B}_\text{ind}(\bm{r}') =
    \bm{B}_\text{orb}(\bm{r}') +
    \bm{B}_{\bm{m}}(\bm{r}') .
\end{equation}
Where $\bm{B}_\text{orb}(\bm{r}')$ is obtained from the Biot-Savart law (Eq.~\ref{eqn:BS}) using the orbital current density as suggested by YPM, i.e.,
\begin{equation}
    \bm{j}_\text{orb}(\bm{r}') 
    = \sum_n^\text{occ}
    \langle \Psi_n |
    \bm{\mathcal{J}}^p(\bm{r}') +
    \bm{\mathcal{J}}^d(\bm{r}')
    | \Psi_n \rangle ,
    \label{eqn:orbcur}
\end{equation}
while $\bm{B}_{\bm{m}}(\bm{r}')$ is obtained using Eq.~\ref{eqn:BSM} with
\begin{equation}
    \bm{m}(\bm{r}')
    = \sum_n^\text{occ}
    \langle \Psi_n |
    \bm{M}(\bm{r}')
    | \Psi_n \rangle,
\end{equation}
where $\bm{M}$ is defined in Eq.~\ref{eqn:M}.

In the next section, we develop this dual approach into a linear response framework using the GIPAW method.

\subsection{\label{ssec:GIPAW}GIPAW}
When a system is translated in a magnetic field, the orbitals acquire a phase proportional to the magnetic field. The GIPAW method is an extension of the original projector augmented wave (PAW) method\cite{Bloechl1994} to account for the acquired phase factor. Here, we introduce the required operator expressions for the Hamiltonian, current density and magnetization employing the GIPAW method.\cite{PM2001}

Within GIPAW the ﬁeld-dependent transformation operator reads (Eq.~(16) in Ref.~\cite{PM2001})
\begin{equation}
    \mathcal{T}_{\bm{B}} = \bm{1} + \sum_{\bm{R},n} e^{\theta} \left[|\phi_{\bm{R},n}\rangle-|\tilde{\phi}_{\bm{R},n}\rangle \right] \left\langle\tilde{p}_{\bm{R},n}\right| e^{-\theta},
    \label{eqn:TB}
\end{equation}
where $\theta = (i/2c)\bm{r} \cdot (\bm{R} \times \bm{B})$, while $|\phi_{\bm{R},n}\rangle$ and $|\tilde{\phi}_{\bm{R},n}\rangle$ are the all-electron and pseudo partial waves, respectively. The $\left\langle\tilde{p}_{\bm{R},n}\right|$ are the usual PAW projector functions.

In all practical implementations of the PAW method, the all-electron partial waves are chosen to be solutions of the spherical scalar relativistic Kohn-Sham equation for a non-spinpolarized atom, at a specific energy $\epsilon_n$, and for a specific angular momentum $\ell_n$. The pseudo partial waves are smooth functions that are identical to the AE partial waves beyond the PAW radius $R_n$.

Using the field-dependent transformation operator (Eq.~\ref{eqn:TB}), any (semi-)local operator $O$ can be transformed to a GIPAW pseudo operator $\bar{O}$ as:
\begin{equation}
    \begin{split}
        \bar{O} &= \mathcal{T}^{\dagger}_{\bm{B}} O \mathcal{T}_{\bm{B}}\\
        &=O + 
        \sum_{\bm{R}, n, m}
            e^{\theta}
            | \tilde{p}_{\bm{R},n} \rangle
                \Big( 
                    \langle \psi_{\bm{R}, n} |
                        e^{-\theta}
                            O
                        e^{\theta}
                    | \psi_{\bm{R},m} \rangle
                    \\ &\qquad-
                    \langle \tilde{\psi}_{\bm{R}, n} |
                        e^{-\theta}
                            O
                        e^{\theta}
                    | \tilde{\psi}_{\bm{R},m} \rangle
                \Big)
            \langle \tilde{p}_{\bm{R},m} |
            e^{-\theta}.
            \label{eqn:GIPAWO}
    \end{split}
\end{equation}

We continue developing our GIPAW linear response framework by applying Eq.~\ref{eqn:GIPAWO} to yield the necessary GIPAW operators.

\subsubsection{GIPAW Hamiltonian}
We expand the AE Hamiltonian (Eq.~\ref{eqn:HamiltAE}) using Dirac's relation:\cite{DiracRelation}
\[\left(\sigma \cdot \bm{u}\right)\left(\sigma \cdot \bm{v}\right) = \bm{u}\cdot\bm{vI}+i\sigma\cdot\left(\bm{u}\times\bm{v}\right),\]
and limit the expansion to terms of zeroth and first order in the applied field $\bm{B}$ ($\bm{A}=\bm{B \times r}/2$):
\begin{align}
	H^{(0)}=\,\,&V+\bm{p \cdot}\frac{K}{2}\bm{p}+\bm{\sigma \cdot}\left(\bm{\nabla}\frac{K}{2}\bm{\times p}\right),\\
	\begin{split}
		H^{(1)}=\,\,& \frac{K}{2c} \bm{\sigma \cdot B}
		+ \frac{1}{4c} \left[K \bm{B \cdot L}+\bm{B \cdot L} K\right]
		\\&+\frac{1}{c}\bm{\sigma \cdot} \left(\bm{\nabla}\frac{K}{2}\bm{\times}\left(\frac{1}{2}\bm{B \times r}\right)\right),
        \label{eqn:H1}
	\end{split}
\end{align}
where $\bm{L} = \bm{r \times p}$, the angular momentum operator.

The zeroth-order Hamiltonian $H^{(0)}$ transforms to a standard PAW Hamiltonian:
\begin{equation}
    \bar{H}^{(0)}
    = V^{\text{loc}}(\bm{r}) +\frac{1}{2}\bm{p \cdot p} + \sum_{\bm{R}} V^{nl}_{\bm{R}} .
    \label{eqn:ppham_0}
\end{equation}
The first term on the right hand side is the local potential, the second is the kinetic energy, and the last is the non-local potential which is given by
\begin{equation}
    V^{nl}_{\bm{R}} = \sum_{n,m} |\tilde{p}_{\bm{R},n}\rangle
        a^{\bm{R}}_{n,m}
    \langle\tilde{p}_{\bm{R},m} | .
\end{equation}
The one-center strengths are:
\begin{equation}
    \begin{split}
        a^{\bm{R}}_{n,m}=&
                \langle \phi_{\bm{R},n} |
                    \left(
                        V(\bm{r})+\bm{p \cdot}\frac{K(\bm{r})}{2}\bm{p}
                        \right.\\&+\left.\frac{K(r)^2}{4c^2 r}\frac{dV(r)}{dr}\bm{\sigma \cdot L_R}
                    \right)
                | \phi_{\bm{R},m} \rangle
                \\&-
                \langle \tilde{\phi}_{\bm{R},n} |
                    \left(
                        V^{\text{loc}}(\bm{r})+\frac{1}{2}\bm{p \cdot p}
                    \right)
                | \tilde{\phi}_{\bm{R},m} \rangle .
    \end{split}
    \label{eqn:one_0}
\end{equation}
Relativistic effects are apparent in the matrix elements of the all-electron partial waves, where the kinetic energy now depends on the ZORA $K$-factor and the spin-orbit interaction is present.\cite{Steiner2016} The relativistic modifications to the operators in the pseudo partial wave strengths only occur close to the nucleus, where $K \rightarrow  0$. In a complete local partial waves basis they therefore completely cancel the relativistic modifications in the Hamiltonian acting directly on the plane waves.\cite{YPPM-ZORA2003} In Eqs.~\ref{eqn:ppham_0} and~\ref{eqn:one_0} this perfect cancellation is assumed, i.e., relativistic effects are only present in the AE partial wave strengths in Eq.~\ref{eqn:one_0}. Notably, the GIPAW transformation of the zeroth-order AE $H^{(0)}$ also generates a term that is first-order in $\bm{B}$:
\begin{equation}
    \bar{H}^{\left(0 \rightarrow 1\right)} = \frac{1}{2c} \left( \sum_{\bm{R}} \bm{R} \times \frac{1}{i} \left[ \bm{r}, V^{nl}_{\bm{R}} \right] \right) \cdot \bm{B}.
    \label{eqn:H0to1}
\end{equation}

Concerning the first order Hamiltonian, $H^{(1)}$ (Eq.~\ref{eqn:H1}), we adopt the terminology of the Pauli approximation. The first term of $H^{(1)}$ is then called the electron spin Zeeman term, the next two the orbital Zeeman interaction, and the last the spin-orbit Zeeman gauge correction.\cite{Ad-EPR} We rewrite the latter into two more convenient separate terms (see Ref.~\onlinecite{Wolff1999-ZORA}):
\begin{gather}
    \begin{gathered}
    \frac{1}{c}\bm{\sigma \cdot} \left(\bm{\nabla}\frac{K}{2}\bm{\times}\left(\frac{1}{2}\bm{B \times r}\right)\right)
    =\\
    \bm{\sigma \cdot B} \left(\bm{r \cdot} \nabla \frac{K-1}{4c}\right)
    -
    \bm{\sigma \cdot r} \left(\bm{B \cdot} \nabla \frac{K-1}{4c}\right)
    =\\
    \left\{\bm{\sigma} \left(\bm{r \cdot} \nabla \frac{K-1}{4c}\right)
    -
    (\bm{\sigma \cdot r}) \nabla \left(\frac{K-1}{4c}\right) \right\} \bm{\cdot B}.
    \end{gathered}
\end{gather}

We then perform the GIPAW transformation of $H^{(1)}$, and include the term first-order in $\bm{B}$ originating from $H^{(0)}$ (Eq.~\ref{eqn:H0to1}):
\begin{equation}
	\begin{split}
		\bar{H}^{(1)}
		=&
		\frac{1}{2c} \left(
			\bm{\sigma}
			+
			\bm{L}
			+
			\sum_{\bm{R}} \bm{R \times} \frac{1}{i} [\bm{r}, V^{nl}_{\bm{R}}]
		\right) \cdot \bm{B}
		\\+& \frac{1}{2c} \sum_{\bm{R}}
				\left[
					V^{nl}_{\bm{\sigma}, \bm{R}}
					+
					V^{nl}_{Q_{\bm{R}}, \bm{R}}
                    +
                    V^{nl}_{\perp, \bm{R}}
				\right]
			\cdot \bm{B}
		\label{eqn:ppham_1} .
	\end{split}
\end{equation}
Where, we have defined three potentials:
\begin{gather}
    \begin{split}
    V^{nl}_{\bm{\sigma}, \bm{R}}
    =
    \sum_{n,m}& | \tilde{p}_{\bm{R},n} \rangle
        a^{\bm{\sigma},\bm{R}}_{n,m}
    \langle \tilde{p}_{\bm{R},m} |
    \end{split},\\
    \begin{split}
    V^{nl}_{Q_{\bm{R}}, \bm{R}}
    =
    \sum_{n,m}& | \tilde{p}_{\bm{R},n} \rangle
        a^{Q_{\bm{R}},\bm{R}}_{n,m}
    \langle \tilde{p}_{\bm{R},m} |
    \label{eqn:Lr}
    \end{split},\\
    \begin{split}
    V^{nl}_{\perp, \bm{R}}
    =
    \sum_{n,m} | \tilde{p}_{\bm{R},n} \rangle
    \left(
        a^{s,\bm{R}}_{n,m} - a^{d,\bm{R}}_{n,m}
    \right)
    \langle \tilde{p}_{\bm{R},m} |
    \end{split}
    \label{eqn:Vperp},
\end{gather}
with one-center strengths
\begin{align}
    a^{\bm{\sigma},\bm{R}}_{n,m}
    &=
    \langle \phi_{\bm{R},n} | 
        K \bm{\sigma}
    | \phi_{\bm{R},m} \rangle
    -
    \langle \tilde{\phi}_{\bm{R},n} |
        \bm{\sigma}
    | \tilde{\phi}_{\bm{R},m} \rangle , \label{eqn:asig}\\
    a^{Q_{\bm{R}},\bm{R}}_{n,m}
    &=
    \langle \phi_{\bm{R},n} | 
    K \bm{L_R}
    | \phi_{\bm{R},m} \rangle
    -
    \langle \tilde{\phi}_{\bm{R},n} |
        \bm{L_R}		
    | \tilde{\phi}_{\bm{R},m} \rangle , \label{eqn:aQ}\\
    a^{s,\bm{R}}_{n,m}
    &=
    \langle \phi_{\bm{R},n} |
        \bm{\sigma} \left((\bm{r}-\bm{R})\bm{\cdot} \nabla \frac{K-1}{4c}\right)
   | \phi_{\bm{R},m} \rangle , \label{eqn:as}\\
    a^{d,\bm{R}}_{n,m}
    &=
    \langle \phi_{\bm{R},n} |
        \left(\bm{\sigma \cdot} (\bm{r}-\bm{R})\right) \nabla \left(\frac{K-1}{4c}\right)
   | \phi_{\bm{R},m} \rangle . \label{eqn:ad}
\end{align}
$\bm{L_R} = (\bm{r}-\bm{R}) \bm{\times p}$ is the angular momentum operator centered on atomic site $\bm{R}$. The three potentials of Eq.~\ref{eqn:asig}~-~\ref{eqn:ad} are linked to the electron spin Zeeman, orbital Zeeman, and spin-orbit Zeeman gauge correction terms, respectively. All potentials are defined with respect to an atomic site $\bm{R}$. Particularly, for Eq.~\ref{eqn:as} and \ref{eqn:ad}, we assume that $K(\bm{r})$ is spherically symmetric. In the non-relativistic limit ($c\rightarrow\infty$ and $K=1$), the contributions of Eq.~\ref{eqn:as} and \ref{eqn:ad} vanish, and Eq.~\ref{eqn:aQ} reduces to its non-relativistic form.\cite{YPM2007} Lastly, Eq.~\ref{eqn:asig} does not have a non-relativistic form.

For linear response, in addition to the Hamiltonian, we also need the overlap operator (Ref.~\onlinecite{YPM2007}, $S = \bar{1}$):
\begin{gather}
    S = 1 + \sum_{\bm{R}} e^{(i/2c)\bm{r \cdot R \times B}} Q_{\bm{R}} e^{(i/2c)\bm{r \cdot R \times B}} .
\end{gather}
The zeroth and first-order terms of the overlap operator are:
\begin{gather}
    S^{(0)} = 1 + \sum_{\bm{R}} Q_{\bm{R}} ,\\
    S^{(1)} = \frac{1}{2c} \sum_{\bm{R}} \bm{R} \frac{1}{i} \left[\bm{r}, Q_{\bm{R}} \right] \bm{\cdot B} .
\end{gather}
Here $Q_{\bm{R}}$ accounts for a non-vanishing augmentation charge:
\begin{gather}
    Q_{\bm{R}} = | \tilde{p}_{\bm{R},n} \rangle q_{\bm{R},nm} \langle \tilde{p}_{\bm{R},m} | ,\\
    q_{\bm{R},nm} = \langle \phi_{\bm{R},n} | \phi_{\bm{R},m} \rangle - \langle \tilde{\phi}_{\bm{R},n} | \tilde{\phi}_{\bm{R},m} \rangle.
\end{gather}
Together, $S^{(1)}$ and $\bar{H}^{(1)}$ are used to calculate the first-order perturbation to the wave function:
\begin{equation}
| \bar{\Psi}^{(1)}_n \rangle = \mathcal{G}(\varepsilon^{(0)}_o)(\bar{H}^{(1)} - \varepsilon^{(0)}_o S^{(1)})| \bar{\Psi}^{(0)}_o \rangle,
\end{equation}
where we use a Green's function:
\begin{equation}
	\mathcal{G}(\varepsilon) = \sum_e \frac{|\bar{\Psi}^{(0)}_e\rangle \langle \bar{\Psi}^{(0)}_e|}{\varepsilon - \varepsilon_e} 
\end{equation}
over empty orbitals, $e$. Rather than performing an explicit sum over empty states, we make use of a Sternheimer equation as demonstrated in Refs.~\onlinecite{PM2001} and \onlinecite{deWijs2017}.

At this point, we have developed the necessary operators to obtain the zeroth and first-order spinors, i.e. $| \bar{\Psi}^{(0)}_n \rangle$ and $| \bar{\Psi}^{(1)}_n \rangle$. Next, we develop the current operators to calculate the induced current.

\subsubsection{GIPAW Current Operators}
We apply the GIPAW transformation to the orbital current density, i.e., we consider Eq.~\ref{eqn:Jrelp} and \ref{eqn:Jreld} as a single operator
$
{\bm{\mathcal{J}}(\bm{r}') = \bm{\mathcal{J}}^p (\bm{r}') + \bm{\mathcal{J}}^d (\bm{r}')},
$
and separate terms into orders of $\bm{B}$:
\begin{align}
    \bar{\bm{J}}^{(0)}(\bm{r}') &=
        \bm{J}^p (\bm{r}')
        +
        \sum_{\bm{R}}
            \Delta \bm{\mathcal{J}}^p_{\bm{R}}(\bm{r}'),
    \label{eqn:J0}\\
    \begin{split}
        \bar{\bm{J}}^{(1)}(\bm{r}') &=
        \bm{J}^d (\bm{r}')
        +
        \sum_{\bm{R}}
            \Delta \bm{\mathcal{J}}^d_{\bm{R}} (\bm{r}')
            \\ & \quad+
            \frac{1}{2ci}
            \left[
                \bm{B \times R \cdot r}, \Delta \bm{\mathcal{J}}^p_{\bm{R}}(\bm{r}')
            \right].
        \label{eqn:J1}
    \end{split}
\end{align}
In addition to the ZORA relativistic operators denoted with $\bm{\mathcal{J}}(\bm{r}')$, we introduce their non-relativistic counterparts:
\begin{gather}
    \bm{J}^p (\bm{r}') = 
        \frac{1}{2} (\bm{p} | \bm{r}' \rangle\langle \bm{r}' | + | \bm{r}' \rangle\langle \bm{r}' | \bm{p}),
    \label{eqn:Jp}\\
    \bm{J}^d (\bm{r}') = 
        -\frac{1}{c} 
            \bm{A}(\bm{r}') | \bm{r}' \rangle\langle \bm{r}' | .
    \label{eqn:Jd}
\end{gather}
In the one-center operators \footnote{Within ultra-soft PAW, we distinguish between one-center terms and augmentation terms. We refer to all contributions involving partial waves in a single PAW sphere as ``one-center''. The term ``augmentation'' is reserved for the subset of contributions that specifically include the augmentation charges. Therefore, we refer to these operators as ``one-center'' and deviate from YPM who refer to these as ``augmentation'' operators.} the ZORA current densities are only kept in the matrix elements with AE partial waves:\cite{YPPM-ZORA2003}
\begin{gather}
    \begin{split}
        \Delta \bm{\mathcal{J}}^p_{\bm{R}}(\bm{r}') =&
            \sum_{n, m}
            | \tilde{p}_{\bm{R},n} \rangle
                \big[
                    \langle \phi_{\bm{R},n} |
                        \bm{\mathcal{J}}^p(\bm{r}')
                    | \phi_{\bm{R},m} \rangle
                \\&-
                    \langle \tilde{\phi}_{\bm{R},n} |
                        \bm{J}^p(\bm{r}')
                    | \tilde{\phi}_{\bm{R},m} \rangle
                \big]
            \langle \tilde{p}_{\bm{R},m} |  ,
    \end{split}\\
    \begin{split}
        \Delta \bm{\mathcal{J}}^d_{\bm{R}} (\bm{r}') =& 
        -\frac{\bm{A}(\bm{r}'-\bm{R}')}{c}
        \sum_{n,m}
            | \tilde{p}_{\bm{R},n} \rangle
                \big[
                \\&\langle \phi_{\bm{R},n} |
                    \bm{r}' \rangle K(\bm{r'}) \langle \bm{r}' 
                | \phi_{\bm{R},m} \rangle
                \\&-
                \langle \tilde{\phi}_{\bm{R},n} |
                                        \bm{r}' \rangle\langle \bm{r}' 
                | \tilde{\phi}_{\bm{R},m} \rangle
                \big]
            \langle \tilde{p}_{\bm{R},m} | .
    \end{split}
\end{gather}
The developed GIPAW current operators reflect the work of YPM.\cite{YPM2007, YPPM-ZORA2003} We reiterate that we treat the spin contributions by means of the magnetization rather than a spin current. Hence, we will now discuss the GIPAW transformation of the magnetization operator.
\subsubsection{\label{sec:Magnetization}GIPAW Magnetization Operator}
Although the magnetization operator is a local operator and could in principle be transformed by a regular PAW, we do a GIPAW transform because it allows for an easier formulation of equations compatible with periodic systems in a unit cell (as discussed in section~\ref{sec:periodic}).

We start from Eq.~\ref{eqn:M}, its GIPAW transformation gives:
\begin{equation}
    \begin{split}
        \bar{\bm{M}} (\bm{r}') = %
            \bm{M}(\bm{r}')
            +\sum_{\bm{R}}
                \Delta \bm{M}_{\bm{R}} (\bm{r}')
                \\
                +
                \frac{1}{2ci}
                \left[
                    \bm{B \times R \cdot r},
                    \Delta \bm{M}_{\bm{R}} (\bm{r}')
                \right],
    \end{split}
    \label{eqn:M-GIPAW}
\end{equation}
where we define a one-center operator:
\begin{equation}
    \begin{split}
    \Delta \bm{M}_{\bm{R}} (\bm{r}')
    =
    \sum_{n,m}&
        | \tilde{p}_{\bm{R},n} \rangle
            \left[
                \langle\phi_{\bm{R},n}|
                    \bm{r}' \rangle \frac{1}{2c} \bm{\sigma} \langle \bm{r}' 
                |\phi_{\bm{R},m}\rangle
            \right.\\&-\left.
                \langle\tilde{\phi}_{\bm{R},n}|
                    \bm{r}' \rangle \frac{1}{2c} \bm{\sigma} \langle \bm{r}' 
                |\tilde{\phi}_{\bm{R},m}\rangle
            \right]
        \langle \tilde{p}_{\bm{R},m}| .
    \end{split}
\end{equation}

We again separate into terms that are zeroth-order and first-order in $\bm{B}$:
\begin{gather}
    \bar{\bm{M}}^{(0)}(\bm{r}') = \bm{M}(\bm{r}') + \sum_{\bm{R}} \Delta \bm{M}_{\bm{R}} (\bm{r}'),
    \label{eqn:M0}\\
    \bar{\bm{M}}^{(1)}(\bm{r}') =
    \sum_{\bm{R}}
    \frac{1}{2ci}
    \left[
        \bm{B \times R \cdot r},
        \Delta \bm{M}_{\bm{R}} (\bm{r}')
    \right].
    \label{eqn:M1}
\end{gather}
Notably, the ZORA $K$-factor is absent in the magnetization operators. For the magnetization, the $K$-factor is only applied when calculating the induced field, see, e.g., Eq.~\ref{eqn:BSM}. 

\subsection{Linear Response}
In the previous subsections we separated the Hamiltonian and property operators into orders with respect to the magnetic field $\bm{B}$ and performed a GIPAW transformation on each of them. Now, we can consider the calculation of the induced magnetic field through linear response by evaluating the first-order current and magnetization.

Before we discuss how to calculate the properties using linear response, we refer to a technicality pointed out by Wolff and Ziegler.\cite{Wolff1998-Pauli} When a magnetic field is applied, the total energy no longer depends only on the zeroth-order electron density but also the first-order electron and current density. Consequently, the total energy should be calculated iteratively to account for these first-order contributions. This leads to what is referred to as ``coupled'' DFT. Neglecting such an iterative aspect is called ``uncoupled'' DFT.\cite{Malkin1993, Malkin1994} In our implementation, we first solve the unperturbed problem including the magnetic field and then apply linear response to obtain our current densities and magnetizations. Thus, we treat the total energy independent of the first-order electron and current density in an ``uncoupled'' DFT fashion.

In developing a linear response framework for the current contributions, we closely follow YPM. For ultra-soft pseudo potentials under GIPAW, the induced orbital current can be calculated as:\cite{YPM2007}
\begin{equation}
    \begin{split}
        &\bm{j}^{(1)}_\text{orb}(\bm{r}') =
            \\&2 \sum_o
            \text{Re} \left\{ \langle \bar{\Psi}^{(0)}_o |
                \bar{\bm{J}}^{(0)}(\bm{r}') \mathcal{G}(\varepsilon^{(0)}_o)(\bar{H}^{(1)} - \varepsilon^{(0)}_o S^{(1)})
            | \bar{\Psi}^{(0)}_o \rangle \right\}
            \\&-
             \sum_{o, o'}
            \langle \bar{\Psi}^{(0)}_o |
                \bar{\bm{J}}^{(0)}(\bm{r}')
            | \bar{\Psi}^{(0)}_{o'} \rangle
            \langle \bar{\Psi}^{(0)}_{o'} |
                S^{(1)}
            | \bar{\Psi}^{(0)}_o \rangle
            \\&+
             \sum_o
            \langle \bar{\Psi}^{(0)}_o |
                \bar{\bm{J}}^{(1)}(\bm{r}')
            | \bar{\Psi}^{(0)}_{o} \rangle,
    \label{eqn:jone}
    \end{split}
\end{equation}
where the sums over $o$ and $o'$ are over the set of occupied spinors. Note, because we sum over spinors instead of doubly occupied orbitals, the factors before our sums over $o$ and $o'$ are a factor two smaller than Ref.~\onlinecite{YPM2007}. 

We insert Eq.~\ref{eqn:J0} and \ref{eqn:J1} into Eq.~\ref{eqn:jone} and split $\bm{j}^{(1)}$ into  contributions:\cite{PM2001}
\begin{equation}
    \bm{j}^{(1)}_\text{orb}(\bm{r}') = \bm{j}^{(1)}_\text{bare} (\bm{r}') + \bm{j}^{(1)}_{\Delta d} (\bm{r}') + \bm{j}^{(1)}_{\Delta p} (\bm{r}'),
\end{equation}
the first term, $\bm{j}^{(1)}_\text{bare}$, is the plane wave contribution:
\begin{equation}
	\begin{split}
		&\bm{j}^{(1)}_\text{bare} (\bm{r}') =
			\\&2 \sum_o
			\text{Re} \left\{ \langle \bar{\Psi}^{(0)}_o |
				\bm{J}^p(\bm{r}') \mathcal{G}(\varepsilon^{(0)}_o)(\bar{H}^{(1)} - \varepsilon^{(0)}_o S^{(1)})
			| \bar{\Psi}^{(0)}_o \rangle \right\}
			\\&+
			\sum_o
			\langle \bar{\Psi}^{(0)}_o |
				-\frac{ \bm{A}(\bm{r}')}{c} | \bm{r}' \rangle\langle \bm{r}' |
			| \bar{\Psi}^{(0)}_{o} \rangle
			\\&-
			  \sum_{o, o'}
			\langle \bar{\Psi}^{(0)}_o |
				\bm{J}^p(\bm{r}')
			| \bar{\Psi}^{(0)}_{o'} \rangle
			\langle \bar{\Psi}^{(0)}_{o'} |
				S^{(1)}
			| \bar{\Psi}^{(0)}_o \rangle,
    \label{eqn:jbare}
	\end{split}
\end{equation}
the second and third term are the diamagnetic and paramagnetic one-center terms, respectively:
\begin{equation}
    \bm{j}^{(1)}_{\Delta d} (\bm{r}') =
        \sum_{\bm{R}', o}
        \langle \bar{\Psi}^{(0)}_o |
            \Delta \bm{\mathcal{J}}^d_{\bm{R}'} (\bm{r}')
        | \bar{\Psi}^{(0)}_{o} \rangle,
    \label{eqn:jdd}
\end{equation}

\begin{equation}
	\begin{split}
		&\bm{j}^{(1)}_{\Delta p} (\bm{r}') =
			\\&2 \sum_{\bm{R}', o}
			\text{Re} \left\{ \langle \bar{\Psi}^{(0)}_o |
				\Delta \bm{\mathcal{J}}^p_{\bm{R}'}(\bm{r}') \mathcal{G}(\varepsilon^{(0)}_o)(\bar{H}^{(1)} - \varepsilon^{(0)}_o S^{(1)})
			| \bar{\Psi}^{(0)}_o \rangle \right\}
			\\&-
			\sum_{\bm{R}', o, o'}
			\langle \bar{\Psi}^{(0)}_o |
				\Delta \bm{\mathcal{J}}^p_{\bm{R}'}(\bm{r}')
			| \bar{\Psi}^{(0)}_{o'} \rangle
			\langle \bar{\Psi}^{(0)}_{o'} |
				S^{(1)}
			| \bar{\Psi}^{(0)}_o \rangle
			\\&+
		      \sum_{\bm{R}', o}
			\langle \bar{\Psi}^{(0)}_o |
				\frac{1}{2ci}
				\left[
					\bm{B \times R' \cdot r}, \Delta \bm{\mathcal{J}}^p_{\bm{R}'}(\bm{r}')
				\right]
			| \bar{\Psi}^{(0)}_{o} \rangle.
    \label{eqn:jdp}
	\end{split}
\end{equation}

Several remarks about these equations are in place. First, $\bm{j}^{(1)}_\text{bare}$ contains both diamagnetic and paramagnetic terms, which experience different rates of convergence with respect to the basis set. Reason being that the diamagnetic term relates the ground state charge density whereas the paramagnetic term contains a sum over unoccupied states.\cite{YPM2007} Second, $\bm{j}^{(1)}_\text{bare}$ and $\bm{j}^{(1)}_{\Delta p}$ contain the expectation value of the position operator which is troublesome in periodic systems. Third, all three terms are individually invariant upon translation. Lastly, $\Delta \bm{\mathcal{J}}^d_{\bm{R}}$ and $\Delta \bm{\mathcal{J}}^p_{\bm{R}}$ have the subscript $\bm{R}'$ instead of $\bm{R}$. We do this to accommodate for further development of the formulas which introduces additional summations over ``$\bm{R}$'' through $\bar{H}^{(1)}$ and $S^{(1)}$.

The first issue can be solved by rewriting the diamagnetic contribution and using the $f$-sum rule (see appendix~\ref{app:fsum}). We can rewrite the diamagnetic contribution using $i \hbar | \bm{r}' \rangle \langle \bm{r}' | = [\bm{r}, \bm{J}^p(\bm{r}')]$:
\begin{equation*}
    \begin{split}
        &\sum_o
            \langle \bar{\Psi}^{(0)}_o |
                -\frac{ \bm{A}(\bm{r}')}{c} | \bm{r}' \rangle\langle \bm{r}' %
            | \bar{\Psi}^{(0)}_{o} \rangle
        \\&=
        \sum_o
            \frac{1}{2c}
            \langle \bar{\Psi}^{(0)}_o |
                \frac{1}{i} [ \bm{B \times r' \cdot r}, \bm{J}^p (\bm{r}') ]
            | \bar{\Psi}^{(0)}_{o} \rangle.
    \end{split}
\end{equation*}
Inserting this rewritten diamagnetic term in the $f$-sum rule transforms the fast converging diamagnetic term into two slower converging paramagnetic terms (Eq.~\ref{eqn:fsumj}). Ultimately, this can also be used to resolve the issue that the position operator is ill-defined for periodic systems. We elaborate on this in the discussion of equations for extended systems (Sec.~\ref{ssec:extended}).

For the linear response of the magnetization, we follow a similar path to the currents (Eq.~\ref{eqn:jone}). We can write for the magnetization:
\begin{equation}
	\begin{split}
		&\bm{m}^{(1)}(\bm{r}') =
			\\&2 \sum_o
			\text{Re} \left\{ \langle \bar{\Psi}^{(0)}_o |
				\bar{\bm{M}}^{(0)}(\bm{r}') \mathcal{G}(\varepsilon^{(0)}_o)(\bar{H}^{(1)} - \varepsilon^{(0)}_o S^{(1)})
			| \bar{\Psi}^{(0)}_o \rangle \right\}
			\\&-
			\sum_{o, o'}
			\langle \bar{\Psi}^{(0)}_o |
				\bar{\bm{M}}^{(0)}(\bm{r}')
			| \bar{\Psi}^{(0)}_{o'} \rangle
			\langle \bar{\Psi}^{(0)}_{o'} |
				S^{(1)}
			| \bar{\Psi}^{(0)}_o \rangle
			\\&+
			\sum_o
			\langle \bar{\Psi}^{(0)}_o |
				\bar{\bm{M}}^{(1)}(\bm{r}')
			| \bar{\Psi}^{(0)}_{o} \rangle.
	\label{eqn:Mone}
	\end{split}
\end{equation}
We insert Eq.~\ref{eqn:M0} and \ref{eqn:M1} and also split the magnetization into contributions:
\begin{equation}
    \bm{m}^{(1)}(\bm{r}') = \bm{m}^{(1)}_\text{bare} (\bm{r}') + \bm{m}^{(1)}_{\Delta M} (\bm{r}').
\end{equation}
With plane wave magnetization density, $\bm{m}^{(1)}_\text{bare}$:
\begin{equation}
    \begin{split}
        &\bm{m}^{(1)}_\text{bare} (\bm{r}')
        =
        \\&2 \sum_o
            \text{Re} \left\{ \langle \bar{\Psi}^{(0)}_o |
                \bm{M}(\bm{r}') \mathcal{G}(\varepsilon^{(0)}_o)(\bar{H}^{(1)} - \varepsilon^{(0)}_o S^{(1)})
            | \bar{\Psi}^{(0)}_o \rangle \right\}
        \\&-
        \sum_{o, o'}
            \langle \bar{\Psi}^{(0)}_o |
                \bm{M}(\bm{r}')
            | \bar{\Psi}^{(0)}_{o'} \rangle
            \langle \bar{\Psi}^{(0)}_{o'} |
                S^{(1)}
            | \bar{\Psi}^{(0)}_o \rangle,
    \label{eqn:mbare}
    \end{split}
\end{equation}
and the one-center contribution, $\bm{m}^{(1)}_{\Delta M}$:
\begin{equation}
    \begin{split}
        &\bm{m}^{(1)}_{\Delta \bm{M}} (\bm{r}')
        =
        \\&2 \sum_{\bm{R}', o}
            \text{Re} \left\{ \langle \bar{\Psi}^{(0)}_o |
            \Delta \bm{M}_{\bm{R}'} (\bm{r}') \mathcal{G}(\varepsilon^{(0)}_o)(\bar{H}^{(1)} - \varepsilon^{(0)}_o S^{(1)})
            | \bar{\Psi}^{(0)}_o \rangle \right\}
        \\&-
        \sum_{\bm{R}', o, o'}
            \langle \bar{\Psi}^{(0)}_o |
                \Delta \bm{M}_{\bm{R}'} (\bm{r}')
            | \bar{\Psi}^{(0)}_{o'} \rangle
            \langle \bar{\Psi}^{(0)}_{o'} |
                S^{(1)}
            | \bar{\Psi}^{(0)}_o \rangle
        \\&+
        \sum_{\bm{R}', o}
        \langle \bar{\Psi}^{(0)}_o |
            \frac{1}{2ci}
            \left[
                \bm{B \times R' \cdot r}, \Delta \bm{M}_{\bm{R}'}(\bm{r}')
            \right]
        | \bar{\Psi}^{(0)}_{o} \rangle.
    \label{eqn:mdm}
    \end{split}
\end{equation}
For the magnetization we experience similar problems regarding the position operators as previously mentioned for the current operators which will be addressed in the next section (Sec.~\ref{ssec:extended}).

However, for both the current and magnetization operators, part of these equations remain valid for extended system as the first-order Hamiltonian $\bar{H}^{(1)}$ contains terms localized within augmentation regions centered on atomic sites $\bm{R}$.\cite{YPM2007} As such, we can separate those terms from $\bar{H}^{(1)}$ (Eq.~\ref{eqn:ppham_0}), e.g.:
\begin{equation}
		\bar{H}^{(1)}
		=
		\frac{1}{2c} \left(
			\bm{L}
			+
			\sum_{\bm{R}} \bm{R \times} \frac{1}{i} [\bm{r}, V^{nl}_{\bm{R}}]
		\right) \bm{\cdot B}
		+ \bar{H}^{(1)}_{Q_{\bm{R}}},
\end{equation}
where we define:
\begin{equation}
	\bar{H}^{(1)}_{Q_{\bm{R}}} =
		\frac{1}{2c} \sigma \bm{\cdot B}
		+
		\frac{1}{2c} \sum_{\bm{R}}
		\left[
			V^{nl}_{\sigma, \bm{R}}
			+
			V^{nl}_{Q_{\bm{R}}, \bm{R}}
            +
            V^{nl}_{\perp, \bm{R}}
		\right]
		\bm{\cdot B} .
\end{equation}

By treating the Hamiltonian terms of $\bar{H}^{(1)}_{Q_{\bm{R}}}$ separately, we can separate parts of Eq.~\ref{eqn:jbare}, \ref{eqn:jdp}, \ref{eqn:mbare}, and \ref{eqn:mdm}. From these we can define $\bm{j}^{(1)}_{\text{bare}, Q_{\bm{R}}} (\bm{r}')$, $\bm{j}^{(1)}_{\Delta p, Q_{\bm{R}}} (\bm{r}')$, $\bm{m}^{(1)}_{bare, Q_{\bm{R}}} (\bm{r}')$, and $\bm{m}^{(1)}_{\Delta M, Q_{\bm{R}}} (\bm{r}')$, respectively. The benefit is that these terms can be evaluated for the atomic sites individually and do not need to account for periodicity.

\begin{equation}
    \bm{j}^{(1)}_{\text{bare}, Q_{\bm{R}}} (\bm{r}')
    =
    2 \sum_o
    \text{Re} \left\{ \langle \bar{\Psi}^{(0)}_o |
        \bm{J}^p(\bm{r}') \mathcal{G}(\varepsilon^{(0)}_o)\bar{H}^{(1)}_{Q_{\bm{R}}}
    | \bar{\Psi}^{(0)}_o \rangle \right\}
    \label{eqn:jbare-qr}
\end{equation}

\begin{equation}
    \bm{j}^{(1)}_{\Delta p, Q_{\bm{R}}} (\bm{r}') =
        2 \sum_{\bm{R}', o}
        \text{Re} \left\{ \langle \bar{\Psi}^{(0)}_o |
            \Delta \bm{\mathcal{J}}^p_{\bm{R}'}(\bm{r}') \mathcal{G}(\varepsilon^{(0)}_o)\bar{H}^{(1)}_{Q_{\bm{R}}}
        | \bar{\Psi}^{(0)}_o \rangle \right\}
        \label{eqn:jdp-qr}
\end{equation}

\begin{equation}
    \bm{m}^{(1)}_{\text{bare}, Q_{\bm{R}}} (\bm{r}')
    =
    2 \sum_o
    \text{Re} \left\{ \langle \bar{\Psi}^{(0)}_o |
        \bm{M}(\bm{r}') \mathcal{G}(\varepsilon^{(0)}_o)\bar{H}^{(1)}_{Q_{\bm{R}}}
    | \bar{\Psi}^{(0)}_o \rangle \right\}
    \label{eqn:mbare-qr}
\end{equation}

\begin{equation}
    \bm{m}^{(1)}_{\Delta \bm{M}, Q_{\bm{R}}} (\bm{r}')
    =
    2 \sum_{\bm{R}', o}
        \text{Re} \left\{ \langle \bar{\Psi}^{(0)}_o |
        \Delta \bm{M}_{\bm{R}'} (\bm{r}') \mathcal{G}(\varepsilon^{(0)}_o)\bar{H}^{(1)}_{Q_{\bm{R}}}
        | \bar{\Psi}^{(0)}_o \rangle \right\}
        \label{eqn:mdm-qr}
\end{equation}

\subsubsection{\label{ssec:extended}Extended Systems}
Transitioning from the previous molecular picture to an extended system, the main problem is that the position operator is ill-defined. However, position differences, e.g., $(\bm{r}-\bm{r}')$ are properly defined. On this basis, we can rewrite our equations to work for extended systems, following Ref.~\onlinecite{PM2001}.

Reformulating our equations to include such position differences is possible by using the $f$-sum rule (see Eq.~\ref{eqn:fsumg}). For $\mathbf{j}^{(1)}_{\text{bare}}$ (Eq.~\ref{eqn:jbare}), we applied the $f$-sum rule to convert the fast-converging diamagnetic term into two slower converging paramagnetic terms. The result is Eq.~\ref{eqn:fsumj}. Simultaneously, the result of this $f$-sum rule also allows to write position differences $(\bm{r}-\bm{r}')$ when expanding the various terms. We insert the $f$-sum rule result of Eq.~\ref{eqn:fsumj} into Eq.~\ref{eqn:jbare}, whilst also inserting the first-order overlap operator, and write:
\begin{widetext}
\begin{equation}
    \begin{split}
        \mathbf{j}^{(1)}_{\text{bare}}(\mathbf{r}') =&
        \ 2 \sum_o
        \text{Re} \left\{ \langle \bar{\Psi}^{(0)}_o |
            \mathbf{J}^p(\mathbf{r}') \mathcal{G}(\varepsilon^{(0)}_o)\left(
                \frac{1}{2c} \left(
                    (\mathbf{r}-\mathbf{r}')\times\mathbf{p}
                    +
                    \sum_\mathbf{R} (\mathbf{R}-\mathbf{r}') \times \frac{1}{i}
                    \left[
                        \mathbf{r},
                        V^{nl}_\mathbf{R}- \varepsilon^{(0)}_o Q_\mathbf{R}
                    \right]
                \right) \cdot \mathbf{B}
            \right)
        | \bar{\Psi}^{(0)}_o \rangle
        \right\}
        \\&-
        \sum_{\mathbf{R}, o, o'}
            \langle \bar{\Psi}^{(0)}_o |
                \mathbf{J}^{p}(\mathbf{r}')
            | \bar{\Psi}^{(0)}_{o'} \rangle
            \langle \bar{\Psi}^{(0)}_{o'} |
                (\mathbf{R} -\mathbf{r}')
                \times
                \frac{1}{2ci}		
                \left[
                    \mathbf{r},
                    Q_\mathbf{R}
                \right]
                \cdot \mathbf{B}
            | \bar{\Psi}^{(0)}_o \rangle.
    \label{eqn:jbarepos}
    \end{split}
\end{equation}
For $\mathbf{j}^{(1)}_{\Delta p} (\mathbf{r}')$ (Eq.~\ref{eqn:jdp}), the commutator $\left[\bm{B \times R' \cdot r}, \Delta \bm{\mathcal{J}}^p_{\bm{R}'}(\bm{r}')\right]/i$ should be used within the $f$-sum rule.
\begin{equation}
    \begin{split}
        \mathbf{j}^{(1)}_{\Delta p} (\mathbf{r}')
        =&
        \ 2 \sum_o
        \text{Re} \left\{ \langle \bar{\Psi}^{(0)}_o |
            \Delta \bm{\mathcal{J}}^p_\mathbf{R}(\mathbf{r}') \mathcal{G}(\varepsilon^{(0)}_o)\left(
                \frac{1}{2c} \left(
                    (\mathbf{r}-\mathbf{R}')\times\mathbf{p}
                    +
                    \sum_\mathbf{R} (\mathbf{R}-\mathbf{R}') \times \frac{1}{i}
                    \left[
                        \mathbf{r},
                        V^{nl}_\mathbf{R}- \varepsilon^{(0)}_o Q_\mathbf{R}
                    \right]
                \right) \cdot \mathbf{B}
            \right)
        | \bar{\Psi}^{(0)}_o \rangle
        \right\}
        \\&-
        \sum_{\mathbf{R}, \mathbf{R}', oo'}
        \langle \bar{\Psi}^{(0)}_o |
            \Delta \bm{\mathcal{J}}^p_\mathbf{R}(\mathbf{r}')
        | \bar{\Psi}^{(0)}_{o'} \rangle
        \langle \bar{\Psi}^{(0)}_{o'} |
            (\mathbf{R}-\mathbf{R}')			
            \frac{1}{2c} \times
            \left[
                \mathbf{r}, Q_\mathbf{R}
            \right]
            \cdot \mathbf{B}
        | \bar{\Psi}^{(0)}_o \rangle.
    \label{eqn:jdppos}
    \end{split}
\end{equation}
\end{widetext}

It is not immediately obvious how to formulate similar equations for the bare magnetization as the terms used within the $f$-sum rule for the bare current have no direct equivalents.
However, as $\bm{r}$ and $\bm{M}(\bm{r}')$ are respectively even and odd under time inversion, we use the commutator $
\left[
    \bm{B \times r' \cdot r},
    \bm{M}(\bm{r'})
\right]
=
0
$ in the $f$-sum rule, resulting in Eq.~\ref{eqn:fsumm}.
We can add the result as a ``$0$'' to the right-hand side of Eq.~\ref{eqn:mbare} to arrive at a difference $(\bm{r}-\bm{r}')$ for $\bm{m}^{(1)}_\text{bare} (\bm{r}')$:
\begin{widetext}
    \begin{equation}
        \begin{split}
            \bm{m}^{(1)}_\text{bare} (\bm{r}')=&
            \ 2 \sum_o
                \text{Re} \left\{ \langle \bar{\Psi}^{(0)}_o |
                    \bm{M}(\bm{r}') \mathcal{G}(\varepsilon^{(0)}_o)
                    \left(
                        (\bm{r}-\bm{r}') \times \bm{p}
                        +
                        \sum_{\bm{R}}
                            (\bm{R}-\bm{r}')
                            \times
                            \frac{1}{i}\left[
                                \bm{r},
                                V^{nl}_{\bm{R}} - \varepsilon^{(0)}_o Q_{\bm{R}}
                            \right]
                    \right)
                | \bar{\Psi}^{(0)}_o \rangle \right\} \bm{\cdot B}
            \\&-
            \sum_{o, o'}
                \langle \bar{\Psi}^{(0)}_o |
                    \bm{M}(\bm{r}')
                | \bar{\Psi}^{(0)}_{o'} \rangle
                \langle \bar{\Psi}^{(0)}_{o'} |
                    \sum_{\bm{R}}
                        (\bm{R}-\bm{r}')
                        \times
                        \frac{1}{2ci} \left[
                            \bm{r}, Q_{\bm{R}}
                        \right]
                | \bar{\Psi}^{(0)}_o \rangle \bm{\cdot B}.
        \end{split}
        \label{eqn:mbarepos}
    \end{equation}
    For $\bm{m}^{(1)}_{\Delta \bm{M}} (\bm{r}')$ we essentially do the same as what we did for $\mathbf{j}^{(1)}_{\Delta p} (\mathbf{r}')$ and use $\left[\bm{B \times R'\cdot r}, \Delta \bm{M}_{\bm{R}'}(\bm{r}')\right]$:
    \begin{equation}
        \begin{split}
            \bm{m}^{(1)}_{\Delta \bm{M}} (\bm{r}') =&
            \ 2 \sum_o
                \text{Re} \left\{ \langle \bar{\Psi}^{(0)}_o |
                    \Delta \bm{M}_{\bm{R}'}(\bm{r}') \mathcal{G}(\varepsilon^{(0)}_o)
                    \left(
                        (\bm{r}-\bm{R}') \times \bm{p}
                        +
                        \sum_{\bm{R}}
                            (\bm{R}-\bm{R}')
                            \times
                            \frac{1}{i}\left[
                                \bm{r},
                                V^{nl}_{\bm{R}} - \varepsilon^{(0)}_o Q_{\bm{R}}
                            \right]
                    \right)
                | \bar{\Psi}^{(0)}_o \rangle \right\} \bm{\cdot B}
            \\&-
            \sum_{o, o'}
                \langle \bar{\Psi}^{(0)}_o |
                    \Delta \bm{M}_{\bm{R}'}(\bm{r}')
                | \bar{\Psi}^{(0)}_{o'} \rangle
                \langle \bar{\Psi}^{(0)}_{o'} |
                    \sum_{\bm{R}}
                        (\bm{R}-\bm{R}')
                        \times
                        \frac{1}{2ci} \left[
                            \bm{r}, Q_{\bm{R}}
                        \right]
                | \bar{\Psi}^{(0)}_o \rangle \bm{\cdot B}.
        \end{split}
        \label{eqn:mdmpos}
    \end{equation}
\end{widetext}

\subsubsection{Periodic Systems\label{sec:periodic}}
In periodic systems, the expressions for extended systems can be reformulated using only the cell-periodic parts of the Bloch states. Following Ref.~\onlinecite{Vanderbilt1998} we can restrict the previous equations to a single unit cell by rewriting the position differences as:
\begin{equation}
    (\bm{r}-\bm{r}')
    =
    \lim_{q \to 0} \frac{1}{2q} \sum_{\iota=x,y,z}
    \left[
        e^{iq\hat{\bm{u}}_\iota \bm{\cdot} (\bm{r}-\bm{r}')}
        -
        e^{-iq\hat{\bm{u}}_\iota \bm{\cdot} (\bm{r}-\bm{r}')}
    \right].
    \label{eqn:real-recip}
\end{equation}

We return to the previously formulated equations for extended systems (Eq.~\ref{eqn:jbarepos}~-~\ref{eqn:mdmpos}), and insert Eq.~\ref{eqn:real-recip}. The wave functions can be written as Bloch functions, i.e., $| \bar{\Psi}^{(0)}_{n, \bm{k}} \rangle = e^{i \bm{k \cdot r}} | \bar{u}^{(0)}_{n, \bm{k}} \rangle$. The exponents of Eq.~\ref{eqn:real-recip} can be combined with the Bloch functions following Ref.~\onlinecite{Vanderbilt1998}. For the bare current, this gives:
\begin{equation}
    \bm{j}^{(1)}_{\text{bare}} (\bm{r}') = 
    \lim_{q \to 0} \frac{1}{2q}
        [ \bm{S^j}_{\text{bare}}(\bm{r}', q) - \bm{S^j}_{\text{bare}}(\bm{r}', -q) ]
    + \bm{j}^{(1)}_{\text{bare}, Q_{\bm{R}}} (\bm{r}')
    \label{eqn:jbareper}
\end{equation}
with
\begin{equation}
    \begin{split}
        \bm{S^j}_{\text{bare}} (\bm{r}', q) &=
        \frac{1}{c N_k} \sum_{\iota=x,y,z} \sum_{o, \bm{k}}
            \text{Re} \bigg\{ 
                \frac{1}{i}
                \langle \bar{u}^{(0)}_{o, \bm{k}} |
                    \bm{J}^p_{\bm{k}, \bm{k}+\bm{q}_\iota} (\bm{r}')
            \\&\times\left.
                    \mathcal{G}_{\bm{k}+\bm{q}_\iota}(\varepsilon_{o,\bm{k}})
                    \bm{B \times \hat{u}}_\iota \bm{\cdot}
                    \bm{v}_{\bm{k}+\bm{q}_\iota, \bm{k}} (\varepsilon_{o,\bm{k}})
                | \bar{u}^{(0)}_{o, \bm{k}} \rangle
            \right.\\&-\left.
            \sum_{o'}
                \langle \bar{u}^{(0)}_{o, \bm{k}} |
                    \bm{J}^p_{\bm{k}, \bm{k}+\bm{q}_\iota} (\bm{r}')
                | \bar{u}^{(0)}_{o', \bm{k}+\bm{q}_\iota} \rangle
                \right.\\&
                \langle \bar{u}^{(0)}_{o', \bm{k}+\bm{q}_\iota} |
                    \bm{B
                    \times
                    \hat{u}}_\iota
                    \bm{\cdot}
                    \bm{s}_{\bm{k}+\bm{q}_\iota, \bm{k}}
                | \bar{u}^{(0)}_{o, \bm{k}} \rangle
            \bigg\}.
    \end{split}
\end{equation}
and where $\bm{j}^{(1)}_{\text{bare}, Q_{\bm{R}}} (\bm{r}'$ is given by Eq.~\ref{eqn:jbare-qr}.
Additionally, we  reformulated the operators to be $\bm{k}$-dependent. $\bm{\tau}$ denotes the internal coordinates of the atoms. We write the augmentation charge as:
\begin{equation}
    Q_{\bm{k}, \bm{k}+\bm{q}_\iota} =
    \sum_{\bm{\tau}}
    \sum_{n,m}
        | \tilde{p}^{\bm{k}}_{\bm{\tau}, n} \rangle
            q^{\bm{\tau}}_{n,m}
        \langle \tilde{p}^{\bm{k}+\bm{q}_\iota}_{\bm{\tau}, m} |,
\end{equation}
for the velocity operator, we write:
\begin{equation}
    \begin{split}
	\bm{v}_{\bm{k}, \bm{k}+\bm{q}_\iota} (\varepsilon^{(0)}_{o, \bm{k}+\bm{q}_\iota})
	=&
	-i\bm{\nabla}
	+\bm{k}+\bm{q}_\iota
	\\&+ \frac{1}{i} \left[
		\bm{r}, V^{nl}_{\bm{k}, \bm{k}+\bm{q}_\iota} - \varepsilon^{(0)}_{o, \bm{k}+\bm{q}_\iota} Q_{\bm{k}, \bm{k}+\bm{q}_\iota}
	\right]
    \end{split},
\end{equation}
and lastly the paramagnetic operator becomes:
\begin{equation}
    \bm{J}^p_{\bm{k}, \bm{k}+\bm{q}_\iota} (\bm{r}')
    =
    - \frac{
		(-i \bm{\nabla} + \bm{k})
		| \bm{r} ' \rangle \langle \bm{r}' |
		+
		| \bm{r} ' \rangle \langle \bm{r}' |
		(-i \bm{\nabla} + \bm{k}+\bm{q}_\iota)
	}{2}.
\end{equation}

Similar to the bare current (Eq.~\ref{eqn:jbareper}), for the paramagnetic one-center current we obtain:
\begin{equation}
    \bm{j}^{(1)}_{\Delta p} (\bm{r}') = 
    \lim_{q \to 0} \frac{1}{2q}
        [ \bm{S^j}_{\Delta p}(\bm{r}', q) - \bm{S^j}_{\Delta p}(\bm{r}', -q) ]
    + \bm{j}^{(1)}_{\Delta p, Q_{\bm{R}}} (\bm{r}'),
    \label{eqn:jdpper}
\end{equation}
where
\begin{equation}
    \begin{split}
        \bm{S^j}_{\Delta p} (\bm{r}', q) &=
        \frac{1}{c N_k} \sum_{\iota=x,y,z} \sum_{o, \bm{k}}
            \text{Re} \left\{
                \frac{1}{i}
                \langle \bar{u}^{(0)}_{o, \bm{k}} |
                    \Delta \bm{\mathcal{J}}^{p}_{\bm{T}, \tau, \bm{k}, \bm{k}+\bm{q}_\iota}
                \right.\\&\times\left.
                    \mathcal{G}_{\bm{k}+\bm{q}_\iota}(\varepsilon_{o,\bm{k}})
                    \bm{B \times \hat{u}}_\iota \bm{\cdot}
                    \bm{v}_{\bm{k}+\bm{q}_\iota, \bm{k}} (\varepsilon_{o,\bm{k}})
                | \bar{u}^{(0)}_{o, \bm{k}} \rangle
            \right.\\&-\left.
            \sum_{o'}
                \langle \bar{u}^{(0)}_{o, \bm{k}} |
                    \Delta \bm{\mathcal{J}}^{p}_{\bm{T}, \tau, \bm{k}, \bm{k}+\bm{q}_\iota}
                | \bar{u}^{(0)}_{o', \bm{k}+\bm{q}_\iota} \rangle
                \right.\\&\left.\langle \bar{u}^{(0)}_{o', \bm{k}+\bm{q}_\iota} |
                    \bm{B
                    \times
                    \hat{u}}_\iota
                    \bm{\cdot}
                    \bm{s}_{\bm{k}+\bm{q}_\iota, \bm{k}}
                | \bar{u}^{(0)}_{o, \bm{k}} \rangle
            \right\}
    \end{split}
\end{equation}
with
\begin{equation}
    \begin{split}
        \Delta \bm{\mathcal{J}}^{p}_{\bm{T}, \tau, \bm{k}, \bm{q}_\iota}
        =&
        \sum_{n,m}
            | \tilde{p}^{\bm{k}}_{\tau, n} \rangle
                \left[
                    \langle \phi_{\bm{T}+\tau, n} |
                        \bm{\mathcal{J}}^p (\bm{r}')
                    |\phi_{\bm{T}+\tau, m} \rangle
                \right.\\&\left.-
                    \langle \tilde{\phi}_{\bm{T}+\tau, n} |
                        \bm{J}^p (\bm{r}')
                    | \tilde{\phi}_{\bm{T}+\tau, m} \rangle
                \right]
            \langle \tilde{p}^{\bm{k}+\bm{q}_\iota}_{\tau, m} |.
    \end{split}
    \label{eqn:Jdp-k}
\end{equation}
and where $\bm{j}^{(1)}_{\Delta p, Q_{\bm{R}}} (\bm{r}')$ is given by Eq.~\ref{eqn:jdp-qr}.
Compared to \citet{YPM2007}, Eq.~\ref{eqn:Jdp-k} presents a relativistic modification through the relativistic operator $\bm{\mathcal{J}}^p (\bm{r}')$.

For the magnetization we follow the same steps, which gives us
\begin{equation}
    \bm{m}^{(1)}_{\text{bare}} (\bm{r}') =
    \lim_{q \to 0} \frac{1}{2q}
    [ \bm{S^m}_{\text{bare}}(\bm{r}', q) - \bm{S^m}_{\text{bare}}(\bm{r}', -q) ]
    + \bm{m}^{(1)}_{\text{bare}, Q_{\bm{R}}} (\bm{r}'),
    \label{eqn:mbareper}
\end{equation}
where
\begin{equation}
    \begin{split}
        \bm{S^m}_{\text{bare}} (\bm{r}', q) &=
        \frac{1}{c N_k} \sum_{\iota=x,y,z} \sum_{o, \bm{k}}
            \text{Re} \Big\{
                \frac{1}{i}
                \langle \bar{u}^{(0)}_{o, \bm{k}} |
                    \bm{M}(\bm{r}')
            \\&\times\left.
                    \mathcal{G}_{\bm{k}+\bm{q}_\iota}(\varepsilon_{o,\bm{k}})
                    \bm{B \times \hat{u}}_\iota \bm{\cdot}
                    \bm{v}_{\bm{k}+\bm{q}_\iota, \bm{k}} (\varepsilon_{o,\bm{k}})
                | \bar{u}^{(0)}_{o, \bm{k}} \rangle
            \right.\\&-\left.
            \sum_{o'}
                \langle \bar{u}^{(0)}_{o, \bm{k}} |
                    \bm{M}(\bm{r}')
                | \bar{u}^{(0)}_{o', \bm{k}+\bm{q}_\iota} \rangle
                \right.\\&\langle \bar{u}^{(0)}_{o', \bm{k}+\bm{q}_\iota} |
                    \bm{B
                    \times
                    \hat{u}}_\iota
                    \bm{\cdot}
                    \bm{s}_{\bm{k}+\bm{q}_\iota, \bm{k}}
                | \bar{u}^{(0)}_{o, \bm{k}} \rangle \Big\}
            .
    \end{split}
\end{equation}
and where $\bm{m}^{(1)}_{\text{bare}, Q_{\bm{R}}} (\bm{r}')$ is given by Eq.~\ref{eqn:mbare-qr}.
Note that the magnetization operator does not attain a $\bm{k}$-dependence, because the phase factors commute with the operator and thus cancel.

Lastly, we do the same for the one-center magnetization:
\begin{equation}
    \bm{m}^{(1)}_{\Delta \bm{M}} (\bm{r}') =
    \lim_{q \to 0} \frac{1}{2q}
    [ \bm{S^m}_{\Delta \bm{M}}(\bm{r}', q) - \bm{S^m}_{\Delta \bm{M}}(\bm{r}', -q) ]
    + \bm{m}^{(1)}_{\Delta \bm{M}, Q_{\bm{R}}} (\bm{r}'),
    \label{eqn:mdmper}
\end{equation}
where
\begin{equation}
    \begin{split}
        \bm{S^m}_{\Delta \bm{M}}(\bm{r}', q) &=
        \frac{1}{c N_k} \sum_{\iota=x,y,z} \sum_{o, \bm{k}}
            \text{Re} \left\{
                \frac{1}{i}
                \langle \bar{u}^{(0)}_{o, \bm{k}} |
                    \Delta \bm{M}_{\bm{T}, \tau, \bm{k}, \bm{k}+\bm{q}_\iota}(\bm{r}')
                \right.\\&\times\left.
                    \mathcal{G}_{\bm{k}+\bm{q}_\iota}(\varepsilon_{o,\bm{k}})
                    \bm{B \times \hat{u}}_\iota \bm{\cdot}
                    \bm{v}_{\bm{k}+\bm{q}_\iota, \bm{k}} (\varepsilon_{o,\bm{k}})
                | \bar{u}^{(0)}_{o, \bm{k}} \rangle
            \right.\\&-\left.
            \sum_{o'}
                \langle \bar{u}^{(0)}_{o, \bm{k}} |
                    \Delta \bm{M}_{\bm{T}, \tau, \bm{k}, \bm{k}+\bm{q}_\iota}(\bm{r}')
                | \bar{u}^{(0)}_{o', \bm{k}+\bm{q}_\iota} \rangle
                \right.\\&\left.\langle \bar{u}^{(0)}_{o', \bm{k}+\bm{q}_\iota} |
                    \bm{B
                    \times
                    \hat{u}}_\iota
                    \bm{\cdot}
                    \bm{s}_{\bm{k}+\bm{q}_\iota, \bm{k}}
                | \bar{u}^{(0)}_{o, \bm{k}} \rangle
            \right\},
    \end{split}
\end{equation}
with
\begin{equation}
    \begin{split}
    \Delta \bm{M}_{\bm{T}, \tau, \bm{k}, \bm{k}+\bm{q}_\iota}(\bm{r}')
    =&
	\sum_{n,m}
		| \tilde{p}^{\bm{k}}_{\tau, n} \rangle
			\left[
				\langle \phi_{\bm{T}+\tau, n} |
                    \frac{\bm{\sigma}}{2c}
				|\phi_{\bm{T}+\tau, m} \rangle
            \right.\\&-\left.
				\langle \tilde{\phi}_{\bm{T}+\tau, n} |
                    \frac{\bm{\sigma}}{2c}
				| \tilde{\phi}_{\bm{T}+\tau, m} \rangle
			\right]
		\langle \tilde{p}^{\bm{k}+\bm{q}_\iota}_{\tau, m} |
    \end{split}.
\end{equation}
and where $\bm{m}^{(1)}_{\Delta \bm{M}, Q_{\bm{R}}} (\bm{r}')$ is given by Eq.~\ref{eqn:mdm-qr}.

For the implementation of Eq.~\ref{eqn:jbareper}, \ref{eqn:jdpper}, \ref{eqn:mbareper}, and \ref{eqn:mdmper} we treat each $k$-point as a ``$k$-point star''. This means we add two additional $k$-points at positive ($+q$) and negative ($-q$) offset in each Cartesian direction.

In short, the equations presented in this section are the final result for our linear response approach. We started from the GIPAW operators and showed how these should be implemented within linear response for ultra-soft PAW. We consecutively developed equations for molecular, extended and periodic systems. Use of the $f$-sum rule is critical to introduce position differences which ultimately enable the periodic description.\cite{Mauri1996-Shift} We will now discuss how the equations are implemented for calculating the induced magnetic field.

\subsection{Induced Magnetic Field}
From the expressions for the first-order induced currents and magnetization, the induced magnetic fields can be calculated. Similar to YPM, we make use of the linearity of Biot-Savart to calculate an induced field for each of contributions separately. This is a convenient approach given that we calculate parts of the induced magnetic fields from the induced current (Eq.~\ref{eqn:BS}) and parts from the magnetization (Eq.~\ref{eqn:BSM}). Thus, we calculate the induced magnetic fields due to each of the terms $\bm{j}^{(1)}_{\text{bare}}$, $\bm{j}^{(1)}_{\Delta d}$, $\bm{j}^{(1)}_{\Delta p}$, $\bm{m}^{(1)}_{\text{bare}}$, and $\bm{m}^{(1)}_{\Delta \bm{M}}$ (Eq.~\ref{eqn:jbareper}, \ref{eqn:jdd}, \ref{eqn:jdpper}, \ref{eqn:mbareper}, and \ref{eqn:mdmper}) on each atomic site and sum to the total induced field as:
\begin{equation}
    \begin{split}
        \bm{B}^{(1)}_{\text{in}} (\bm{R})
        =&
        \bm{B}^{(1)}_{\bm{j}_\text{bare}} (\bm{R})
        +
        \bm{B}^{(1)}_{\Delta d} (\bm{R})
        +
        \bm{B}^{(1)}_{\Delta p} (\bm{R})
        \\&+
        \bm{B}^{(1)}_{\bm{m}_\text{bare}} (\bm{R})
        +
        \bm{B}^{(1)}_{\Delta \bm{M}} (\bm{R}).
    \end{split}\label{eqn:Bcontribs}
\end{equation}
Contributions to the induced magnetic fields that stem from PAW one-center current- and magnetization densities, are treated in the so-called on-site approximation: the chemical shielding at a particular site is assumed to only be affected by the one-center contributions centered at the site itself.\cite{PM2001} Although the on-site approximation is less appropriate for ultra-soft PAW compared to norm-conserving PAW, it remains a very good approximation in general, as is demonstrated by close agreement with all-electron results.\cite{YPM2007} In specific cases going beyond the on-site approximation offers a systematic improvement (it can be achieved {\it via} augmentation currents on the plane wave grid).\cite{LLRAUG, Vasconcelos2013} We will first discuss how the current contributions are calculated and then cover the contributions from the magnetization.

\subsubsection{\label{sssec:Bcurr}Induced by current density}
In the aforementioned on-site approximation one may combine Eqs.~\ref{eqn:BS} and \ref{eqn:jdd} to write $\bm{B}^{(1)}_{\Delta d} (\bm{R})$ as:
\begin{equation}
    \bm{B}^{(1)}_{\Delta d} (\bm{R}) =
    \sum_{o, n, m}
        \langle \bar{\Psi}^{(0)}_o | \tilde{p}_{\bm{R},n} \rangle
            \bm{e}^{\bm{R}}_{m,n}
        \langle \tilde{p}_{\bm{R},m} | \bar{\Psi}^{(0)}_o \rangle,
\end{equation}
where the sum runs over the occupied ground-state spinors, and the one-center coefficients $\bm{e}^{\bm{R}}_{m,n}$ are
\begin{equation}
    \begin{split}
        \bm{e}^{\bm{R}}_{m,n} =&
        \langle \phi_{\bm{R},n} |
            \frac{
                K(\bm{r})(\bm{R}-\bm{r})\bm{\times}\left[\bm{B \times}(\bm{R}-\bm{r}) \right]
            }{
                2c^2 |\bm{R}-\bm{r}|^3
            }
        | \phi_{\bm{R},m} \rangle
        \\&-
        \langle \tilde{\phi}_{\bm{R},n} |
            \frac{
                (\bm{R}-\bm{r})\bm{\times}\left[\bm{B \times}(\bm{R}-\bm{r}) \right]
            }{
                2c^2 |\bm{R}-\bm{r}|^3
            }
        | \tilde{\phi}_{\bm{R},m} \rangle.
    \end{split}
\end{equation}

Similar on-site approximations are used for the induced magnetic field originating from the paramagnetic one-center current $\bm{B}^{(1)}_{\Delta p} (\bm{R})$. We illustrate this for the contribution to $\bm{B}^{(1)}_{\Delta p} (\bm{R})$ from $\bm{B}^{(1)}_{\Delta p, Q_R} (\bm{R})$ (Eq.~\ref{eqn:jdp-qr}):
\begin{equation}
    \bm{B}^{(1)}_{\Delta p, Q_R} (\bm{R}) =
    \sum_{o, n, m}
        \langle \bar{\Psi}^{(0)}_o | \tilde{p}_{\bm{R},n} \rangle
            \bm{f}^{\bm{R}}_{m,n}
        \langle \tilde{p}_{\bm{R},m} | \bar{\Psi}^{(1)}_o \rangle + \text{c.c.},
\end{equation}
where the one-center strength is
\begin{equation}
        \bm{f}^{\bm{R}}_{m,n} =%
        \langle \phi_{\bm{R},n} |
            \frac{
                K(\bm{r}) \bm{L_R}
            }{
                |\bm{r}-\bm{R}|^3
            }
        | \phi_{\bm{R},m} \rangle
        -
        \langle \tilde{\phi}_{\bm{R},n} |
            \frac{
                \bm{L_R}
            }{
                |\bm{r}-\bm{R}|^3
            }
        | \tilde{\phi}_{\bm{R},m} \rangle.
\end{equation}
In contrast to the diamagnetic contribution, where the operator is first-order in $\bm{B}$, the paramagnetic operator is zeroth-order in $\bm{B}$. As such, the coefficients $\bm{f}^{\bm{R}}_{m,n}$ interact with first-order spinors rather than zeroth-order as we saw for the diamagnetic contribution. Within the on-site approximation the coefficients $\bm{e}^{\bm{R}}_{m,n}$ and $\bm{f}^{\bm{R}}_{m,n}$ need only be calculated once for every atomic species.

The plane-wave contribution $\bm{B}^{(1)}_{\bm{j}_\text{bare}} (\bm{R})$ is evaluated by using Biot-Savart in reciprocal space:\cite{PM2001, YPM2007}
\begin{equation}
    \bm{B}^{(1)}_{\bm{j}_\text{bare}} (\bm{G}) = \frac{4\pi}{c}
    \frac{i\bm{G \times} \bm{j}^{(1)}_{\text{bare}}(\bm{G})}{G^2}.
    \label{eqn:Bbare}
\end{equation}
When evaluating this expression, a problem arises for the $\bm{G}=0$ component as this is not a bulk property.\cite{Mauri1996-Sus} In fact, the $\bm{G}=0$ component is affected by the surface currents of the sample. Generally, this surface integral should be treated by the proper microscopic surface current. %
For solids, we can account for this surface integral through the macroscopic susceptibility.\cite{PM2001} For a more elaborate discussion on the calculation of the macroscopic susceptibility in VASP see Ref.~\onlinecite{LLRAUG}.

\subsubsection{\label{sssec:Bmag}Induced by magnetization density}
The induced magnetic field contributions, $\bm{B}^{(1)}_{\bm{m}_\text{bare}} (\bm{R})$ and $\bm{B}^{(1)}_{\Delta \bm{M}} (\bm{R})$, are computed from $\bm{m}^{(1)}_{\text{bare}}(\bm{r}')$ and $\bm{m}^{(1)}_{\Delta \bm{M}}(\bm{r}')$ (Eqs.~\ref{eqn:mbareper} and \ref{eqn:mdmper}) using Eq.~\ref{eqn:BSM}. This is done in a manner that closely follows the method introduced by \citet{Bluegel1987} and \citet{Bloechl1994} for the evaluation of hyperfine parameters within the PAW method.

The particulars of the implementation and evaluation of hyperfine parameters in VASP are described by \citet{Szasz2013}.\footnote{Note they write the spin density as $\sigma$, and hence their $\sigma$ is related to our $\bm{M}(\bm{r})$ (Eq.~\ref{eqn:M}) by a mere unit conversion.}
We will not go into further detail here. Suffice it to say we calculate the Fermi-contact and dipolar-field contributions of Eq.~\ref{eqn:BSM} to $\bm{B}^{(1)}_{\bm{m}_\text{bare}} (\bm{R})$ and $\bm{B}^{(1)}_{\Delta \bm{M}} (\bm{R})$, from $\bm{m}^{(1)}_{\text{bare}}(\bm{r}')$ and $\bm{m}^{(1)}_{\Delta \bm{M}}(\bm{r}')$, respectively, in the manner described in Ref.~\onlinecite{Szasz2013}.

In principle, the bare magnetization $\bm{B}^{(1)}_{\bm{m}_\text{bare}} (\bm{R})$ also contributes to the macroscopic susceptibility through its $\bm{G} = \bm{0}$ component which would be the spin susceptibility.\cite{dAvezac2007} %
Our implementation does not include the spin susceptibility. In the supplementary material (Tab.~S22), we compare the macroscopic susceptibilities originating from the bare current for a selection of our calculated compounds to their experimental values.\cite{SusceptCRC105} Ultimately, contributions of the macroscopic susceptibility to the chemical shielding are negligible for these systems. Therefore, the lack of spin susceptibility should not be problematic in most cases.

At this point, we have explained how the derived theory is implemented to calculate the induced magnetic field and thereby the chemical shielding. In the following section, we discuss the relativistic nature of the PAW dataset, which proves crucial for the calculation chemical shielding.

\subsection{\label{sec:PAW}PAW Dataset}
As mentioned before, the all-electron partial waves in the (GI)PAW method (see Sec.~\ref{ssec:GIPAW}) are the solution to the spherical scalar relativistic Kohn-Sham equation for non-spin-polarized atoms. The neglect of spin-orbit relativistic effects in the construction of these partial waves has some important consequences in the context of this work.

In the presence of SOC, mixing of orbital and spin angular momentum splits the levels with quantum number $\ell$. Each level with $\ell>0$ is split into levels $j = \ell-\frac{1}{2}$ and $j = \ell + \frac{1}{2}$. All orbitals with $j=\ell-\frac{1}{2}$ and those with $\ell=0$ exhibit a divergence near the nucleus.

In standard (GI)PAW, the AE partial waves do not discriminate between $j=\ell-\frac{1}{2}$ and $j=\ell+\frac{1}{2}$, i.e., there is only a single radial partial wave $R_\ell(r)$ for each channel~$n$, obtained from solving the scalar relativistic Kohn-Sham equation.

For $\ell=0$ this partial wave exhibits the correct divergence for $r \rightarrow 0$. For each $\ell > 0$ channel, the scalar relativistic partial waves do not diverge for $r \rightarrow 0$. The divergent partial waves with $(\ell,j)=(\ell,\ell-\frac{1}{2})$ are not present. Therefore, divergences are only included for $s$-orbitals and not for higher $\ell$ orbitals.

\begin{figure}[h!]
    \centering
    \includegraphics[width=\linewidth]{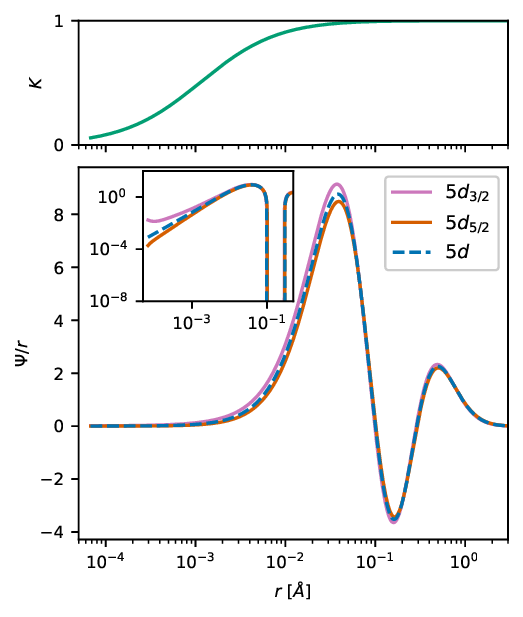}
    \caption{Upper panel: ZORA $K$-factor for Hg atom, plotted with $r$ on a logarithmic scale. Lower panel: scalar relativistic all-electron partial wave (dashed) versus SOC relativistic orbitals (solid). Plotted on a logarithmic radial scale. The insert shows the divergence of the $5d_{3/2}$ function near the nucleus.}
    \label{fig:HgrelOrb}
\end{figure}

Besides the difference in the divergent behavior for $r \rightarrow 0$ of the $(\ell,j)=(\ell,\ell-\frac{1}{2})$ states compared to their scalar relativistic counterparts.
The $j$-states (both $j = \ell + \frac{1}{2}$ as well as $j = \ell - \frac{1}{2}$ differ from the corresponding $l$-state outwards from the nucleus as well.
An example of this is shown in Fig.~\ref{fig:HgrelOrb}. This figure shows the radial part of the $(n=5, \ell=2)$ solutions for the scalar relativistic ($5d$) and spin-orbit coupled ($5d_{3/2}$ and $5d_{5/2}$) Kohn-Sham equations for an Hg atom as well as the corresponding ZORA $K$-factor. 

As one can see, the radial behavior of the Hg $5d_{3/2}$ and $5d_{5/2}$ states differs appreciably from the scalar relativistic $5d$ state for $10^{-3} < r < 10^{-1}$\AA.
Notably, the $5d_{3/2}$ partial wave has a small divergence near the origin which is not present in the default $5d$ partial wave nor in the SOC $5d_{5/2}$ partial wave. However, the ZORA $K$-factor, also shown in Fig.~\ref{fig:HgrelOrb}, dampens any divergence near the origin. The most significant differences between the partial waves are found between $10^{-3}$ and $10^{-1}$~\AA. Although the partial waves have very similar shape, relative to the default $5d^1$, $5d_{3/2}$ has a higher (lower) maximum (minimum) and $5d_{5/2}$ a lower (higher) minimum (maximum). The effects in this region are not completely dampened by the ZORA $K$-factor and could thus be reflected in the calculated chemical shielding.
Essentially, this means our partial wave basis is incomplete in a way that makes it not particularly well suited for spin-orbit coupled calculations.

Typically, the (GI)PAW method is not very sensitive to the incompleteness of the PAW partial wave basis: the one-center contributions within the (GI)PAW approach represent smallish corrections to the plane-wave part. However, for the calculation of chemical shielding tensors the one-center contributions are far more significant than is usually the case. As such, these calculations are more sensitive to the particulars of the partial wave basis and more affected by any incompleteness.

Ideally, the PAW partial wave basis would be made up of relativistic partial waves (i.e. $(\ell,j)=(\ell,\ell-\frac{1}{2})$ and $(\ell,j)=(\ell,\ell+\frac{1}{2})$). However, the current GIPAW implementation in VASP does not facilitate such a $j$-dependent basis and the necessary changes are not easily implemented. Nonetheless, we can take measures to improve the description available from our basis set.

As discussed above, the scalar relativistic partial waves in the PAW one-center basis of our PAW datasets differ from their relativistic counterpart in two ways: i) For $l > 0$ they do not show a divergence $r \rightarrow 0$, whereas $\ell,j)=(\ell,\ell-\frac{1}{2})$ states do, and ii) their radial behavior is generally different outward from (but still in close proximity to) the nucleus.

One might expect that the former (i) be of consequence to the description of hyperfine parameters:
\citet{Bluegel1987} and \citet{Bloechl1994} pay special attention to the divergent behavior of orbital and spin densities near the nucleus in their hyperfine implementations. There, the divergent $s$-orbitals give the dominant contribution to the Fermi-contact interaction. In order to adequately capture this divergence, \citet{Bloechl2000} extrapolates the spin density between the first radial grid point and the origin. Fortunately, the appropriate divergent $s$-type partial waves are included in our PAW dataset. We use these partial waves within Blöchl's extrapolation for the Fermi-contact contribution to the induced field originating from the magnetization.

Regarding the latter (ii):
we found that the paramagnetic one-center contribution ($\bm{B}^{(1)}_{\Delta p} (\bm{R})$) is particularly sensitive to details of the $\ell > 0$ partial waves away from the nucleus. Furthermore, there appears to be a considerable interplay with the ZORA $K$-factor. In section~\ref{ssec:OrbK} we show the effect of the different $j$-type partial waves on the paramagnetic one-center contribution. We also demonstrate how we can essentially introduce a cancellation of error by omitting the ZORA $K$-factor.

A final point of attention is the frozen core approximation. Although (GI)PAW does not necessitate the use of a frozen-core, virtually all implementations only include a limited number of valence electrons. Concerning chemical shielding calculations, Ref.~\onlinecite{Gregor1999} shows that the contribution from the frozen core is essentially rigid, i.e., one can add a constant $\sigma_\text{core}$ that only depends on the chemical species to the GIPAW shielding and obtain the total. It is not apparent that this result remains valid with SOC. We point out that GIPAW allows for unfreezing of shallow core states and include core-valence excitations that might be important. In VASP, the core shielding is only calculated for a frozen core at a scalar-relativistic level of theory.

\section{\label{sec:results}Results and Discussion}
Before we discuss the chemical shielding tensors calculated by VASP, we summarize its properties to facilitate the discussion. The chemical shielding tensor is a second-rank tensor which we calculate in a Cartesian axes system:
\begin{equation}
    {\overset \leftrightarrow {\bm{\sigma}}}\left[ \bm{R} \right]=
    \begin{bmatrix}
    \sigma_{xx} & \sigma_{xy} & \sigma_{xz}\\
    \sigma_{yx} & \sigma_{yy} & \sigma_{yz}\\
    \sigma_{zx} & \sigma_{zy} & \sigma_{zz}
    \end{bmatrix}.
\end{equation}
For every nucleus $\bm{R}$ we symmetrize and diagonalize the tensor to obtain the principal components. Under the frequency-ordered convention these are labeled as: $\sigma_{11} \leq \sigma_{22} \leq \sigma_{33}$. We follow the Herzfeld-Berger convention and define the isotropic shielding $\sigma_\text{iso}$ and span $\Omega$ as:\cite{Herzfeld-Berger}
\begin{gather}
    \sigma_\text{iso} = \frac{\sigma_{11}+\sigma_{22}+\sigma_{33}}{3}, \label{eqn:iso}\\
    \Omega = \sigma_{33}-\sigma_{11}. \label{eqn:span}
\end{gather}

In the following, we first discuss a molecular test series and compare to shieldings obtained with molecular codes. Following that, we elaborate on the (in)completeness of the PAW dataset with SOC with a focus on contributions of the paramagnetic one-center currents and demonstrate how the inaccuracies arising from incompleteness can be substantially reduced by suppressing the ZORA $K$-factor in these currents. We conclude by comparing periodic calculations of crystalline systems with VASP to experiment and cluster approximations from literature.

\subsection{\label{ssec:mol}Molecules}
The first benchmark for our implementation is a series of molecules. We compare to results of two molecular quantum chemical codes: Amsterdam Density Functional (ADF)\cite{ADF2001} and {\sc Dirac}.\cite{DiracMain, DIRAC19}

ADF and VASP evaluate chemical shieldings at the same ZORA level of theory and should give comparable results.
In VASP, we use a PAW basis and have a frozen core. The ADF calculations are all-electron and use Slater-type orbital basis sets.

{\sc Dirac} is a four-component relativistic code and represents our ``golden standard'' of relativistic effects. In {\sc Dirac} chemical shieldings are calculated through ``simple magnetic balance''.\cite{Olejniczak2012} {\sc Dirac} does not necessitate a frozen core. Its basis set is comprised of Gaussian-type orbitals.

The first molecular series we consider is the Hg series as used by \citet{Wolff1999-ZORA} Additionally, we computed two series of Sn and Pb halogen compounds. VASP results are for a plane-wave kinetic energy cutoff of 500~eV, using the recommended PAW datasets from the PBE.54 library. For Sn this is the ``Sn\_d'' dataset which has a valence shell with components $4d^{10}5s^{2}5p^{2}4f^{0}$. The ``Hg'' dataset contains $5d^{10}6d^{2}6p^{0}$ components, and ``Pb\_d'' $5d^{10}6s^{2}6p^{2}5f^{0}$. The uncoupled ADF results are for the QZ4P-J Slater-type basis set.\cite{STO2003} {\sc Dirac} results use the Dyall.4z Gaussian-type basis.\cite{Dyall2023} All calculations used the PBE functional.\cite{PBE1996,PBE1997}
For VASP we placed the molecules in $18 \times 18 \times 18$~\AA$^3$ supercells.

As discussed in the previous section, in VASP the shielding of the frozen core is calculated at the scalar relativistic level. This would result in a large discrepancy when comparing isotropic shieldings with ADF and {\sc Dirac} results which do have a relativistic description of the core. To remedy this, we reference each of the series by subtracting the average of the series from each individual value, i.e., $\sigma_\text{iso} \leftarrow \sigma_\text{iso} - \langle \sigma_\text{iso} \rangle$. For the span, assuming there are no relativistic core (polarization) effects, the lack of a relativistic treatment of the core is not a problem.

\begin{figure*}
    \centering
    \includegraphics[]{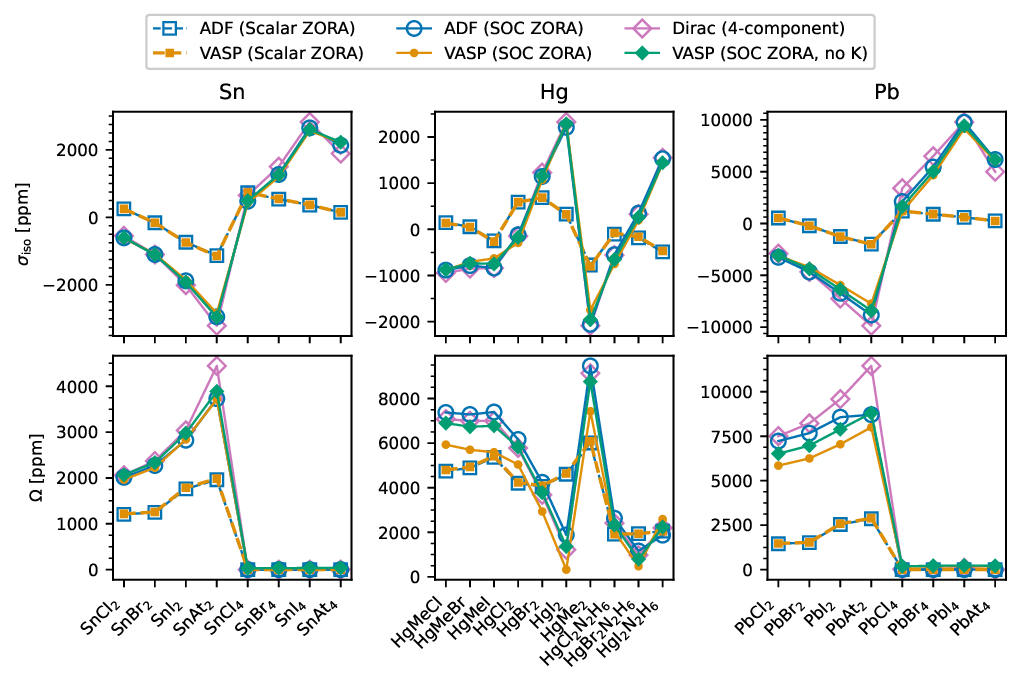}
    \caption{Isotropic shielding (top row) and span (bottom row) for a series of molecules containing Sn (left column), Hg (center column) and Pb (right column). Results generated on the level of scalar ZORA (squares) and with SOC ZORA (circles) DFT (i) with Amsterdam Density Functional (ADF)\cite{ADF2001} code (blue), (ii) with four-component theory in {\sc Dirac} code \cite{DiracMain, DIRAC19} (pink diamonds), and (iii) with VASP using the presently developed theory (orange and green). The green data points do not apply the ZORA~$K$ factor in the one-center current contributions.}
    \label{fig:molsys}
\end{figure*}

Figure~\ref{fig:molsys} shows the isotropic shielding $\sigma_{\mathrm{iso}}$ (Eq.~\ref{eqn:iso}) and the span $\Omega$ (Eq.~\ref{eqn:span}) obtained for the Sn, Hg and Pb series using {\sc Dirac}, ADF and our VASP implementation. In particular, note the ``VASP (SOC ZORA no~$K$)'' series. For this (spin-orbit coupled) series, the ZORA $K$-factor is excluded in the one-center current operators (Eq.~\ref{eqn:J1}). We find that this gives much closer agreement with the SOC-ZORA ADF results. Section~\ref{ssec:OrbK} further discusses this finding.

Table~\ref{tab:molMAE} lists the mean absolute errors (MAE) between the VASP and ADF ZORA results from Fig.~\ref{fig:molsys}.
For the scalar relativistic results, we find excellent agreement between VASP and ADF with only small MAE for both the isotropic shielding and span.
Including SOC, figure~\ref{fig:molsys} already shows a considerable effect for the relatively light Sn nucleus. For Hg and Pb the effect of SOC is even more pronounced. Although the agreement between spin-orbit coupled VASP and ADF results is good, we see an increase in MAE in Tab.~\ref{tab:molMAE}. However, the MAE are still small compared to the variations in isotropic shielding (20000 ppm for Pb) and span.
If we remove the ZORA~$K$ from the one-center currents (``SOC ZORA (no K)'' in the table) we observe a significant improvement of the MAE. Only the MAE in the span of Sn increases, from 32.0~ppm to 70.5~ppm, but it remains relatively small.
\begin{table*}
\caption{Mean Absolute Error (MAE) of the isotropic shielding and span comparing VASP scalar and spin-orbit coupled molecular test series and a reference. First column: VASP (Scalar ZORA) - ADF (Scalar ZORA), second column: VASP (SOC ZORA) - ADF (SOC ZORA), third column: VASP (SOC ZORA no K) - ADF (SOC ZORA), fourth column: VASP (SOC ZORA no K) - Dirac (4-component).}
\label{tab:molMAE}
\begin{tabular}{ccddcddcddcdd}
\hline
\hline
VASP
 && \multicolumn{2}{c}{Scalar ZORA}
 && \multicolumn{2}{c}{SOC ZORA}
 && \multicolumn{2}{c}{SOC ZORA (no $K$)}
 && \multicolumn{2}{c}{SOC ZORA (no $K$)} \\
Reference
 && \multicolumn{2}{c}{ADF Scalar ZORA}
 && \multicolumn{2}{c}{ADF SOC ZORA}
 && \multicolumn{2}{c}{ADF SOC ZORA}
 && \multicolumn{2}{c}{4-component {\sc Dirac}}\\
 \cline{3-4}
 \cline{6-7}
 \cline{9-10}
 \cline{12-13}
 && \multicolumn{1}{c}{$\sigma_\text{iso}\ [\text{ppm}]$}
 &  \multicolumn{1}{c}{$\Omega\ [\text{ppm}]$}
 && \multicolumn{1}{c}{$\sigma_\text{iso}\ [\text{ppm}]$}
 &  \multicolumn{1}{c}{$\Omega\ [\text{ppm}]$}
 && \multicolumn{1}{c}{$\sigma_\text{iso}\ [\text{ppm}]$}
 &  \multicolumn{1}{c}{$\Omega\ [\text{ppm}]$}
 && \multicolumn{1}{c}{$\sigma_\text{iso}\ [\text{ppm}]$}
 &  \multicolumn{1}{c}{$\Omega\ [\text{ppm}]$} \\ \hline
Sn &&  5.6 &  8.7 &&  54.4 &   32.0 &&  27.8 &  70.5 && 169.9 &  99.3\\
Hg && 13.0 & 32.0 && 141.1 & 1285.1 &&  67.5 & 469.7 &&  82.6 & 163.1\\
Pb &&  7.0 & 12.1 && 623.5 &  718.2 && 313.4 & 379.0 && 943.8 & 909.3\\
\hline
\hline
\end{tabular}
\end{table*}

The results show that the lack of a relativistic core contribution is not problematic when referencing appropriately.

Comparing to results from 4-component {\sc Dirac}, larger discrepancies appear. Logically, we do not expect to match these results as they come from a much more rigorous relativistic treatment. Nonetheless, it is very promising to see that we do share the qualitative features of the Dirac series, i.e., that SOC-ZORA captures all trends well.

\begin{figure*}
    \centering
    \includegraphics[width=\linewidth]{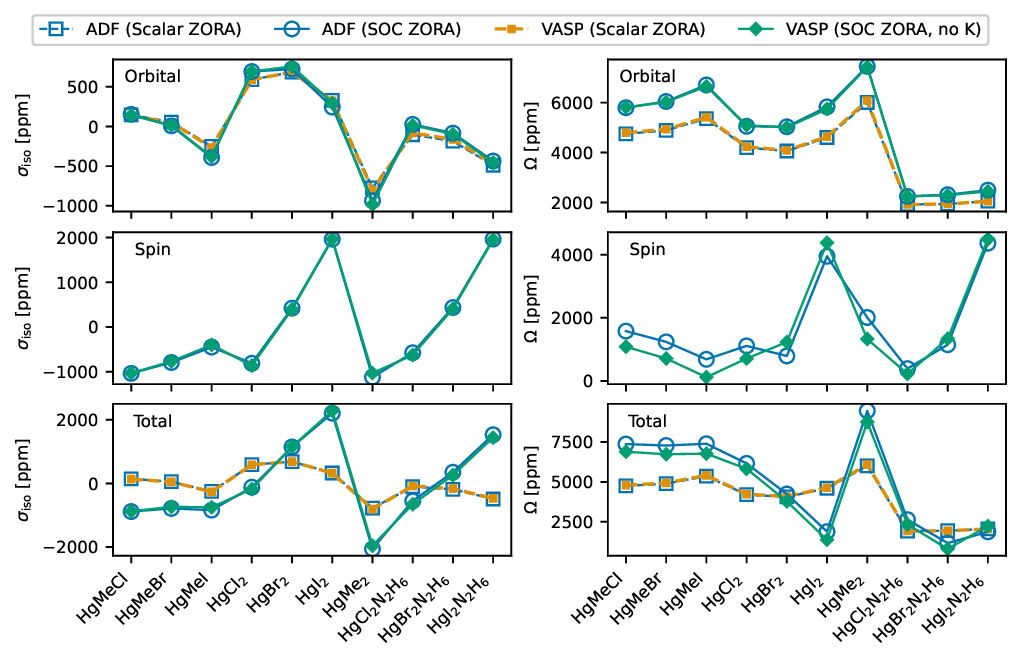}
    \caption{Isotropic value (left column) and span (right column) of the chemical shielding tensor for a series of molecules containing Hg. Top row shows the contribution from the diamagnetic and paramagnetic currents (orbital), the middle row those from the magnetization (spin), and the bottom row the total (orbital + spin) isotropic shielding and span. Scalar (squares) and spin-orbit coupled (circles) ZORA DFT with Amsterdam Density Functional (ADF)\cite{ADF2001} (blue), and VASP results with the presently developed theory (orange and green).  The green data points do not apply the ZORA~$K$ factor in the one-center current contributions.}
    \label{fig:molHg}
\end{figure*}

We would like to emphasize the importance of the contributions resulting from the SOC effects. Fig.~\ref{fig:molHg} shows ZORA orbital and magnetization (spin) contributions to the isotropic shielding and span of the chemical shielding separately for the Hg molecular series. In the orbital contribution trends are not affected, but the size of, e.g, the span changes by $\sim$10~\%. The magnetization contribution is entirely absent without SOC. Another striking feature is the systematic good agreement between the ADF SOC and VASP (SOC ZORA no~$K$, i.e., $K=1$) orbital current-derived contributions. This is discussed in the following section~\ref{ssec:OrbK}. Similar figures are available for Sn and Pb in the supplementary material (Fig.~S1 and S2).

\subsection{\label{ssec:OrbK} Relativity and PAW}
This subsection focuses on the difference between the ``VASP (SOC ZORA)'' and ``VASP (SOC ZORA no K)'' series. In section~\ref{sec:PAW} we discussed the relativistic nature of the PAW dataset and commented on the incompleteness for SOC calculations.

As discussed before, the VASP results show better agreement with ADF and Dirac when we omit the ZORA $K$-factor from the first-order current operators (i.e. the ``VASP (SOC ZORA no K)'' series). Upon closer inspection, we found that the paramagnetic operator attains the largest contributions from terms involving $p$ and $d$-type partial waves. For these $\ell > 0$ partial waves, the divergent $j=\ell-\frac{1}{2}$ partial waves are not present in the PAW datasets.

In order to investigate the effect of the divergent partial waves, we carried out a test where we modified the AE partial waves used in the calculation of the first-order paramagnetic one-center current (Eq.~\ref{eqn:jdp}). Specifically, we considered Sn and Hg atoms and inserted the correct $(\ell,j) = (\ell,\ell-\frac{1}{2})$ or $(\ell,\ell+\frac{1}{2})$ partial wave for each band. %
This substitution is only possible for atoms as each band much correspond to a distinct partial wave. %

\begin{figure}[h!]
    \centering
    \includegraphics[width=\linewidth]{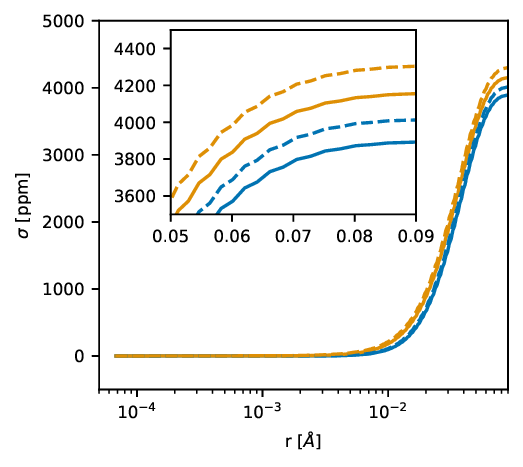}
    \caption{Chemical shielding resulting from a radial Biot-Savart integration of the paramagnetic one-center current for a spin-orbit coupled Hg atom with different PAW AE partial waves. Results generated with the default scalar relativistic PAW dataset in blue, and modified SOC all-electron partial waves in orange. Dashed lines omit the ZORA $K$-factor for the paramagnetic one-center current operator, solid lines include the $K$-factor. Large plot on a logarithmic radius, insert on a linear radius.}
    \label{fig:HgparaK}
\end{figure}

In Fig.~\ref{fig:HgrelOrb} we compared the $(n,\ell,j)=(5,2,\frac{3}{2})$ ``$5d_{3/2}$'' and  $(n,\ell,j)=(5,2,\frac{5}{2})$ ``$5d_{5/2}$'' non-standard partial waves to that present in the default Hg PAW dataset ($5d$). 
For the Hg atom test, we inserted the ``$5d_{3/2}$'' and  ``$5d_{5/2}$'' partial waves for the appropriate $n=5$ bands. The modified partial waves basis was used to calculate the first-order paramagnetic one-center current (i.e., only for evaluating the current in Eq.~\ref{eqn:jdp}) and the resulting magnetic field from the Biot-Savart law (Eq.~\ref{eqn:BS}), where we integrate from the nucleus to a distance~$r$. Thus we can explore the region where differences in the partial waves occur, before reaching the PAW sphere boundary. The resulting function of~$r$, converted to chemical shielding is shown in Fig.~\ref{fig:HgparaK}, using the modified and using the default partial waves.

For both the scalar relativistic default PAW dataset and the modified SOC relativistic AE partial waves, application of the ZORA $K$-factor decreases the calculated chemical shielding. Differences between the integrals appear around $10^{-2}$\,\AA, which is the region where we saw differences between the partial waves. Crucially, we find that the results for the default PAW dataset without ZORA $K$-factor are close to the SOC modified results with $K$-factor. This aligns with our previous observation in section~\ref{ssec:mol} that the ``VASP (SOC ZORA no K)'' series (using the default PAW dataset and omitting the $K$-factor from the one-center currents) agrees better with SOC ZORA ADF than the VASP series where we do apply the $K$-factor in the one-center currents.

Although the effects of the ZORA $K$-factor for Sn are less pronounced (Fig.~\ref{fig:molsys}), we can perform a similar demonstration substituting the default AE partial waves with their appropriate relativistic counterparts. Where for Hg we considered $\ell=2$ $d$-type partial waves, for Sn this involves $\ell=1$ $p$-type partial waves. Notably, the $5p_{3/2}$ orbitals of a Sn atom and thus the AE partial waves are unoccupied. Therefore, these were more cumbersome to acquire and interpret. Nonetheless, we found similar effects to the demonstration for Hg as can be seen in the~supplementary material (Fig.~S3 and S4).

Ultimately, it appears that we can exploit a cancellation of errors: omitting the ZORA $K$-factor compensates for the incorrect shape of the default AE partial waves. It is a rather unique situation for PAW that we see such a large effect of the one-center contributions. Typically, properties are dominated by the plane wave contribution and the one-center parts are really a correction. NMR shielding, however, forms an exception as it is a property that heavily depends on the electronic structure near to the nuclei and therefore presents a ``border-line'' case for the PAW approach. It is for this reason that we also experienced considerable dependence of the results on the PAW dataset used.

The shape of the AE partial waves is also crucial for the magnetization contribution to the shielding. Here the dominant contribution comes from the smeared-out delta function (second and third terms of Eq.~\ref{eqn:BSM}), i.e., arises from the $s$-type ($\ell=0$) magnetization density. On the nucleus this density is dominated by $s$-type partial waves, that exhibit strong divergences that are already accounted for in scalar relativistic partial waves, i.e., the standard PAW data sets. Some $\ell \ne 0$ partial waves also contribute, but the
$j=\ell-1/2$ partial waves typically have much weaker divergences. The lack of these may explain the remaining differences in the spin panels of Fig.~\ref{fig:molHg}.

The final part of our discussion covers solid-state systems. For the spin-orbit coupled results we omit the ZORA $K$-factor following the above considerations.

\subsection{Solids}
Evidently, the real application for our implementation lies with periodic systems. We demonstrate its effectiveness by comparing periodic calculations with VASP to experiment and a ADF cluster calculations by Alkan {\it et al.}, see, Ref.~\onlinecite{Alkan-Sn} (Sn), Ref.~\onlinecite{Alkan-Hg} (Hg), and Ref.~\onlinecite{Alkan-Pb} (Pb) and references therein.

VASP calculations again used the Sn\_d, Hg, and Pb\_d PAW PBE.54 datasets with a 500 eV cut-off energy and the PBE functional. The Sn ADF cluster calculations used the PBE functional, whereas the Hg and Pb cluster calculations were done with BP86.\cite{Perdew1986, Becke1988} Note that we have selected subsets
\footnote{The original series of Refs.~\onlinecite{Alkan-Sn}, \onlinecite{Alkan-Hg}, and \onlinecite{Alkan-Pb} include ambiguous structures, with, e.g., partially occupied crystallographic sites for hydrogen atoms. We omit these structures from the series presented in this article and refitted the data accordingly. Plots of the refitted subsets are available in the supplementary material (Fig.~S5-S7).}
from the sets of systems Refs.~\onlinecite{Alkan-Sn}, \onlinecite{Alkan-Hg}, and \onlinecite{Alkan-Pb}. An overview of the compounds included, their structures, and $k$-point meshes can be found in the supplementary material (Sec.~III.A).

For each series we consider the correlation of the principal components of the calculated chemical shielding $\sigma_{ii}$ and experimental chemical shift $\delta_{ii}$.
Ideally shielding and shift are related as:
\[
    \delta_{ii} = 
    \frac{\sigma_\text{iso}^\text{ref} - \sigma_{ii}}{1 - \sigma_\text{iso}^\text{ref}}.
\]
We carry out a linear fit using:
\begin{equation}
    \sigma_{ii}^\text{calc}=a\delta_{ii}^\text{exp}+\sigma_\text{ref}.
\label{eqn:fitshift}
\end{equation}
For an exact theoretical description, the slope $a = \sigma_\text{iso}^\text{ref} - 1 \approx -1$, and the vertical offset would give $\sigma_\text{ref}$. As DFT is not exact we do not expect such slopes and vertical offsets but it should be able to come close.
Our incomplete relativistic treatment of the core is accommodated by the fit as a different offset $\sigma_\text{ref}$ in as far as it is rigid.
Following the demonstration of the previous section (Sec.~\ref{ssec:OrbK}), spin-orbit coupled results presented here are generated omitting the ZORA $K$-factor from the first-order current operators. Scalar relativistic results do include the ZORA $K$-factor.

\begin{figure*}
    \centering
    \includegraphics[]{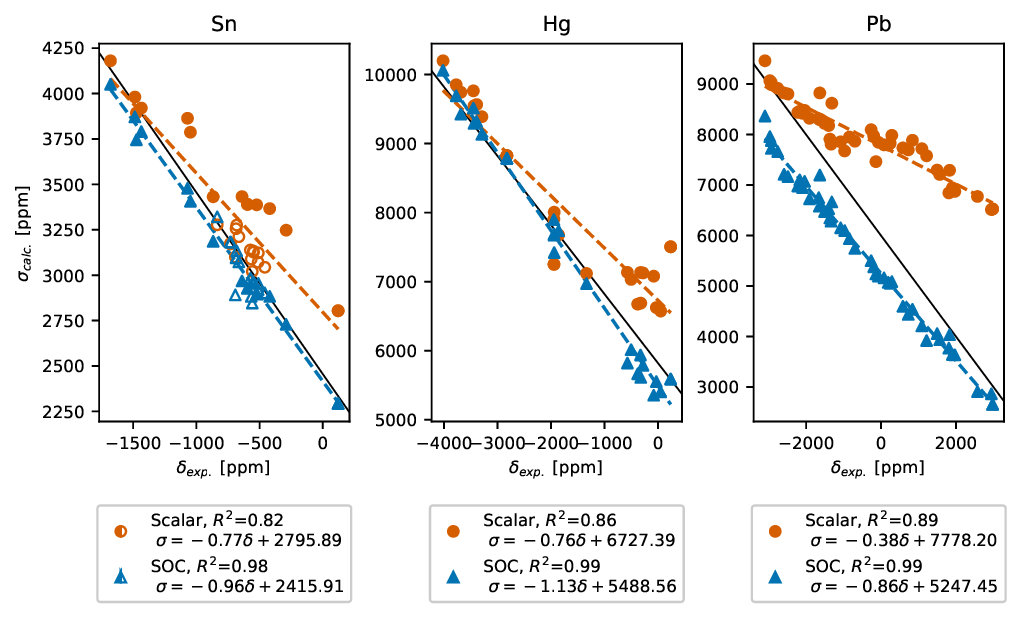}
    \caption{Calculated principal components of chemical shielding using VASP correlated to experimental chemical shift principal components for Sn, Hg and Pb compounds. Closed symbols represent M(II), open symbols M(IV), fits represented by dashed lines. Solid lines depict the ideal slope of ``$-1$''.}
    \label{fig:corrplot}
\end{figure*}

Figure \ref{fig:corrplot} clearly illustrates the improved agreement with experiment when using the SOC implementation.
Table~\ref{tab:corrtab} summarizes the fits to experiment of our data and those of the subsets of the Alkan \textit{et al.} series. We assess the fits using the coefficient of determination, $R^2$:
\[
    R^2 = 1 - \frac{SS_\text{res}}{SS_\text{tot}}
\]
where $SS_\text{res}$ is the sum of squares of residuals, and $SS_\text{tot}$ the total sum of squares.

\begin{table*}
\caption{Summary of fits for series of solids with scalar relativistic and SOC level of theory. $a$ and $\sigma_\text{ref}$ are the fit parameters of Eq.~\protect \ref{eqn:fitshift}. Fits are assessed using the coefficient of determination, $R^2 = 1 - {SS_\text{res}}/{SS_\text{tot}}$ where $SS_\text{res}$ is the sum of squares of residuals, and $SS_\text{tot}$ the total sum of squares.}
\label{tab:corrtab}
\begin{tabular}{lcdddcdddcdddcddd}
\hline \hline
\multirow{3}{*}{\begin{tabular}[c]{@{}l@{}}Compound\\ Series\end{tabular}} & & \multicolumn{7}{c}{Scalar relativistic} & & \multicolumn{7}{c}{SOC} \\
\cline{3-9}
\cline{11-17}
 && \multicolumn{3}{c}{Alkan \textit{et al.}\footnotemark[1]\footnotemark[2]\footnotemark[3]}
 && \multicolumn{3}{c}{VASP}
 && \multicolumn{3}{c}{Alkan \textit{et al.}\footnotemark[1]\footnotemark[2]\footnotemark[3]}
 && \multicolumn{3}{c}{VASP} \\
 \cline{3-5}
 \cline{7-9}
 \cline{11-13}
 \cline{15-17}
 && \multicolumn{1}{c}{$a$}
 &  \multicolumn{1}{c}{$\sigma_\text{ref}\ [\text{ppm}]$}
 &  \multicolumn{1}{c}{$R^2$}
 && \multicolumn{1}{c}{$a$}
 &  \multicolumn{1}{c}{$\sigma_\text{ref}\ [\text{ppm}]$}
 &  \multicolumn{1}{c}{$R^2$}
 && \multicolumn{1}{c}{$a$}
 &  \multicolumn{1}{c}{$\sigma_\text{ref}\ [\text{ppm}]$}
 &  \multicolumn{1}{c}{$R^2$}
 && \multicolumn{1}{c}{$a$}
 &  \multicolumn{1}{c}{$\sigma_\text{ref}\ [\text{ppm}]$}
 &  \multicolumn{1}{c}{$R^2$} \\
 \hline
Sn(II)\footnotemark[1] & & -0.78 & 2714.09 & 0.93 & & -0.70 & 2969.45 & 0.96 & & -1.01 & 2825.00 & 0.99 & & -0.96 & 2404.63 & 0.99 \\
Sn(IV)\footnotemark[1]  & & -0.53 & 2628.89 & 0.38 & & -0.58 & 2788.15 & 0.57 & & -0.99 & 2869.33 & 0.63 & & -1.10 & 2333.05 & 0.71 \\
Sn\footnotemark[1]  & & -0.84 & 2554.08 & 0.88 & & -0.77 & 2795.89 & 0.82 & & -1.00 & 2847.32 & 0.97 & & -0.96 & 2415.91 & 0.98 \\
Hg\footnotemark[2]  & & -0.71 & 5847.25 & 0.85 & & -0.76 & 6727.39 & 0.86 & & -1.04 & 7973.09 & 0.99 & & -1.13 & 5486.56 & 0.99 \\
Pb\footnotemark[3]  & & -0.37 & 7059.72 & 0.88 & & -0.38 & 7778.20 & 0.89 & & -0.87 & 8642.73 & 0.98 & & -0.86 & 5247.45 & 0.99 \\ \hline \hline
\end{tabular}\\
\footnotemark[1]Ref.~\onlinecite{Alkan-Sn}, \footnotemark[2]Ref.~\onlinecite{Alkan-Hg}, \footnotemark[3]Ref.~\onlinecite{Alkan-Pb}
\end{table*}

In all cases, both slope~$a$ and $R^2$ are significantly improved by SOC over scalar relativistic results.
On the scalar relativistic level, the VASP and ADF cluster results agree well. For Sn discrepancies are a bit larger. Here the PAW partial waves are, in principle, a complete local basis set. We speculate that remaining discrepancies are due to finite size effects of the cluster model.

Briefly comparing this to results obtained when including the ZORA $K$-factor for SOC VASP results (plots and fits are available in the supplementary material (Fig.~S10)); For Sn we find a slope of $-0.91$ and $R^2$ of 0.98, for Hg $a=-0.99$ and $R^2=0.98$, and for Pb $a=-0.76$ and $R^2=0.99$. The slopes obtained for Sn and Pb in this fashion are worse, as is the $R^2$ value for Hg. These results are in line with our proposal to omit the ZORA $K$-factor from the one-center currents.

Including SOC, both codes yield comparable results for all three series. With GIPAW VASP $R^2$ is consistently better, although this is a small effect. This, again, could be due the fact that VASP uses a real, infinite crystal. The slopes, however, show a bit more variation with VASP. This might be attributed to the incompleteness of the AE PAW partial waves. However, already at the scalar relativistic level similar relative differences in slope are apparent. An important feature for the Sn series is the improved agreement between the sets of Sn(II) and Sn(IV) nuclei when going from scalar relativistic to spin-orbit coupled results, as also pointed out by \citet{Alkan-Sn}

\section{\label{sec:conclusion}Conclusion}
In this work, we have extended the GIPAW linear response formalism for the calculation of NMR chemical shielding for periodic systems to include SOC for the valence electrons at the ZORA level. We demonstrated the equivalence of the induced magnetic field generated by a spin current and that resulting from the corresponding magnetization density. In fact, the effect of the magnetization density, that arises as a consequence of the SOC, is described as induced dipolar and Fermi-contact hyperfine-like contributions. Orbital contributions to the chemical shielding are calculated from the induced currents.

SOC mixes orbital and spin angular momentum such that the $j$-quantum number associated with the total angular momentum becomes a good quantum number. However, the standard (GI)PAW approach does not accommodate $j$-quantum numbers. Indeed, our analysis of the relativistic nature of the PAW dataset reveals discrepancies in the shape of the default AE partial waves from their proper relativistic counterparts. To counteract this, we propose omitting the ZORA $K$-factor in the one-center current operators. This leads to a partial cancellation of errors, and we can achieve good agreement with other ZORA implementations.

In particular, we demonstrated excellent agreement with ZORA ADF calculations and qualitative agreement with four-component {\sc Dirac} results based on a series of molecular systems (Tab.~\ref{tab:molMAE}).

For crystalline systems, we benchmarked our method with experimental values and with values obtained based on a cluster approximation from the literature for a series of systems containing heavy elements, namely Sn, Hg, and Pb. The inclusion of SOC using either our method or a cluster approximation significantly improves the agreement with the experiment (Tab.~\ref{tab:corrtab}).
Compared to cluster approximations, our calculations are obtained employing only a unit cell of the crystalline systems and are, hence, much less computationally demanding.

Future improvement of our developed theory could be (a) the inclusion of the spin contribution to the macroscopic susceptibility, (b) a spin-orbit coupled relativistic treatment of the core shielding, (c) going from ``uncoupled'' to ``coupled'' DFT, (d) developing treatment of $j$-quantum numbers in GIPAW.

\begin{acknowledgements}
We thank Prof.~J.~R.~Yates for insightful discussions and Prof.~A.~P.~M.~Kentgens for his support.
\end{acknowledgements}
\appendix
\section{\label{app:equivalence}Equivalence of Approaches}
Within our developed theory, we propose to calculate the spin contribution to the chemical shielding through the magnetization rather than the spin current explicitly. Essentially, the spin current is the curl of the magnetization, and thus the resulting induced magnetic field should be equivalent. We demonstrate the equivalence of calculating the induced magnetic field by a current versus a magnetization by transforming the current approach to the magnetization approach.

\begin{widetext}
We start from  Biot-Savart (Eq.~\ref{eqn:BS}) and insert the expression for the spin current (Eq.~\ref{eqn:Jsm}):
\begin{equation}
        \bm{B}(\bm{r})
         = \frac{1}{c} \int 
        \left[c K(\bm{r}') \bm{\nabla}' \times \bm{m}(\bm{r'})\right] \times \frac{\bm{r}-\bm{r}'}{|\bm{r}-\bm{r}'|^3}
        \ d^3r' %
         = \int
        \left[ \bm{\nabla'\times m}(\bm{r}') \right]\bm{\times} K(\bm{r}')\frac{\bm{r}-\bm{r}'}{|\bm{r}-\bm{r}'|^3}
        \ d^3 r' .
\end{equation}

We can expand the integral into three terms:
\begin{equation}
    \bm{B}(\bm{r})
	=
    \int
    \left[ \bm{m}(\bm{r}') \bm{\cdot\nabla'} \right] K(\bm{r}')\frac{\bm{r}-\bm{r}'}{|\bm{r}-\bm{r}'|^3}
	- \bm{\nabla'} \left[ \bm{m}(\bm{r}') \bm{\cdot} K(\bm{r}')\frac{\bm{r}-\bm{r}'}{|\bm{r}-\bm{r}'|^3} \right]
	+ \left[ K(\bm{r}')\frac{\bm{r}-\bm{r}'}{|\bm{r}-\bm{r}'|^3} \bm{\cdot \nabla'} \right] \bm{m}(\bm{r}')
    d^3r'
   \label{eqn:substZORA}.
\end{equation}

We focus our attention on the last term on the RHS. We consider an individual Cartesian component and apply a vector identity twice:
\begin{equation}
    \nabla\cdot(\psi\bm{a}) = \bm{a}\cdot\nabla\psi+\psi\nabla\cdot\bm{a}.
\end{equation}
First with, $\psi = m_i(\bm{r})$ and $\bm{a} = K(\bm{r}') \frac{\bm{r}-\bm{r}'}{|\bm{r}-\bm{r}'|^3}$:

\begin{equation}
    \begin{split}
    \int
    \left[K(\bm{r'})\frac{\bm{r}-\bm{r}'}{|\bm{r}-\bm{r}'|^3}\right]\cdot\nabla'm_i(\bm{r}')
    d^3r'
    =
    \int
    \nabla'\cdot\left[m_i(\bm{r'})K(\bm{r}')\frac{\bm{r}-\bm{r}'}{|\bm{r}-\bm{r}'|^3}\right]
    +
    m_i(\bm{r'})\nabla'\cdot\left[K(\bm{r'})\frac{\bm{r}-\bm{r}'}{|\bm{r}-\bm{r}'|^3}\right]
    d^3r'.
    \end{split}
\end{equation}

The first term can be converted into a surface integral, which can later be generalized for the whole vector. For the second term, we apply the vector identity again with $\psi = K(\bm{r})$ and $\bm{a} = \frac{\bm{r}-\bm{r}'}{|\bm{r}-\bm{r}'|^3}$. 

\begin{equation}
  \begin{split}
  \int
    \left[K(\bm{r}') \frac{\bm{r}-\bm{r}'}{|\bm{r}-\bm{r}'|^3} \bm{\cdot \nabla'}\right]m_i(\bm{r}')
  d^3 r'
  =&
  \int
    K(\bm{r}') m_i(\bm{r}') \bm{\nabla' \cdot} \frac{\bm{r}-\bm{r}'}{|\bm{r}-\bm{r}'|^3}
    +
    m_i (\bm{r}') \frac{\bm{r}-\bm{r}'}{|\bm{r}-\bm{r}'|^3} \bm{\cdot\nabla'} K(\bm{r}')
  d^3 r'
  \\&\ +
  \int
  m_i(\bm{r'})\left[K(\bm{r}')\frac{\bm{r}-\bm{r}'}{|\bm{r}-\bm{r}'|^3} \cdot \bm{n}'\right]
  da\\
  \int
    \left[K(\bm{r}') \frac{\bm{r}-\bm{r}'}{|\bm{r}-\bm{r}'|^3} \bm{\cdot \nabla'}\right]\bm{m}(\bm{r}')
  d^3 r'
  =&  
  \int
    4\pi K(\bm{r}')\bm{m}(\bm{r}') \delta (\bm{r}-\bm{r'})
    +
    \bm{m} (\bm{r}') \frac{\bm{r}-\bm{r}'}{|\bm{r}-\bm{r}'|^3} \bm{\cdot\nabla'} K(\bm{r}')
  d^3 r'
  \\&\ +
  \int
  \bm{m}(\bm{r'})\left[K(\bm{r}')\frac{\bm{r}-\bm{r}'}{|\bm{r}-\bm{r}'|^3} \cdot \bm{n}'\right]
  da
  \end{split}\label{eqn:vecitwo}
\end{equation}

Continuing the derivation, we consider the first RHS term of Eq.~\ref{eqn:substZORA} twice. Again, we consider a single Cartesian component:
\begin{equation}
  \begin{split}
    &\left[ \bm{m}(\bm{r}') \bm{\cdot\nabla'} \right] K(\bm{r}')\frac{x-x'}{|\bm{r}-\bm{r}'|^3}
    =
    \left[m_x(\bm{r'})\frac{\partial}{\partial x'}+m_y(\bm{r'})\frac{\partial}{\partial y'}+m_z(\bm{r'})\frac{\partial}{\partial z'}\right]K(\bm{r}')\frac{x-x'}{|\bm{r}-\bm{r}'|^3}
    \\&=
    \frac{K(\bm{r}')m_x(\bm{r}')}{|\bm{r}-\bm{r}'|^3}
	  +
    3K(\bm{r}') \left[ \bm{m}(\bm{r}')\bm{\cdot}(\bm{r}-\bm{r}') \right]\frac{x-x'}{|\bm{r}-\bm{r}'|^5}
    \ +
    \left[ \bm{m}(\bm{r'}) \cdot \nabla' K(\bm{r'}) \right] \frac{\bm{r}-\bm{r'}}{|\bm{r}-\bm{r'}|^3}
  \end{split}
\end{equation}

However, we can also consider each Cartesian component in an arbitrarily small volume around $\bm{r}$, so that $\bm{m}$ is effectively constant, and $\bm{m}(\bm{r'})\approx\bm{m}(\bm{r})$. 
\begin{equation}
    \begin{split}
    \left[ \bm{m}(\bm{r}') \bm{\cdot\nabla'} \right]K(\bm{r}')\frac{x-x'}{|\bm{r}-\bm{r}'|^3}
    \approx
    \left[m_x(\bm{r})\frac{\partial}{\partial x'}+m_y(\bm{r})\frac{\partial}{\partial y'}+m_z(\bm{r})\frac{\partial}{\partial z'}\right]K(\bm{r}')\frac{x-x'}{|\bm{r}-\bm{r}'|^3}
    \end{split}
\end{equation}
which we can integrate over a sphere centred at $\bm{r}$ to give a contribution: $(4\pi/3)K(\bm{r'})m_x(\bm{r}')\delta(\bm{r}-\bm{r}')$. Combining these two approaches, and generalizing for a complete vector, we find:
\begin{equation}
    \begin{split}
        \left[ \bm{m}(\bm{r}') \bm{\cdot\nabla'} \right] K(\bm{r}')\frac{\bm{r}-\bm{r}'}{|\bm{r}-\bm{r}'|^3}
        =&
        -\frac{4\pi}{3}K(\bm{r'})\bm{m}(\bm{r}')\delta(\bm{r}-\bm{r}')
        +\frac{K(\bm{r}')\bm{m}(\bm{r}')}{|\bm{r}-\bm{r}'|^3}
        +3K(\bm{r}') \left[\bm{m}(\bm{r}')\bm{\cdot}(\bm{r}-\bm{r}') \right]\frac{\bm{r}-\bm{r}'}{|\bm{r}-\bm{r}'|^5}
        \\&+\left[ \bm{m}(\bm{r'}) \cdot \nabla' K(\bm{r'}) \right] \frac{\bm{r}-\bm{r'}}{|\bm{r}-\bm{r'}|^3}
    \end{split}\label{eqn:funny4pi}
\end{equation}

At this point, we have considered both the first and last term of Eq.~\ref{eqn:substZORA}. We note that the second term can be written as a surface integral. By inserting Eq.~\ref{eqn:vecitwo} and \ref{eqn:funny4pi}, and reordering the terms, we can write:
\begin{equation}
  \begin{split}
      \bm{B}(\bm{r})
      =&
      \int
      \frac{8\pi}{3}K(\bm{r'})\bm{m}(\bm{r}')\delta(\bm{r}-\bm{r}')
      +\frac{K(\bm{r}')\bm{m}(\bm{r}')}{|\bm{r}-\bm{r}'|^3}
      +3K(\bm{r}') \left[ \bm{m}(\bm{r}')\bm{\cdot}(\bm{r}-\bm{r}') \right]\frac{\bm{r}-\bm{r}'}{|\bm{r}-\bm{r}'|^5}
      d^3 r'
      \\&+  
      \int
        \left[ \bm{m}(\bm{r'}) \cdot \nabla' K(\bm{r'}) \right] \frac{\bm{r}-\bm{r'}}{|\bm{r}-\bm{r'}|^3}
        d^3 r'
        +
        \bm{m} (\bm{r}') \frac{\bm{r}-\bm{r}'}{|\bm{r}-\bm{r}'|^3} \bm{\cdot\nabla'} K(\bm{r}')
      d^3 r'
      \\&+
      \int
        \bm{m}(\bm{r'})\left[K(\bm{r}')\frac{\bm{r}-\bm{r}'}{|\bm{r}-\bm{r}'|^3} \cdot \bm{n}'\right]
        -
        \left[ \bm{m}(\bm{r}') \bm{\cdot} K(\bm{r}')\frac{\bm{r}-\bm{r}'}{|\bm{r}-\bm{r}'|^3} \right] \bm{n'}
      da
  \end{split}
\end{equation}
\end{widetext}

The two terms in the surface integral can be combined into a single term as:
\begin{equation*}
    \begin{split}
        &\bm{m}(\bm{r'})\left[K(\bm{r}')\frac{\bm{r}-\bm{r}'}{|\bm{r}-\bm{r}'|^3} \cdot \bm{n}'\right]
        -
        \left[ \bm{m}(\bm{r}') \bm{\cdot} K(\bm{r}')\frac{\bm{r}-\bm{r}'}{|\bm{r}-\bm{r}'|^3} \right]
        \\&=
        K(\bm{r'})\frac{\bm{r}-\bm{r}'}{|\bm{r}-\bm{r}'|^3} \bm{\times} \left( \bm{m}(\bm{r}') \bm{\times n'} \right).
    \end{split}
\end{equation*}

This means, that in the end we can write:
\begin{equation}
    \begin{split}
        \bm{B}(\bm{r})
        =&
        \int
            \frac{8\pi}{3} K(\bm{r}') \bm{m}(\bm{r}') \delta(\bm{r}-\bm{r}')
            - \frac{K(\bm{r}')\bm{m}(\bm{r}')}{|\bm{r}-\bm{r}'|^3}
            \\&+
            3K(\bm{r}') \left[ \bm{m}(\bm{r}')\bm{\cdot}(\bm{r}-\bm{r}') \right]
            \frac{\bm{r}-\bm{r}'}{|\bm{r}-\bm{r}'|^5}
            \\&+
            \left[\bm{m}(\bm{r}')\bm{\cdot \nabla'} K(\bm{r}')\right] \frac{\bm{r}-\bm{r}'}{|\bm{r}-\bm{r}'|^3}
            \\&-
            \bm{m} (\bm{r}') \left[\frac{\bm{r}-\bm{r}'}{|\bm{r}-\bm{r}'|^3} \bm{\cdot\nabla'} K(\bm{r}')\right]
        \ d^3 r'
        \\&+\int
            K(\bm{r'})\frac{\bm{r}-\bm{r}'}{|\bm{r}-\bm{r}'|^3} \bm{\times} \left( \bm{m}(\bm{r}') \bm{\times n'} \right)
        da.
    \label{eqn:curtomag}
    \end{split}
\end{equation}
Within the volume integral, we find the ``regular'' Fermi-contact term, the two dipolar terms, and the relativistic Fermi-contact terms. This result matches with what has been discussed for molecules.\cite{Autschbach2000, Wolff1999-ZORA} However, we now also have an additional surface integral. This surface integral is essentially Biot-Savart for a surface current.

\section{\label{app:fsum}The generalized $f$-sum rule}
In the original work of \citet{PM2001} on chemical shielding within GIPAW for norm-conserving pseudo potentials they introduced the $f$-sum rule to deal with the difference in convergence rate of the diamagnetic and paramagnetic terms as well as to convert the equations for extended systems into equations for periodic systems. \citet{YPM2007} adapted the $f$-sum rule to make it suitable for ultra-soft pseudo potentials. Within the present work, the $f$-sum rule is also required in the same ways.

For ultra-soft pseudo potentials, the $f$-sum rule is:\cite{YPM2007}
\begin{equation}
    \begin{split}
        &\sum_o
            \langle \bar{\Psi}^{(0)}_o |
                \frac{1}{i} [ \mathcal{E}, \mathcal{O} ]
            | \bar{\Psi}^{(0)}_o \rangle
        =
        \\&-2 \sum_o
            \text{Re}\left[
                \langle \bar{\Psi}^{(0)}_o |
                    \mathcal{O} \mathcal{G}(\varepsilon^{(0)}_o) \frac{1}{i} \left[ \mathcal{E}, \bar{H}^{(0)} - \varepsilon^{(0)}_o S^{(0)} \right]
                | \bar{\Psi}^{(0)}_o \rangle
            \right]\\
        &+1 \sum_{o, o'} \frac{1}{i}
            \langle \bar{\Psi}^{(0)}_o |
                \mathcal{O}
            | \bar{\Psi}^{(0)}_{o'} \rangle
            \langle \bar{\Psi}^{(0)}_{o'} |
                [ \mathcal{E}, S^{(0)} ]
            | \bar{\Psi}^{(0)}_o \rangle,
    \end{split}
    \label{eqn:fsumg}
\end{equation}
where $\mathcal{O}$ and $\mathcal{E}$ are odd and even operators under time inversion, respectively.

We highlight two of the most important $f$-sum rules used within this article. First, it is used for the currents to convert the fast converging diamagnetic term into two slower converging paramagnetic terms:
\begin{widetext}
\begin{equation}
	\begin{split}
		-\sum_o
		\frac{1}{2c}
		\langle \bar{\Psi}^{(0)}_o |
			\frac{1}{i} [ \bm{B \times r' \cdot r}, \bm{J}^p (\bm{r}') ]
		| \bar{\Psi}^{(0)}_o \rangle
		=&
		2 \sum_o \text{Re} \left[
			\langle \bar{\Psi}^{(0)}_o | \bm{J}^p (\bm{r}') \mathcal{G}(\varepsilon^{(0)}_o)
			\frac{1}{2ci}
            \left[
                \bm{B \times r' \cdot r}, 
                (\bar{H}^{(0)} - \varepsilon^{(0)}_o S^{(0)})
            \right]
            | \bar{\Psi}^{(0)}_o \rangle
		\right]
		\\&-
		\sum_{o, o'}
			\langle \bar{\Psi}^{(0)}_o |
				\bm{J}^{p}(\bm{r}')
			| \bar{\Psi}^{(0)}_{o'} \rangle
			\langle \bar{\Psi}^{(0)}_{o'} |
				\frac{1}{2ci}		
				\left[
					\bm{B \times r' \cdot r},
					S^{(0)}
				\right]
			| \bar{\Psi}^{(0)}_o \rangle
	\end{split}.
    \label{eqn:fsumj}
\end{equation}

Second, we make use of the following $f$-sum rule to generate differences $(\bm{r}-\bm{r'})$ for the magnetization:
\begin{equation}
    \begin{split}
    \sum_o \frac{1}{2ci}
        \langle \bar{\Psi}^{(0)}_o |
            \left[
                \bm{B \times r' \cdot r},
                \bm{M}(\bm{r'})
            \right]
        | \bar{\Psi}^{(0)}_o \rangle
    =&
    -2 \sum_o \text{Re} \left[
        \langle \bar{\Psi}^{(0)}_o |
            \bm{M}(\bm{r}') \mathcal{G}(\varepsilon^{(0)}_o)
            \frac{1}{2ci} \left[
                \bm{B \times r' \cdot r},
                \bar{H}^{(0)} - \varepsilon^{(0)}_o S^{(0)}
            \right]
        | \bar{\Psi}^{(0)}_o \rangle
    \right]
    \\&+
    \sum_{o, o'} \frac{1}{2ci}
        \langle \bar{\Psi}^{(0)}_o |
            \bm{M}(\bm{r}')
        | \bar{\Psi}^{(0)}_{o'} \rangle
        \langle \bar{\Psi}^{(0)}_{o'} |
            \left[
                \bm{B \times r' \cdot r},
                S^{(0)}
            \right]
        | \bar{\Psi}^{(0)}_o \rangle
    \end{split}.
    \label{eqn:fsumm}
\end{equation}
\end{widetext}

The position differences $\left(\bm{r}-\bm{r}'\right)$ can be generated because the last commutator in the equations above can be rewritten as:
\begin{equation}
\left[\bm{B \times r' \cdot r},S^{(0)}\right]
=
\sum_{\bm{R}}\bm{r'}\times\left[\bm{r},Q_{\bm{r}}\right]\cdot\bm{B}.
\end{equation}

\bibliography{deriv}

\appendix
\input{SI_submit}
\end{document}

%% file: SI_submit.tex
\pagebreak
\widetext
\begin{center}
\textbf{\large Supplemental Materials: NMR chemical shielding for solid-state systems using spin-orbit coupled ZORA GIPAW}
\end{center}
\setcounter{equation}{0}
\setcounter{figure}{0}
\setcounter{table}{0}
\setcounter{page}{1}
\makeatletter
\renewcommand{\thetable}{S\arabic{table}}
\renewcommand{\theequation}{S\arabic{equation}}
\renewcommand{\thefigure}{S\arabic{figure}}
\renewcommand{\bibnumfmt}[1]{[S#1]}
\renewcommand{\citenumfont}[1]{S#1}
\renewcommand{\thesection}{\Roman{section}}
\renewcommand{\thesubsection}{\alph{subsection}}
\renewcommand{\thesubsubsection}{\arabic{subsubsection}}

\widetext

\section{Molecular Test Series}
\subsection{Geometries}
\subsubsection{Sn}

\paragraph{$\text{SnCl}_2$}
\begin{Verbatim}[samepage=true]
3
  
Sn   0.000000000   0.000000000   0.000000000
Cl   0.000000075   1.816669784   1.560274496
Cl  -0.000000015  -1.816669983   1.560274393
\end{Verbatim}

\FloatBarrier
\paragraph{$\text{SnBr}_2$}
\begin{Verbatim}[samepage=true]
3
  
Sn   0.000000000   0.000000000   0.000000000
Br  -0.000000126   1.945031423   1.630705862
Br  -0.000000165  -1.945031696   1.630705759
\end{Verbatim}

\FloatBarrier
\paragraph{$\text{SnI}_2$}
\begin{Verbatim}[samepage=true]
3
  
Sn   0.000000000   0.000000000   0.000000000
 I   0.000000000   2.133932361   1.757560456
 I   0.000000012  -2.133932375   1.757560406
\end{Verbatim}

\FloatBarrier
\paragraph{$\text{SnAt}_2$}
\begin{Verbatim}[samepage=true]
3
  
Sn   0.000000000   0.000000000   0.000000000
At  -0.000000009   2.207404925   1.838709243
At  -0.000000000  -2.207404932   1.838709240
\end{Verbatim}

\FloatBarrier
\paragraph{$\text{SnCl}_4$}
\begin{Verbatim}[samepage=true]
5
  
Sn  -0.000000000  -0.000083092  -0.000058755
Cl  -0.000000000  -1.892206576  -1.337992101
Cl  -0.000005826   1.892131258  -1.337826830
Cl   1.892054965   0.000082119   1.337934723
Cl  -1.892049139   0.000076292   1.337942963
\end{Verbatim}

\FloatBarrier
\paragraph{$\text{SnBr}_4$}
\begin{Verbatim}[samepage=true]
5
  
Sn  -0.000000000  -0.000016167  -0.000011432
Br  -0.000000000  -2.023025105  -1.430494770
Br   0.000000271   2.023090810  -1.430576260
Br   2.023115444  -0.000024905   1.430541422
Br  -2.023115715  -0.000024634   1.430541040
\end{Verbatim}

\FloatBarrier
\paragraph{$\text{SnI}_4$}
\begin{Verbatim}[samepage=true]
5
  
Sn   0.000000000   0.000012198   0.000008625
 I  -0.000000000  -2.215032951  -1.566264820
 I   0.000000043   2.215044670  -1.566290018
 I   2.215056607  -0.000011980   1.566273137
 I  -2.215056649  -0.000011937   1.566273076
\end{Verbatim}

\FloatBarrier
\paragraph{$\text{SnAt}_4$}
\begin{Verbatim}[samepage=true]
5
  
Sn  -0.000000000  -0.000006279  -0.000004440
At  -0.000000000  -2.336911723  -1.652446127
At   0.000000014   2.336924037  -1.652459100
At   2.336927047  -0.000003024   1.652454843
At  -2.336927061  -0.000003010   1.652454824
\end{Verbatim}

\subsubsection{Hg}
\FloatBarrier
\paragraph{HgMeCl}
\begin{Verbatim}[samepage=true]
6
  
Hg   0.000000000   0.000000000   0.000000000
Cl  -2.282000000   0.000000000   0.000000000
 C   2.061000000   0.000000000   0.000000000
 H   2.404797847   1.044893794   0.000000000
 H   2.404797847  -0.522446897   0.904904570
 H   2.404797847  -0.522446897  -0.904904570
\end{Verbatim}

\FloatBarrier
\paragraph{HgMeBr}
\begin{Verbatim}[samepage=true]
6
  
Hg   0.000000000   0.000000000   0.000000000
Br  -2.406000000   0.000000000   0.000000000
 C   2.074000000   0.000000000   0.000000000
 H   2.417797847   1.044893794   0.000000000
 H   2.417797847  -0.522446897   0.904904570
 H   2.417797847  -0.522446897  -0.904904570
\end{Verbatim}

\FloatBarrier
\paragraph{HgMeI}
\begin{Verbatim}[samepage=true]
6
  
Hg   0.000000000   0.000000000   0.000000000
 I  -2.528000000   0.000000000   0.000000000
 C   2.087000000   0.000000000   0.000000000
 H   2.430797847   1.044893794   0.000000000
 H   2.430797847  -0.522446897   0.904904570
 H   2.430797847  -0.522446897  -0.904904570
\end{Verbatim}

\FloatBarrier
\paragraph{$\text{HgCl}_2$}
\begin{Verbatim}[samepage=true]
3
  
Hg   0.000000000   0.000000000   0.000000000
Cl  -2.252000000   0.000000000   0.000000000
Cl   2.252000000   0.000000000   0.000000000
\end{Verbatim}

\FloatBarrier
\paragraph{$\text{HgBr}_2$}
\begin{Verbatim}[samepage=true]
3
  
Hg   0.000000000   0.000000000   0.000000000
Br  -2.410000000   0.000000000   0.000000000
Br   2.410000000   0.000000000   0.000000000
\end{Verbatim}

\FloatBarrier
\paragraph{$\text{HgI}_2$}
\begin{Verbatim}[samepage=true]
3
  
Hg   0.000000000   0.000000000   0.000000000
 I  -2.554000000   0.000000000   0.000000000
 I   2.554000000   0.000000000   0.000000000
\end{Verbatim}

\FloatBarrier
\paragraph{$\text{HgMe}_2$}
\begin{Verbatim}[samepage=true]
9
  
Hg   0.000000000   0.000000000   0.000000000
 C  -2.083000000   0.000000000   0.000000000
 C   2.083000000   0.000000000   0.000000000
 H  -2.471740000   1.028670000   0.000000000
 H  -2.471740000  -0.514335000   0.890854352
 H  -2.471740000  -0.514335000  -0.890854352
 H   2.471740000   1.028670000  -0.000000000
 H   2.471740000  -0.514335000   0.890854352
 H   2.471740000  -0.514335000  -0.890854352
\end{Verbatim}

\FloatBarrier
\paragraph{$\text{HgCl}_2\text{N}_2\text{H}_6$}
\begin{Verbatim}[samepage=true]
11
  
Hg   0.000000000   0.000000000   0.000000000
Cl   0.000000000   2.294070000   0.614700000
Cl   0.000000000  -2.294070000   0.614700000
 N   1.999510000   0.000000000  -1.450120000
 N  -1.999510000   0.000000000  -1.450120000
 H   2.793340000   0.000000000  -0.804630000
 H   2.083030000   0.831750000  -2.040160000
 H   2.083030000  -0.831750000  -2.040160000
 H  -2.793340000   0.000000000  -0.804630000
 H  -2.083030000   0.831750000  -2.040160000
 H  -2.083030000  -0.831750000  -2.040160000
\end{Verbatim}

\FloatBarrier
\paragraph{$\text{HgBr}_2\text{N}_2\text{H}_6$}
\begin{Verbatim}[samepage=true]
11
  
Hg   0.000000000   0.000000000   0.000000000
Br   0.000000000   2.417460000   0.625200000
Br   0.000000000  -2.417460000   0.625200000
 N   1.944660000   0.000000000  -1.490230000
 N  -1.944660000   0.000000000  -1.490230000
 H   2.757970000   0.000000000  -0.869130000
 H   2.012070000   0.832030000  -2.082500000
 H   2.012070000  -0.832030000  -2.082500000
 H  -2.757970000   0.000000000  -0.869130000
 H  -2.012070000   0.832030000  -2.082500000
 H  -2.012070000  -0.832030000  -2.082500000
\end{Verbatim}

\FloatBarrier
\paragraph{$\text{HgI}_2\text{N}_2\text{H}_6$}
\begin{Verbatim}[samepage=true]
11
  
Hg   0.000000000   0.000000000   0.000000000
 I   0.000000000   2.527280000   0.845620000
 I   0.000000000  -2.527280000   0.845620000
 N   1.948430000   0.000000000  -1.452080000
 N  -1.948430000   0.000000000  -1.452080000
 H   2.741400000   0.000000000  -0.804940000
 H   2.035000000   0.832050000  -2.042250000
 H   2.035000000  -0.832050000  -2.042250000
 H  -2.741400000   0.000000000  -0.804940000
 H  -2.035000000   0.832050000  -2.042250000
 H  -2.035000000  -0.832050000  -2.042250000
\end{Verbatim}

\clearpage
\subsubsection{Pb}
\FloatBarrier
\paragraph{$\text{PbCl}_2$}
\begin{Verbatim}[samepage=true]
3
  
Pb   0.000000000   0.000000000   0.000000000
Cl   0.000000071   1.818307733   1.553744881
Cl  -0.000000012  -1.818307878   1.553744771
\end{Verbatim}

\FloatBarrier
\paragraph{$\text{PbBr}_2$}
\begin{Verbatim}[samepage=true]
3
  
Pb   0.000000000   0.000000000   0.000000000
Br  -0.000000125   1.951232448   1.633846442
Br  -0.000000164  -1.951232513   1.633846285
\end{Verbatim}

\FloatBarrier
\paragraph{$\text{PbI}_2$}
\begin{Verbatim}[samepage=true]
3
  
Pb   0.000000000   0.000000000   0.000000000
 I   0.000000002   2.133882147   1.757405157
 I   0.000000014  -2.133882189   1.757405117
\end{Verbatim}

\FloatBarrier
\paragraph{$\text{PbAt}_2$}
\begin{Verbatim}[samepage=true]
3
  
Pb   0.000000000   0.000000000   0.000000000
At  -0.000000007   2.209921398   1.835891518
At   0.000000012  -2.209921410   1.835891502
\end{Verbatim}

\FloatBarrier
\paragraph{$\text{PbCl}_4$}
\begin{Verbatim}[samepage=true]
5
  
Pb  -0.000000000  -0.000083092  -0.000058755
Cl  -0.000000000  -1.892206576  -1.337992101
Cl  -0.000005826   1.892131258  -1.337826830
Cl   1.892054965   0.000082119   1.337934723
Cl  -1.892049139   0.000076292   1.337942963
\end{Verbatim}

\FloatBarrier
\paragraph{$\text{PbBr}_4$}
\begin{Verbatim}[samepage=true]
5
  
Pb   0.000000000   0.000034898   0.000024677
Br  -0.000000000  -2.023061492  -1.430520500
Br   0.000000268   2.023073304  -1.430561881
Br   2.023096525  -0.000023489   1.430529042
Br  -2.023096793  -0.000023221   1.430528662
\end{Verbatim}

\FloatBarrier
\paragraph{$\text{PbI}_4$}
\begin{Verbatim}[samepage=true]
5
  
Pb   0.000000000   0.000012198   0.000008625
 I  -0.000000000  -2.215032951  -1.566264820
 I   0.000000043   2.215044670  -1.566290018
 I   2.215056607  -0.000011980   1.566273137
 I  -2.215056649  -0.000011937   1.566273076
\end{Verbatim}

\FloatBarrier
\paragraph{$\text{PbAt}_4$}
\begin{Verbatim}[samepage=true]
5
  
Pb  -0.000000000  -0.000006279  -0.000004440
At  -0.000000000  -2.336911723  -1.652446127
At   0.000000014   2.336924037  -1.652459100
At   2.336927047  -0.000003024   1.652454843
At  -2.336927061  -0.000003010   1.652454824
\end{Verbatim}

\newpage
\subsection{Results}
\subsubsection{Sn}
\begin{table}[h!]
\caption{Calculated principal components in ppm for a series of molecules containing Sn using scalar ZORA VASP.
Excluding core shielding.}
\begin{tabular}{lrrrrrrrrr}
\hline
\hline
\multirow{2}{*}{\begin{tabular}[c]{@{}l@{}}Compound\end{tabular}} & \multicolumn{3}{c}{Orbital} & \multicolumn{3}{c}{Spin} & \multicolumn{3}{c}{Total} \\
 & $\sigma_{11}$ & $\sigma_{22}$ & $\sigma_{33}$ & $\sigma_{11}$ & $\sigma_{22}$ & $\sigma_{33}$ & $\sigma_{11}$ & $\sigma_{22}$ & $\sigma_{33}$\\
\hline
$\text{SnCl}_2$ & -3685.7 & -2806.8 & -2468.5 & 0 & 0 & 0 & -3685.7 & -2806.8 & -2468.5 \\
$\text{SnBr}_2$ & -4201.4 & -3044.1 & -2945.8 & 0 & 0 & 0 & -4201.4 & -3044.0 & -2945.8 \\
$\text{SnI}_2$ & -5037.9 & -3658.6 & -3250.5 & 0 & 0 & 0 & -5037.9 & -3658.6 & -3250.6 \\
$\text{SnAt}_2$ & -5498.3 & -4139.9 & -3512.2 & 0 & 0 & 0 & -5498.3 & -4139.9 & -3512.2 \\
$\text{SnCl}_4$ & -2500.1 & -2500.1 & -2500.1 & 0 & 0 & 0 & -2500.1 & -2500.1 & -2500.0 \\
$\text{SnBr}_4$ & -2689.7 & -2689.7 & -2689.7 & 0 & 0 & 0 & -2689.7 & -2689.7 & -2689.7 \\
$\text{SnI}_4$ & -2868.1 & -2868.1 & -2868.1 & 0 & 0 & 0 & -2868.1 & -2868.1 & -2868.1 \\
$\text{SnAt}_4$ & -3089.5 & -3089.5 & -3089.5 & 0 & 0 & 0 & -3089.6 & -3089.5 & -3089.5 \\
\hline
\hline
\end{tabular}
\end{table}

\begin{table}[h!]
\caption{Calculated principal components in ppm for a series of molecules containing Sn using SOC ZORA VASP. Excluding ZORA $K$ factor in the one-center currents.
Excluding core shielding.}
\begin{tabular}{lrrrrrrrrr}
\hline
\hline
\multirow{2}{*}{\begin{tabular}[c]{@{}l@{}}Compound\end{tabular}} & \multicolumn{3}{c}{Orbital} & \multicolumn{3}{c}{Spin} & \multicolumn{3}{c}{Total} \\
 & $\sigma_{11}$ & $\sigma_{22}$ & $\sigma_{33}$ & $\sigma_{11}$ & $\sigma_{22}$ & $\sigma_{33}$ & $\sigma_{11}$ & $\sigma_{22}$ & $\sigma_{33}$\\
\hline
$\text{SnCl}_2$ & -3974.3 & -3018.0 & -2642.5 & -738.1 & -358.9 & -52.7 & -4761.7 & -3377.0 & -2695.2 \\
$\text{SnBr}_2$ & -4534.2 & -3275.8 & -3159.6 & -919.2 & -353.6 & -30.3 & -5525.1 & -3629.3 & -3189.9 \\
$\text{SnI}_2$ & -5506.1 & -3973.9 & -3521.0 & -1264.0 & -386.2 & -14.7 & -6884.0 & -3988.6 & -3907.2 \\
$\text{SnAt}_2$ & -6684.3 & -4972.6 & -4075.0 & -1576.1 & -473.7 & 63.9 & -8437.3 & -4908.7 & -4548.7 \\
$\text{SnCl}_4$ & -2664.5 & -2664.5 & -2664.4 & 159.4 & 159.4 & 159.4 & -2534.0 & -2505.1 & -2505.1 \\
$\text{SnBr}_4$ & -2873.9 & -2873.9 & -2873.9 & 1137.8 & 1137.9 & 1137.9 & -1769.0 & -1736.1 & -1736.0 \\
$\text{SnI}_4$ & -3089.0 & -3089.0 & -3088.9 & 2697.5 & 2697.8 & 2697.8 & -426.8 & -391.5 & -391.2 \\
$\text{SnAt}_4$ & -3561.7 & -3561.7 & -3561.7 & 2791.4 & 2792.2 & 2792.2 & -806.6 & -770.3 & -769.4 \\
\hline
\hline
\end{tabular}
\end{table}

\begin{table}[h!]
\caption{Calculated principal components in ppm for a series of molecules containing Sn using SOC ZORA VASP. Including ZORA $K$ factor in the one-center currents.
Excluding core shielding.}
\begin{tabular}{lrrrrrrrrr}
\hline
\hline
\multirow{2}{*}{\begin{tabular}[c]{@{}l@{}}Compound\end{tabular}} & \multicolumn{3}{c}{Orbital} & \multicolumn{3}{c}{Spin} & \multicolumn{3}{c}{Total} \\
 & $\sigma_{11}$ & $\sigma_{22}$ & $\sigma_{33}$ & $\sigma_{11}$ & $\sigma_{22}$ & $\sigma_{33}$ & $\sigma_{11}$ & $\sigma_{22}$ & $\sigma_{33}$\\
\hline
$\text{SnCl}_2$ & -3710.9 & -2827.6 & -2468.5 & -725.2 & -341.8 & -51.0 & -4482.9 & -3169.4 & -2519.5 \\
$\text{SnBr}_2$ & -4234.7 & -3068.2 & -2952.4 & -904.0 & -332.8 & -29.4 & -5206.5 & -3401.0 & -2981.7 \\
$\text{SnI}_2$ & -5144.3 & -3714.0 & -3296.2 & -1246.5 & -361.6 & -18.0 & -6498.1 & -3732.0 & -3657.8 \\
$\text{SnAt}_2$ & -6246.8 & -4647.3 & -3812.7 & -1564.7 & -450.3 & 45.1 & -7977.8 & -4602.1 & -4263.0 \\
$\text{SnCl}_4$ & -2507.6 & -2507.6 & -2507.5 & 164.4 & 164.4 & 164.4 & -2370.8 & -2343.1 & -2343.1 \\
$\text{SnBr}_4$ & -2701.1 & -2701.1 & -2701.1 & 1142.8 & 1142.9 & 1143.0 & -1589.7 & -1558.3 & -1558.2 \\
$\text{SnI}_4$ & -2899.8 & -2899.7 & -2899.7 & 2700.9 & 2701.1 & 2701.2 & -232.5 & -198.9 & -198.5 \\
$\text{SnAt}_4$ & -3337.8 & -3337.8 & -3337.8 & 2789.2 & 2790.0 & 2790.0 & -582.6 & -548.7 & -547.8 \\
\hline
\hline
\end{tabular}
\end{table}

\begin{table}[h!]
\caption{Calculated principal components in ppm for a series of molecules containing Sn using scalar ZORA ADF.\cite{SIADF2001}}
\begin{tabular}{lrrrrrrrrr}
\hline
\hline
\multirow{2}{*}{\begin{tabular}[c]{@{}l@{}}Compound\end{tabular}} & \multicolumn{3}{c}{Orbital} & \multicolumn{3}{c}{Spin} & \multicolumn{3}{c}{Total} \\
 & $\sigma_{11}$ & $\sigma_{22}$ & $\sigma_{33}$ & $\sigma_{11}$ & $\sigma_{22}$ & $\sigma_{33}$ & $\sigma_{11}$ & $\sigma_{22}$ & $\sigma_{33}$\\
\hline
$\text{SnCl}_2$ & 1239.7 & 2102.2 & 2446.5 & 0 & 0 & 0 & 1239.7 & 2102.2 & 2446.5 \\
$\text{SnBr}_2$ & 719.7 & 1866.4 & 1968.0 & 0 & 0 & 0 & 719.7 & 1866.4 & 1968.0 \\
$\text{SnI}_2$ & -100.1 & 1263.6 & 1661.1 & 0 & 0 & 0 & -100.1 & 1263.6 & 1661.1 \\
$\text{SnAt}_2$ & -558.5 & 786.2 & 1402.2 & 0 & 0 & 0 & -558.5 & 786.2 & 1402.2 \\
$\text{SnCl}_4$ & 2406.1 & 2406.1 & 2406.2 & 0 & 0 & 0 & 2406.1 & 2406.1 & 2406.2 \\
$\text{SnBr}_4$ & 2217.4 & 2217.4 & 2217.5 & 0 & 0 & 0 & 2217.4 & 2217.4 & 2217.5 \\
$\text{SnI}_4$ & 2042.3 & 2042.4 & 2042.4 & 0 & 0 & 0 & 2042.3 & 2042.4 & 2042.4 \\
$\text{SnAt}_4$ & 1824.2 & 1824.2 & 1824.2 & 0 & 0 & 0 & 1824.2 & 1824.2 & 1824.2 \\
\hline
\hline
\end{tabular}
\end{table}

\begin{table}[h!]
\caption{Calculated principal components in ppm for a series of molecules containing Sn using SOC ZORA ADF.}
\begin{tabular}{lrrrrrrrrr}
\hline
\hline
\multirow{2}{*}{\begin{tabular}[c]{@{}l@{}}Compound\end{tabular}} & \multicolumn{3}{c}{Orbital} & \multicolumn{3}{c}{Spin} & \multicolumn{3}{c}{Total} \\
 & $\sigma_{11}$ & $\sigma_{22}$ & $\sigma_{33}$ & $\sigma_{11}$ & $\sigma_{22}$ & $\sigma_{33}$ & $\sigma_{11}$ & $\sigma_{22}$ & $\sigma_{33}$\\
\hline
$\text{SnCl}_2$ & 1111.1 & 2002.8 & 2392.6 & -161.0 & 246.0 & 571.0 & 950.1 & 2248.8 & 2963.6 \\
$\text{SnBr}_2$ & 562.8 & 1753.7 & 1895.0 & -356.4 & 248.5 & 582.4 & 206.4 & 2002.2 & 2477.4 \\
$\text{SnI}_2$ & -367.0 & 1121.3 & 1515.3 & -724.0 & 216.9 & 581.5 & -1091.0 & 1702.8 & 1732.1 \\
$\text{SnAt}_2$ & -1591.8 & 106.3 & 942.8 & -1078.4 & 117.3 & 648.2 & -2670.2 & 754.5 & 1060.1 \\
$\text{SnCl}_4$ & 2347.9 & 2347.9 & 2347.9 & 789.9 & 789.9 & 790.0 & 3137.8 & 3137.8 & 3137.9 \\
$\text{SnBr}_4$ & 2149.2 & 2149.2 & 2149.3 & 1781.6 & 1781.7 & 1781.7 & 3930.9 & 3930.9 & 3931.0 \\
$\text{SnI}_4$ & 1947.2 & 1947.2 & 1947.2 & 3364.9 & 3364.9 & 3364.9 & 5312.0 & 5312.1 & 5312.1 \\
$\text{SnAt}_4$ & 1457.7 & 1457.7 & 1457.7 & 3338.8 & 3338.9 & 3338.9 & 4796.6 & 4796.6 & 4796.7 \\
\hline
\hline
\end{tabular}
\end{table}

\begin{table}[h!]
\caption{Calculated isotropic chemical shielding and span for a series of molecules containing Sn using {\sc Dirac}.\cite{SIDiracMain}}
\begin{tabular}{lrr}
\hline
\hline
Compound & $\sigma_{\text{iso}}\ [\text{ppm}]$ & $\Omega\ [\text{ppm}]$ \\
\hline
$\text{SnCl}_2$ & -548.1 & 2059.9 \\
$\text{SnBr}_2$ & -1100.4 & 2371.0 \\
$\text{SnI}_2$ & -2007.4 & 3039.5 \\
$\text{SnAt}_2$ & -3210.4 & 4444.5 \\
$\text{SnCl}_4$ & 651.9 & 0.3 \\
$\text{SnBr}_4$ & 1501.1 & 1.0 \\
$\text{SnI}_4$ & 2824.3 & 0.1 \\
$\text{SnAt}_4$ & 1888.9 & 0.1 \\
\hline
\hline
\end{tabular}
\end{table}

\begin{figure}
    \centering
    \includegraphics[width=\linewidth]{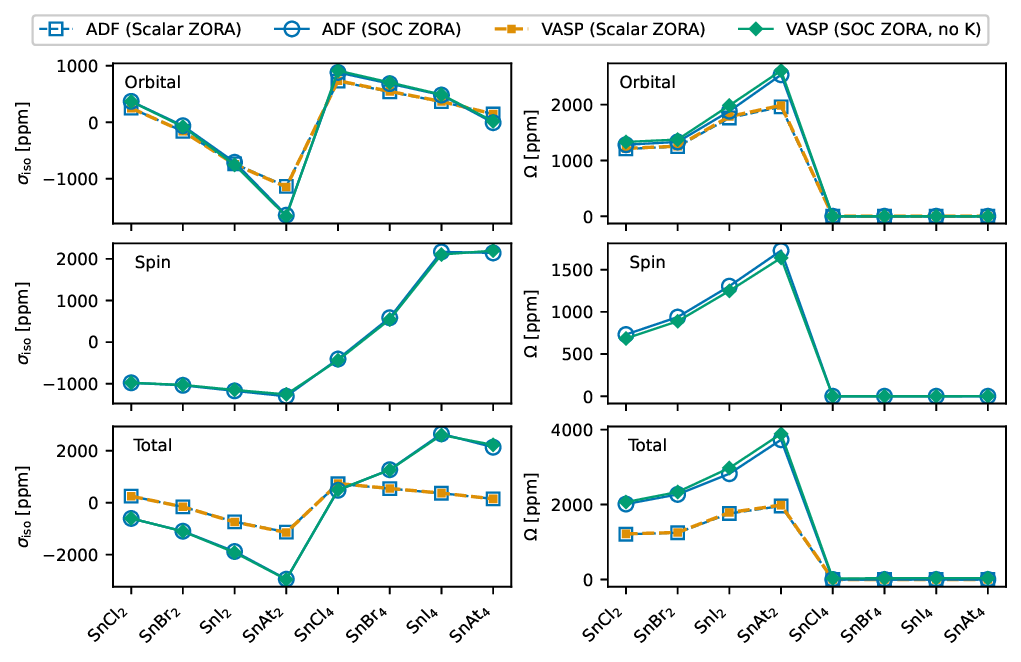}
    \caption{Isotropic value (left column) and span (right column) of the chemical shielding tensor for a series of molecules containing Sn. Top row shows the contribution from the diamagnetic and paramagnetic currents (orbital), the middle row those from the magnetization (spin), and the bottom row the total (orbital + spin) isotropic shielding and span. Scalar (squares) and spin-orbit coupled (circles) ZORA DFT with Amsterdam Density Functional (ADF)\cite{SIADF2001} (blue), and VASP results with the presently developed theory (orange and green).  The green data points do not apply the ZORA~$K$ factor in the one-center current contributions.}
\end{figure}

\FloatBarrier
\clearpage
\subsubsection{Hg}
\begin{table}[h!]
\caption{Calculated principal components in ppm for a series of molecules containing Hg using scalar ZORA VASP,
 excluding core shielding.}
\begin{tabular}{lrrrrrrrrr}
\hline
\hline
\multirow{2}{*}{\begin{tabular}[c]{@{}l@{}}Compound\end{tabular}} & \multicolumn{3}{c}{Orbital} & \multicolumn{3}{c}{Spin} & \multicolumn{3}{c}{Total} \\
 & $\sigma_{11}$ & $\sigma_{22}$ & $\sigma_{33}$ & $\sigma_{11}$ & $\sigma_{22}$ & $\sigma_{33}$ & $\sigma_{11}$ & $\sigma_{22}$ & $\sigma_{33}$\\
\hline
HgMeCl & -4687.9 & -4687.9 & 105.6 & 0 & 0 & 0 & -4687.9 & -4687.9 & 105.6 \\
HgMeBr & -4822.8 & -4822.8 & 110.0 & 0 & 0 & 0 & -4822.8 & -4822.8 & 110.0 \\
HgMeI & -5297.3 & -5297.3 & 115.7 & 0 & 0 & 0 & -5297.3 & -5297.3 & 115.7 \\
$\text{HgCl}_2$ & -4047.4 & -4047.4 & 184.2 & 0 & 0 & 0 & -4047.4 & -4047.4 & 184.2 \\
$\text{HgBr}_2$ & -3902.9 & -3902.9 & 184.9 & 0 & 0 & 0 & -3902.9 & -3902.9 & 184.9 \\
$\text{HgI}_2$ & -4449.1 & -4449.1 & 187.0 & 0 & 0 & 0 & -4449.2 & -4449.2 & 187.0 \\
$\text{HgMe}_2$ & -6057.0 & -6056.9 & 26.4 & 0 & 0 & 0 & -6057.0 & -6057.0 & 26.4 \\
$\text{HgCl}_2\text{N}_2\text{H}_6$ & -3957.1 & -3923.4 & -2041.3 & 0 & 0 & 0 & -3957.1 & -3923.4 & -2041.3 \\
$\text{HgBr}_2\text{N}_2\text{H}_6$  & -4057.4 & -3991.1 & -2120.2 & 0 & 0 & 0 & -4057.4 & -3991.1 & -2120.2 \\
$\text{HgI}_2\text{N}_2\text{H}_6$  & -4418.7 & -4300.1 & -2372.7 & 0 & 0 & 0 & -4418.7 & -4300.2 & -2372.7 \\
\hline
\hline
\end{tabular}
\end{table}

\begin{table}[h!]
\caption{Calculated principal components in ppm for a series of molecules containing Hg using SOC ZORA VASP. Excluding ZORA $K$ factor in the one-center currents.
 Excluding core shielding.}
\begin{tabular}{lrrrrrrrrr}
\hline
\hline
\multirow{2}{*}{\begin{tabular}[c]{@{}l@{}}Compound\end{tabular}} & \multicolumn{3}{c}{Orbital} & \multicolumn{3}{c}{Spin} & \multicolumn{3}{c}{Total} \\
 & $\sigma_{11}$ & $\sigma_{22}$ & $\sigma_{33}$ & $\sigma_{11}$ & $\sigma_{22}$ & $\sigma_{33}$ & $\sigma_{11}$ & $\sigma_{22}$ & $\sigma_{33}$\\
\hline
HgMeCl & -5670.4 & -5670.4 & 142.4 & -1042.0 & -1042.0 & 40.3 & -6712.5 & -6672.9 & 182.7 \\
HgMeBr & -5868.8 & -5868.8 & 155.8 & -657.0 & -657.0 & 54.0 & -6525.7 & -6475.4 & 209.8 \\
HgMeI & -6473.3 & -6473.3 & 182.5 & -89.7 & -89.7 & 28.7 & -6563.0 & -6479.8 & 211.2 \\
$\text{HgCl}_2$ & -4879.1 & -4879.1 & 213.9 & -751.2 & -751.2 & -41.8 & -5653.5 & -5630.4 & 172.1 \\
$\text{HgBr}_2$ & -4786.3 & -4786.3 & 221.8 & -74.3 & 1156.0 & 1156.1 & -3631.3 & -3630.2 & 147.6 \\
$\text{HgI}_2$ & -5491.8 & -5491.8 & 251.5 & -603.2 & 3774.2 & 3774.3 & -1717.6 & -1664.7 & -351.6 \\
$\text{HgMe}_2$ & -7335.8 & -7335.8 & 97.0 & -1123.5 & -1123.5 & 203.3 & -8459.3 & -8290.4 & 300.3 \\
$\text{HgCl}_2\text{N}_2\text{H}_6$  & -4626.8 & -4566.8 & -2387.3 & -363.1 & -325.3 & -139.2 & -4989.9 & -4892.1 & -2661.4 \\
$\text{HgCl}_2\text{N}_2\text{H}_6$  & -4764.1 & -4669.3 & -2476.0 & -112.6 & 1155.8 & 1225.5 & -3538.6 & -3513.5 & -2741.9 \\
$\text{HgCl}_2\text{N}_2\text{H}_6$ & -5196.0 & -5041.2 & -2751.8 & -445.9 & 3311.2 & 4048.2 & -3378.6 & -1730.0 & -1147.7 \\
\hline
\hline
\end{tabular}
\end{table}

\begin{table}[h!]
\caption{Calculated principal components in ppm for a series of molecules containing Hg using SOC ZORA VASP. Including ZORA $K$ factor in the one-center currents.
Excluding core shielding.}
\begin{tabular}{lrrrrrrrrr}
\hline
\hline
\multirow{2}{*}{\begin{tabular}[c]{@{}l@{}}Compound\end{tabular}} & \multicolumn{3}{c}{Orbital} & \multicolumn{3}{c}{Spin} & \multicolumn{3}{c}{Total} \\
 & $\sigma_{11}$ & $\sigma_{22}$ & $\sigma_{33}$ & $\sigma_{11}$ & $\sigma_{22}$ & $\sigma_{33}$ & $\sigma_{11}$ & $\sigma_{22}$ & $\sigma_{33}$\\
\hline
HgMeCl & -4839.9 & -4839.9 & 142.9 & -933.7 & -933.7 & 14.4 & -5773.5 & -5753.0 & 157.3 \\
HgMeBr & -4986.4 & -4986.4 & 154.4 & -533.7 & -533.7 & 26.1 & -5520.1 & -5489.8 & 180.4 \\
HgMeI & -5476.9 & -5476.9 & 177.0 & -3.8 & 59.3 & 59.3 & -5417.6 & -5358.5 & 173.3 \\
$\text{HgCl}_2$ & -4187.8 & -4187.8 & 209.4 & -667.2 & -667.2 & -60.6 & -4889.0 & -4855.1 & 148.8 \\
$\text{HgBr}_2$ & -4056.5 & -4056.5 & 215.5 & -93.8 & 1255.9 & 1255.9 & -2812.6 & -2800.7 & 121.7 \\
$\text{HgI}_2$ & -4621.4 & -4621.4 & 240.3 & -627.8 & 3908.3 & 3908.4 & -713.1 & -675.5 & -387.5 \\
$\text{HgMe}_2$ & -6206.2 & -6206.2 & 100.1 & -963.1 & -963.0 & 158.1 & -7169.3 & -7040.2 & 258.2 \\
$\text{HgCl}_2\text{N}_2\text{H}_6$ & -4052.6 & -4012.4 & -2081.0 & -308.4 & -271.2 & -128.0 & -4360.9 & -4283.6 & -2321.6 \\
$\text{HgCl}_2\text{N}_2\text{H}_6$ & -4153.5 & -4081.6 & -2156.4 & -101.2 & 1213.6 & 1285.3 & -2868.2 & -2868.0 & -2389.3 \\
$\text{HgCl}_2\text{N}_2\text{H}_6$ & -4521.0 & -4398.9 & -2401.5 & -434.7 & 3374.0 & 4115.8 & -2995.4 & -1024.9 & -405.2 \\
\hline
\hline
\end{tabular}
\end{table}

\begin{table}[h!]
\caption{Calculated principal components in ppm for a series of molecules containing Hg using scalar relativistic ZORA ADF.}
\begin{tabular}{lrrrrrrrrr}
\hline
\hline
\multirow{2}{*}{\begin{tabular}[c]{@{}l@{}}Compound\end{tabular}} & \multicolumn{3}{c}{Orbital} & \multicolumn{3}{c}{Spin} & \multicolumn{3}{c}{Total} \\
 & $\sigma_{11}$ & $\sigma_{22}$ & $\sigma_{33}$ & $\sigma_{11}$ & $\sigma_{22}$ & $\sigma_{33}$ & $\sigma_{11}$ & $\sigma_{22}$ & $\sigma_{33}$\\
\hline
HgMeCl & 5460.7 & 5460.7 & 10207.4 & 0 & 0 & 0 & 5460.7 & 5460.7 & 10207.4 \\
HgMeBr & 5323.4 & 5323.4 & 10212.7 & 0 & 0 & 0 & 5323.4 & 5323.4 & 10212.7 \\
HgMeI & 4858.9 & 4858.9 & 10219.9 & 0 & 0 & 0 & 4858.9 & 4858.9 & 10219.9 \\
$\text{HgCl}_2$ & 6087.7 & 6087.7 & 10286.9 & 0 & 0 & 0 & 6087.7 & 6087.7 & 10286.9 \\
$\text{HgBr}_2$ & 6229.3 & 6229.3 & 10289.6 & 0 & 0 & 0 & 6229.3 & 6229.3 & 10289.6 \\
$\text{HgI}_2$ & 5692.3 & 5692.3 & 10294.2 & 0 & 0 & 0 & 5692.3 & 5692.3 & 10294.2 \\
$\text{HgMe}_2$ & 4121.8 & 4121.8 & 10128.1 & 0 & 0 & 0 & 4121.8 & 4121.8 & 10128.1 \\
$\text{HgCl}_2\text{N}_2\text{H}_6$  & 6146.8 & 6170.7 & 8063.8 & 0 & 0 & 0 & 6146.8 & 6170.7 & 8063.8 \\
$\text{HgBr}_2\text{N}_2\text{H}_6$  & 6048.0 & 6104.9 & 7986.7 & 0 & 0 & 0 & 6048.0 & 6104.9 & 7986.7 \\
$\text{HgI}_2\text{N}_2\text{H}_6$  & 5688.4 & 5799.2 & 7737.9 & 0 & 0 & 0 & 5688.4 & 5799.2 & 7737.9 \\
\hline
\hline
\end{tabular}
\end{table}

\begin{table}[h!]
\caption{Calculated principal components in ppm for a series of molecules containing Hg using SOC ZORA ADF.}
\begin{tabular}{lrrrrrrrrr}
\hline
\hline
\multirow{2}{*}{\begin{tabular}[c]{@{}l@{}}Compound\end{tabular}} & \multicolumn{3}{c}{Orbital} & \multicolumn{3}{c}{Spin} & \multicolumn{3}{c}{Total} \\
 & $\sigma_{11}$ & $\sigma_{22}$ & $\sigma_{33}$ & $\sigma_{11}$ & $\sigma_{22}$ & $\sigma_{33}$ & $\sigma_{11}$ & $\sigma_{22}$ & $\sigma_{33}$\\
\hline
HgMeCl & 4574.2 & 4574.2 & 10371.1 & 3128.3 & 3128.3 & 4706.4 & 7702.5 & 7702.5 & 15077.5 \\
HgMeBr & 4346.1 & 4346.1 & 10392.6 & 3484.9 & 3484.9 & 4722.4 & 7831.0 & 7831.0 & 15115.1 \\
HgMeI & 3726.7 & 3726.7 & 10436.7 & 4013.7 & 4013.7 & 4698.5 & 7740.4 & 7740.4 & 15135.2 \\
$\text{HgCl}_2$ & 5362.5 & 5362.5 & 10415.1 & 3509.2 & 3509.2 & 4619.3 & 8871.6 & 8871.6 & 15034.4 \\
$\text{HgBr}_2$ & 5400.5 & 5400.5 & 10433.1 & 4584.8 & 5376.0 & 5376.0 & 10776.5 & 10776.5 & 15018.0 \\
$\text{HgI}_2$ & 4654.6 & 4654.6 & 10487.0 & 4020.9 & 7970.9 & 7970.9 & 12625.4 & 12625.4 & 14507.9 \\
$\text{HgMe}_2$ & 2934.0 & 2934.0 & 10387.3 & 2898.8 & 2898.8 & 4908.5 & 5832.8 & 5832.8 & 15295.8 \\
$\text{HgCl}_2\text{N}_2\text{H}_6$  & 5616.3 & 5667.6 & 7858.4 & 3969.5 & 4015.9 & 4359.6 & 9585.8 & 9683.5 & 12218.0 \\
$\text{HgBr}_2\text{N}_2\text{H}_6$  & 5471.3 & 5559.2 & 7780.9 & 4379.4 & 5469.9 & 5527.0 & 10998.4 & 11029.1 & 12160.3 \\
$\text{HgI}_2\text{N}_2\text{H}_6$  & 5036.7 & 5187.6 & 7529.8 & 3992.0 & 7620.4 & 8356.4 & 11521.8 & 12808.1 & 13393.1 \\
\hline
\hline
\end{tabular}
\end{table}

\begin{table}[h!]
\caption{Calculated isotropic chemical shielding and span for a series of molecules containing Hg using {\sc Dirac}.}
\begin{tabular}{lrr}
\hline
\hline
Compound & $\sigma_{\text{iso}}\ [\text{ppm}]$ & $\Omega\ [\text{ppm}]$ \\
\hline
HgMeCl & -946.4 & 7075.5 \\
HgMeBr & -851.1 & 7000.2 \\
HgMeI & -840.9 & 6997.9 \\
$\text{HgCl}_2$ & -147.5 & 5781.8 \\
$\text{HgBr}_2$ & 1229.2 & 3681.4 \\
$\text{HgI}_2$ & 2324.8 & 1218.6 \\
$\text{HgMe}_2$ & -2089.2 & 9133.9 \\
$\text{HgCl}_2\text{N}_2\text{H}_6$ & -557.6 & 2413.6 \\
$\text{HgBr}_2\text{N}_2\text{H}_6$ & 327.3 & 977.3 \\
$\text{HgI}_2\text{N}_2\text{H}_6$ & 1551.3 & 2197.4 \\
\hline
\hline
\end{tabular}
\end{table}

\FloatBarrier
\clearpage
\subsubsection{Pb}
\begin{table}[h!]
\caption{Calculated principal components in ppm for a series of molecules containing Pb using scalar ZORA VASP,
 excluding core shielding.}
\begin{tabular}{lrrrrrrrrr}
\hline
\hline
\multirow{2}{*}{\begin{tabular}[c]{@{}l@{}}Compound\end{tabular}} & \multicolumn{3}{c}{Orbital} & \multicolumn{3}{c}{Spin} & \multicolumn{3}{c}{Total} \\
 & $\sigma_{11}$ & $\sigma_{22}$ & $\sigma_{33}$ & $\sigma_{11}$ & $\sigma_{22}$ & $\sigma_{33}$ & $\sigma_{11}$ & $\sigma_{22}$ & $\sigma_{33}$\\
\hline
$\text{PbCl}_2$ & -6751.5 & -5643.4 & -5280.7 & 0 & 0 & 0 & -6751.5 & -5643.4 & -5280.7 \\
$\text{PbBr}_2$ & -7634.7 & -6178.8 & -6103.2 & 0 & 0 & 0 & -7634.7 & -6178.8 & -6103.2 \\
$\text{PbI}_2$ & -9070.0 & -7503.7 & -6495.8 & 0 & 0 & 0 & -9070.0 & -7503.7 & -6495.8 \\
$\text{PbAt}_2$ & -9890.0 & -8382.4 & -6996.5 & 0 & 0 & 0 & -9890.0 & -8382.4 & -6996.5 \\
$\text{PbCl}_4$ & -5219.6 & -5219.5 & -5219.4 & 0 & 0 & 0 & -5219.6 & -5219.5 & -5219.4 \\
$\text{PbBr}_4$ & -5521.6 & -5521.6 & -5521.5 & 0 & 0 & 0 & -5521.6 & -5521.6 & -5521.5 \\
$\text{PbI}_4$ & -5817.6 & -5817.5 & -5817.5 & 0 & 0 & 0 & -5817.6 & -5817.5 & -5817.5 \\
$\text{PbAt}_4$ & -6173.8 & -6173.8 & -6173.7 & 0 & 0 & 0 & -6173.8 & -6173.8 & -6173.7 \\
\hline
\hline
\end{tabular}
\end{table}

\begin{table}[h!]
\caption{Calculated principal components in ppm for a series of molecules containing Pb using SOC ZORA VASP. Excluding ZORA $K$ factor in the one-center currents.
 Excluding core shielding.}
\begin{tabular}{lrrrrrrrrr}
\hline
\hline
\multirow{2}{*}{\begin{tabular}[c]{@{}l@{}}Compound\end{tabular}} & \multicolumn{3}{c}{Orbital} & \multicolumn{3}{c}{Spin} & \multicolumn{3}{c}{Total} \\
 & $\sigma_{11}$ & $\sigma_{22}$ & $\sigma_{33}$ & $\sigma_{11}$ & $\sigma_{22}$ & $\sigma_{33}$ & $\sigma_{11}$ & $\sigma_{22}$ & $\sigma_{33}$\\
\hline
$\text{PbCl}_2$ & -8679.2 & -7168.1 & -6418.4 & -5065.5 & -2902.8 & -1181.9 & -14115.1 & -10070.9 & -7600.3 \\
$\text{PbBr}_2$ & -9762.3 & -7733.8 & -7475.1 & -5717.6 & -3029.6 & -1511.8 & -15955.6 & -10763.4 & -8986.8 \\
$\text{PbI}_2$ & -11553.3 & -9025.6 & -8195.4 & -6906.5 & -3195.6 & -2164.1 & -19076.4 & -11391.0 & -11189.7 \\
$\text{PbAt}_2$ & -13417.0 & -10656.6 & -9117.8 & -7475.3 & -3606.3 & -2883.1 & -21525.6 & -13539.7 & -12724.1 \\
$\text{PbCl}_4$ & -6306.5 & -6306.5 & -6306.4 & 483.6 & 483.7 & 483.7 & -6011.1 & -5822.8 & -5822.7 \\
$\text{PbBr}_4$ & -6702.4 & -6702.4 & -6702.4 & 4282.9 & 4282.9 & 4283.0 & -2633.4 & -2419.6 & -2419.4 \\
$\text{PbI}_4$ & -7103.0 & -7103.0 & -7103.0 & 9095.1 & 9095.6 & 9095.7 & 1759.3 & 1992.1 & 1992.7 \\
$\text{PbAt}_4$ & -7965.7 & -7965.7 & -7965.7 & 6680.9 & 6683.4 & 6683.4 & -1493.8 & -1284.8 & -1282.2 \\
\hline
\hline
\end{tabular}
\end{table}

\begin{table}[h!]
\caption{Calculated principal components in ppm for a series of molecules containing Pb using SOC ZORA VASP. Including ZORA $K$ factor in the one-center currents.
 Excluding core shielding.}
\begin{tabular}{lrrrrrrrrr}
\hline
\hline
\multirow{2}{*}{\begin{tabular}[c]{@{}l@{}}Compound\end{tabular}} & \multicolumn{3}{c}{Orbital} & \multicolumn{3}{c}{Spin} & \multicolumn{3}{c}{Total} \\
 & $\sigma_{11}$ & $\sigma_{22}$ & $\sigma_{33}$ & $\sigma_{11}$ & $\sigma_{22}$ & $\sigma_{33}$ & $\sigma_{11}$ & $\sigma_{22}$ & $\sigma_{33}$\\
\hline
$\text{PbCl}_2$ & -7130.6 & -5968.3 & -5301.5 & -4841.7 & -2571.2 & -1135.6 & -12283.8 & -8539.5 & -6437.1 \\
$\text{PbBr}_2$ & -8017.2 & -6425.2 & -6166.0 & -5460.7 & -2650.2 & -1459.7 & -13876.2 & -9075.4 & -7625.7 \\
$\text{PbI}_2$ & -9487.3 & -7436.8 & -6797.0 & -6612.6 & -2775.0 & -2130.2 & -16613.9 & -9571.9 & -9567.0 \\
$\text{PbAt}_2$ & -11014.3 & -8766.5 & -7540.5 & -7187.1 & -3188.0 & -2921.0 & -18729.2 & -11687.5 & -10728.5 \\
$\text{PbCl}_4$ & -5335.5 & -5335.4 & -5335.4 & 580.2 & 580.3 & 580.3 & -4917.1 & -4755.2 & -4755.2 \\
$\text{PbBr}_4$ & -5643.1 & -5643.1 & -5643.1 & 4386.3 & 4386.3 & 4386.4 & -1439.8 & -1256.9 & -1256.7 \\
$\text{PbI}_4$ & -5955.9 & -5955.8 & -5955.8 & 9193.1 & 9193.6 & 9193.6 & 3039.2 & 3237.2 & 3237.8 \\
$\text{PbAt}_4$ & -6635.7 & -6635.7 & -6635.7 & 6761.4 & 6763.9 & 6764.0 & -48.3 & 125.7 & 128.2 \\
\hline
\hline
\end{tabular}
\end{table}

\begin{table}[h!]
\caption{Calculated principal components in ppm for a series of molecules containing Pb using scalar ZORA ADF.}
\begin{tabular}{lrrrrrrrrr}
\hline
\hline
\multirow{2}{*}{\begin{tabular}[c]{@{}l@{}}Compound\end{tabular}} & \multicolumn{3}{c}{Orbital} & \multicolumn{3}{c}{Spin} & \multicolumn{3}{c}{Total} \\
 & $\sigma_{11}$ & $\sigma_{22}$ & $\sigma_{33}$ & $\sigma_{11}$ & $\sigma_{22}$ & $\sigma_{33}$ & $\sigma_{11}$ & $\sigma_{22}$ & $\sigma_{33}$\\
\hline
$\text{PbCl}_2$ & 3715.1 & 4802.1 & 5171.6 & 0 & 0 & 0 & 3715.1 & 4802.1 & 5171.6 \\
$\text{PbBr}_2$ & 2822.7 & 4270.8 & 4344.3 & 0 & 0 & 0 & 2822.7 & 4270.8 & 4344.3 \\
$\text{PbI}_2$ & 1416.4 & 2963.9 & 3952.8 & 0 & 0 & 0 & 1416.4 & 2963.9 & 3952.8 \\
$\text{PbAt}_2$ & 598.8 & 2091.8 & 3457.6 & 0 & 0 & 0 & 598.8 & 2091.8 & 3457.6 \\
$\text{PbCl}_4$ & 5230.1 & 5230.2 & 5230.2 & 0 & 0 & 0 & 5230.1 & 5230.2 & 5230.2 \\
$\text{PbBr}_4$ & 4925.9 & 4926.0 & 4926.0 & 0 & 0 & 0 & 4925.9 & 4926.0 & 4926.0 \\
$\text{PbI}_4$ & 4631.4 & 4631.4 & 4631.5 & 0 & 0 & 0 & 4631.4 & 4631.4 & 4631.5 \\
$\text{PbAt}_4$ & 4281.0 & 4281.0 & 4281.0 & 0 & 0 & 0 & 4281.0 & 4281.0 & 4281.0 \\
\hline
\hline
\end{tabular}
\end{table}

\begin{table}[h!]
\caption{Calculated principal components in ppm for a series of molecules containing Pb using SOC ZORA ADF.}
\begin{tabular}{lrrrrrrrrr}
\hline
\hline
\multirow{2}{*}{\begin{tabular}[c]{@{}l@{}}Compound\end{tabular}} & \multicolumn{3}{c}{Orbital} & \multicolumn{3}{c}{Spin} & \multicolumn{3}{c}{Total} \\
 & $\sigma_{11}$ & $\sigma_{22}$ & $\sigma_{33}$ & $\sigma_{11}$ & $\sigma_{22}$ & $\sigma_{33}$ & $\sigma_{11}$ & $\sigma_{22}$ & $\sigma_{33}$\\
\hline
$\text{PbCl}_2$ & 1568.2 & 2897.3 & 4296.1 & -1123.2 & 1277.7 & 3373.6 & 445.0 & 4175.0 & 7669.7 \\
$\text{PbBr}_2$ & 408.3 & 2269.9 & 3262.7 & -1908.3 & 1084.3 & 2908.9 & -1500.0 & 3354.2 & 6171.6 \\
$\text{PbI}_2$ & -1413.4 & 1769.5 & 1827.6 & -3263.4 & 885.9 & 2062.4 & -4676.7 & 2655.4 & 3890.1 \\
$\text{PbAt}_2$ & -3309.7 & 345.6 & 898.4 & -3909.8 & 421.1 & 1143.2 & -7219.5 & 1319.5 & 1488.7 \\
$\text{PbCl}_4$ & 4263.3 & 4263.4 & 4263.4 & 5213.1 & 5213.1 & 5213.1 & 9476.5 & 9476.5 & 9476.5 \\
$\text{PbBr}_4$ & 3860.3 & 3860.3 & 3860.4 & 8944.1 & 8944.1 & 8944.1 & 12804.4 & 12804.4 & 12804.5 \\
$\text{PbI}_4$ & 3473.1 & 3473.2 & 3473.2 & 13653.0 & 13653.0 & 13653.1 & 17126.1 & 17126.2 & 17126.3 \\
$\text{PbAt}_4$ & 2582.0 & 2582.0 & 2582.1 & 10947.4 & 10948.2 & 10948.3 & 13529.5 & 13530.2 & 13530.4 \\
\hline
\hline
\end{tabular}
\end{table}

\begin{table}[h!]
\caption{Calculated isotropic chemical shielding and span for a series of molecules containing Pb using {\sc Dirac}.}
\begin{tabular}{lrr}
\hline
\hline
Compound & $\sigma_{\text{iso}}\ [\text{ppm}]$ & $\Omega\ [\text{ppm}]$ \\
\hline
$\text{PbCl}_2$ & -2902.4 & 7501.1 \\
$\text{PbBr}_2$ & -4664.0 & 8220.0 \\
$\text{PbI}_2$& -7244.8 & 9584.2 \\
$\text{PbAt}_2$ & -9863.9 & 11449.6 \\
$\text{PbCl}_4$ & 3396.4 & 30.2 \\
$\text{PbBr}_4$ & 6497.1 & 28.3 \\
$\text{PbI}_4$ & 9791.2 & 96.8 \\
$\text{PbAt}_4$ & 4990.4 & 0.7 \\
\hline
\hline
\end{tabular}
\end{table}

\begin{figure}
    \centering
    \includegraphics[width=\linewidth]{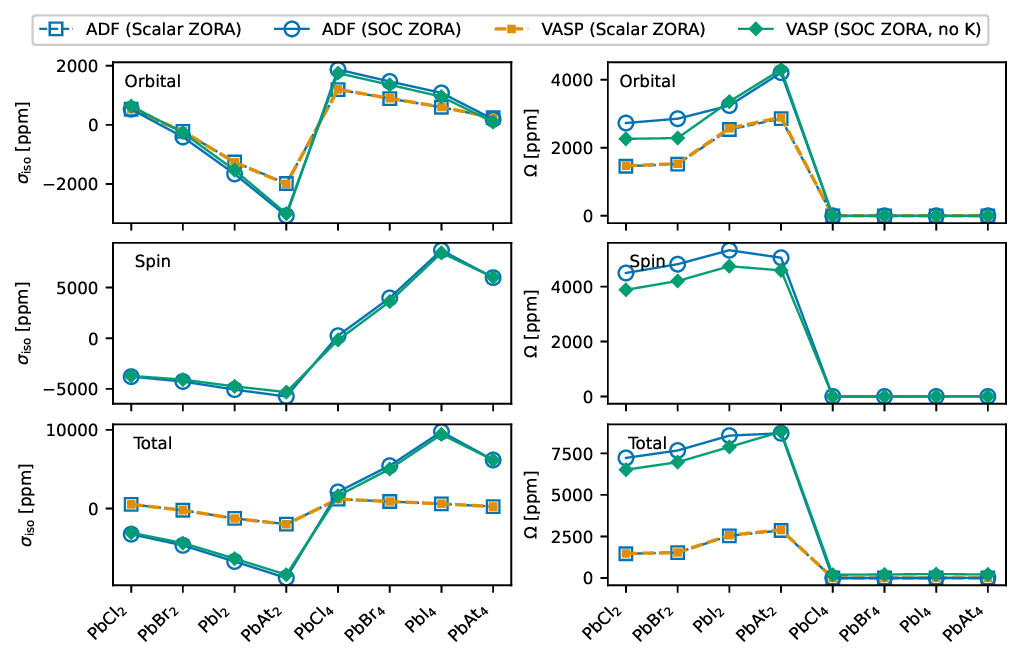}
    \caption{Isotropic value (left column) and span (right column) of the chemical shielding tensor for a series of molecules containing Pb. Top row shows the contribution from the diamagnetic and paramagnetic currents (orbital), the middle row those from the magnetization (spin), and the bottom row the total (orbital + spin) isotropic shielding and span. Scalar (squares) and spin-orbit coupled (circles) ZORA DFT with Amsterdam Density Functional (ADF)\cite{SIADF2001} (blue), and VASP results with the presently developed theory (orange and green).  The green data points do not apply the ZORA~$K$ factor in the one-center current contributions.}
\end{figure}

\FloatBarrier
\newpage
\section{Relativistic Orbitals}
\subsection{Sn\_d atom}
\begin{figure}[h!]
    \centering
    \includegraphics[]{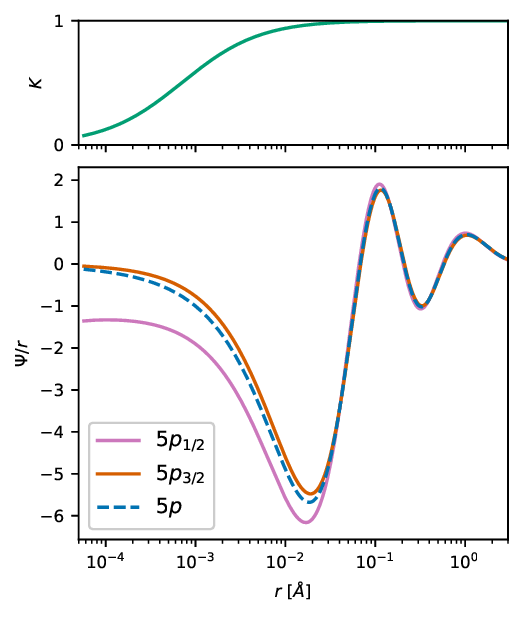}
    \caption{Upper panel: ZORA $K$-factor for Sn atom, plotted with $r$ on a logarithmic scale. Lower panel: scalar relativistic all-electron partial wave (dashed) versus SOC relativistic orbitals (solid).}
\end{figure}

\begin{figure}[h!]
    \centering
    \includegraphics[]{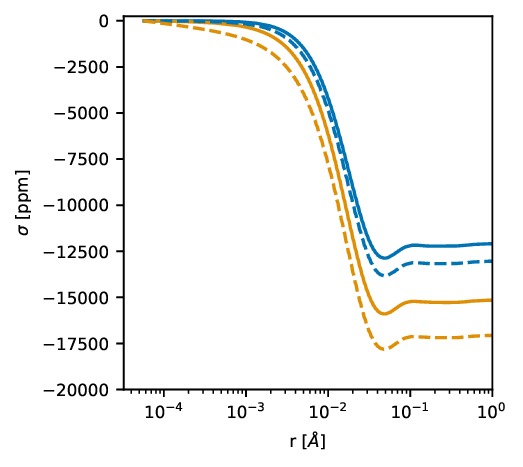}
    \caption{Chemical shielding resulting from a radial Biot-Savart integration of the paramagnetic one-center current for a spin-orbit coupled Sn atom with different PAW AE partial waves. Results generated with the default scalar relativistic PAW dataset in blue, and modified SOC all-electron partial waves in orange. Dashed lines omit the ZORA K-factor for the paramagnetic one-center current operator, solid lines include the K-factor. Plotted on a logarithmic radial scale.}
\end{figure}

\FloatBarrier
\newpage
\clearpage
\section{Solid-State Systems}
\subsection{Crystal Structures}
\begin{table}[h!]
\caption{Overview of Sn compounds with CCDC identifier of structure and the k-point mesh.}
\begin{tabular}{@{}rrr@{}}
\hline
\hline
Compound & CCDC & $\Gamma$ k-point mesh \\ 
\hline
SnO\footnotemark[1] &  1597057 & $4 \times 4 \times 3$ \\
$\text{SnC}_2\text{O}_4$\footnotemark[2] & 1271166 & $2 \times 3 \times 2$ \\
$\text{SnSO}_4$\footnotemark[3] & 1592555 & $2 \times 3 \times 2$ \\
$\text{BaSnF}_4$\footnotemark[4] & 1682198 & $3 \times 3 \times 2$ \\
$\text{SnHPO}_3$\footnotemark[5] & 1601025 & $9 \times 6 \times 9$ \\
$\text{SnO}_2$\footnotemark[6] & 1597192 & $4 \times 4 \times 3$ \\
$\text{Ca}_2\text{SnO}_4$\footnotemark[7] & 1688551 & $3 \times 2 \times 4$ \\
$\text{SnS}_2$\footnotemark[8] & 1659742 & $4 \times 4 \times 3$ \\
$\text{Pb}_2\text{SnO}_4$\footnotemark[9] & 1605543 & $2 \times 2 \times 2$ \\
$\text{Sr}_2\text{SnO}_4$\footnotemark[10] & 1623553 & $3 \times 3 \times 1$ \\
\hline
\hline
\end{tabular}\\
\footnotemark[1]Ref.~\onlinecite{SnO}, \footnotemark[2]Ref.~\onlinecite{SnC2O4}, \footnotemark[3]Ref.~\onlinecite{SnSO4}, \footnotemark[4]Ref.~\onlinecite{BaSnF4}, \footnotemark[5]Ref.~\onlinecite{SnHPO3}, \footnotemark[6]Ref.~\onlinecite{SnO2}, \footnotemark[7]Ref.~\onlinecite{Ca2SnO4}, \footnotemark[8]Ref.~\onlinecite{SnS2}, \footnotemark[9]Ref.~\onlinecite{Pb2SnO4}, \footnotemark[10]Ref.~\onlinecite{Sr2SnO4}
\end{table}

\FloatBarrier

\FloatBarrier 
\begin{table}[h!]
\caption{Overview of Hg compounds with CCDC identifier of structure and the k-point mesh.}
\begin{tabular}{@{}rrr@{}}
\hline
\hline
Compound & CCDC & $\Gamma$ k-point mesh \\
\hline
$\text{Hg}(\text{SCN})_2$\footnotemark[1] & 1595156 & $2 \times 4 \times 2$ \\
$\text{Hg}(\text{CN})_2$\footnotemark[2] & 1727652 & $2 \times 2 \times 2$ \\
$\text{Hg}(\text{SeCN})_2$\footnotemark[3] & 1645673 & $3 \times 3 \times 2$ \\
$\text{Hg}(\text{CO}_2\text{CH}_3)_2$\footnotemark[4] & 1175795 & $2 \times 1 \times 3$ \\
$\text{HgF}_2$\footnotemark[5] & 1606371 & $3 \times 3 \times 3$ \\
$\text{HgCl}_2$\footnotemark[6] & 1599405 & $1 \times 3 \times 3$ \\
$\text{HgBr}_2$\footnotemark[7] & 1609313 & $3 \times 2 \times 2$ \\
$\text{Hg}_2\text{Cl}_2$\footnotemark[8] & 1599804 & $3 \times 3 \times 2$ \\
$\text{K}[\text{Hg}(\text{SCN})_3]$\footnotemark[3] & 1645674 & $2 \times 4 \times 1$ \\
\hline
\hline
\end{tabular}\\
\footnotemark[1]Ref.~\onlinecite{HgSCN2}, \footnotemark[2]Ref.~\onlinecite{HgCN2}, \footnotemark[3]Ref.~\onlinecite{HgSeCN2}, \footnotemark[4]Ref.~\onlinecite{HgCO2CH3}, \footnotemark[5]Ref.~\onlinecite{HgF2}, \footnotemark[6]Ref.~\onlinecite{HgCl2}, \footnotemark[7]Ref.~\onlinecite{HgBr2}, \footnotemark[8]Ref.~\onlinecite{Hg2Cl2}
\end{table}

\FloatBarrier

\FloatBarrier 
\begin{table}[h!]
\caption{Overview of Pb compounds with CCDC identifier of structure and the k-point mesh.}
\begin{tabular}{@{}rrr@{}}
\hline
\hline
Compound & CCDC & $\Gamma$ k-point mesh \\
\hline
$\alpha - \text{PbO}$\footnotemark[1] & 1596211 & $8 \times 8 \times 8$ \\
$\beta - \text{PbO}$\footnotemark[2] & 1610087 & $8 \times 8 \times 8$ \\
$\text{Pb}_3\text{O}_4$\footnotemark[3] & 1592812 & $6 \times 6 \times 6$ \\
$\text{Pb}_2\text{SnO}_4$\footnotemark[4] & 1605543 & $6 \times 6 \times 6$ \\
$\text{PbF}_2$\footnotemark[5] & 1671614 & $3 \times 6 \times 3$ \\
$\text{PbCl}_2$\footnotemark[6] & 1602940 & $4 \times 8 \times 4$ \\
$\text{PbBr}_2$\footnotemark[7] & 1706726 & $3 \times 3 \times 6$ \\
$\text{PbClOH}_2$\footnotemark[8] & 1725105 & $4 \times 8 \times 4$ \\
$\text{PbBrOH}_2$\footnotemark[8] & 1725106 & $3 \times 6 \times 3$ \\
$\text{PbIOH}_2$\footnotemark[8] & 1725108 & $4 \times 8 \times 4$ \\
$\text{PbSiO}_3$\footnotemark[9] & 1602191 & $2 \times 4 \times 2$ \\
$\text{Pb}_3(\text{PO}_4)_2$\footnotemark[10] & 1595646 & $4 \times 4 \times 4$ \\
\hline
\hline
\end{tabular}\\
\footnotemark[1]Ref.~\onlinecite{aPbO}, \footnotemark[2]Ref.~\onlinecite{bPbO}, \footnotemark[3]Ref.~\onlinecite{Pb3O4}, \footnotemark[4]Ref.~\onlinecite{Pb2SnO4}, \footnotemark[5]Ref.~\onlinecite{PbF2}, \footnotemark[6]Ref.~\onlinecite{PbCl2}, \footnotemark[7]Ref.~\onlinecite{PbBr2}, \footnotemark[8]Ref.~\onlinecite{PbXOH}, \footnotemark[9]Ref.~\onlinecite{PbSiO3}, \footnotemark[10]Ref.~\onlinecite{PbP2O8}
\end{table}

\FloatBarrier

\FloatBarrier
\clearpage
\subsection{Data on cluster models by Alkan {\it et al.}}
\begin{figure}[h!]
    \centering
    \includegraphics[width=\linewidth]{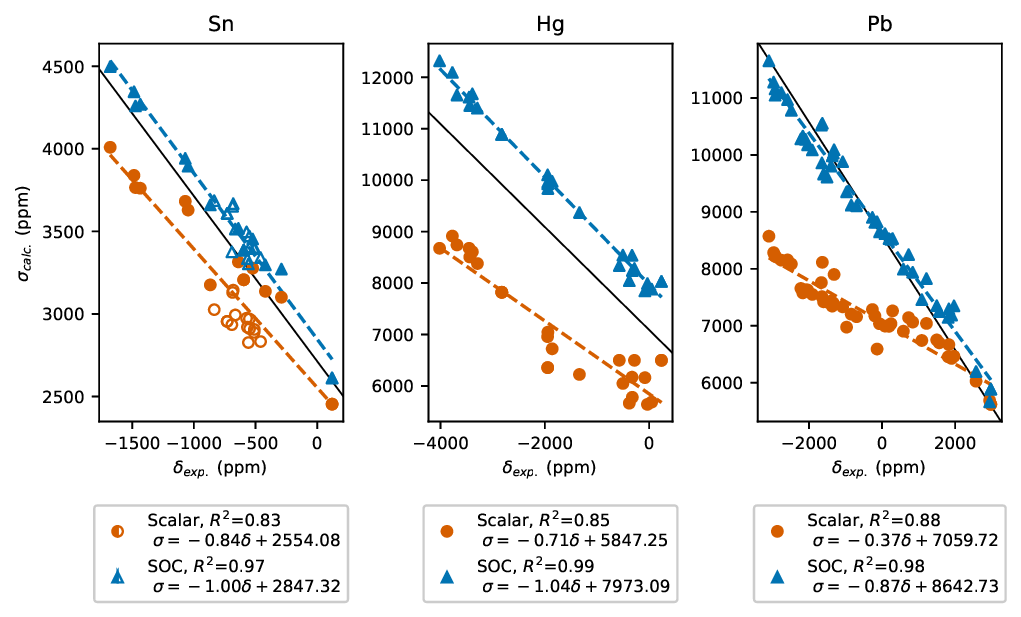}
    \caption{Selection of Alkan calculated principal components of chemical shielding correlated to experimental chemical shift principal components for Sn, Hg and Pb compounds. Closed symbols represent M(II), open symbols M(IV), fits represented by dashed lines. Solid lines depict the ideal slope of ``$-1$''.}
    \label{SI:fig:CustomAlkan}
\end{figure}

\begin{figure}[h!]
    \centering
    \includegraphics[width=\linewidth]{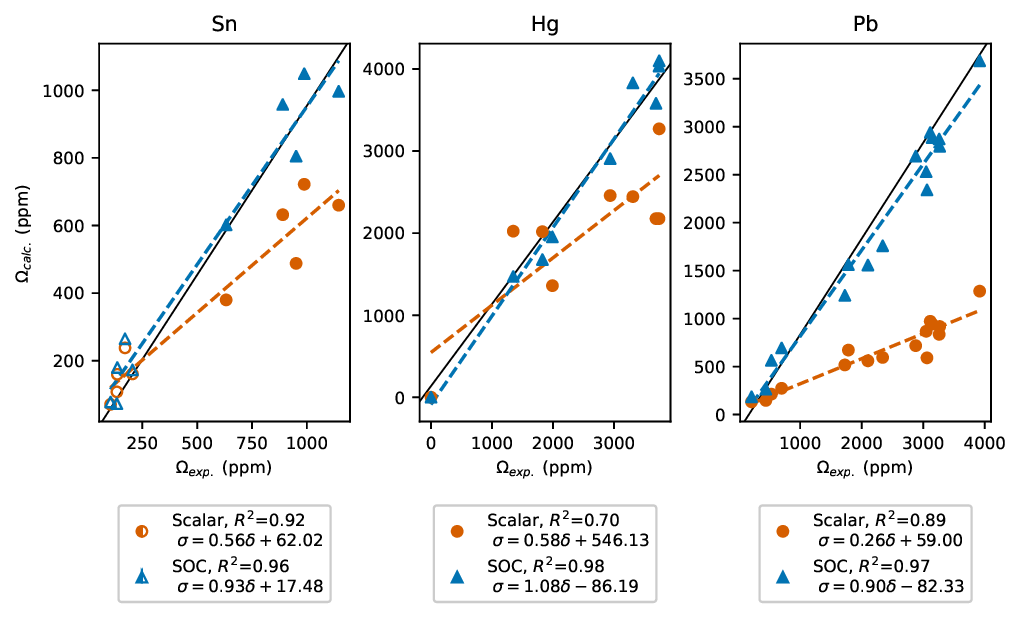}
    \caption{Selection of Alkan calculated span correlated to experimental span for Sn, Hg and Pb compounds. Closed symbols represent M(II), open symbols M(IV), fits represented by dashed lines. Solid lines depict the ideal slope of ``$-1$''.}
    \label{SI:fig:CustomAlkanSpan}
\end{figure}

\begin{figure}[h!]
    \centering
    \includegraphics[width=\linewidth]{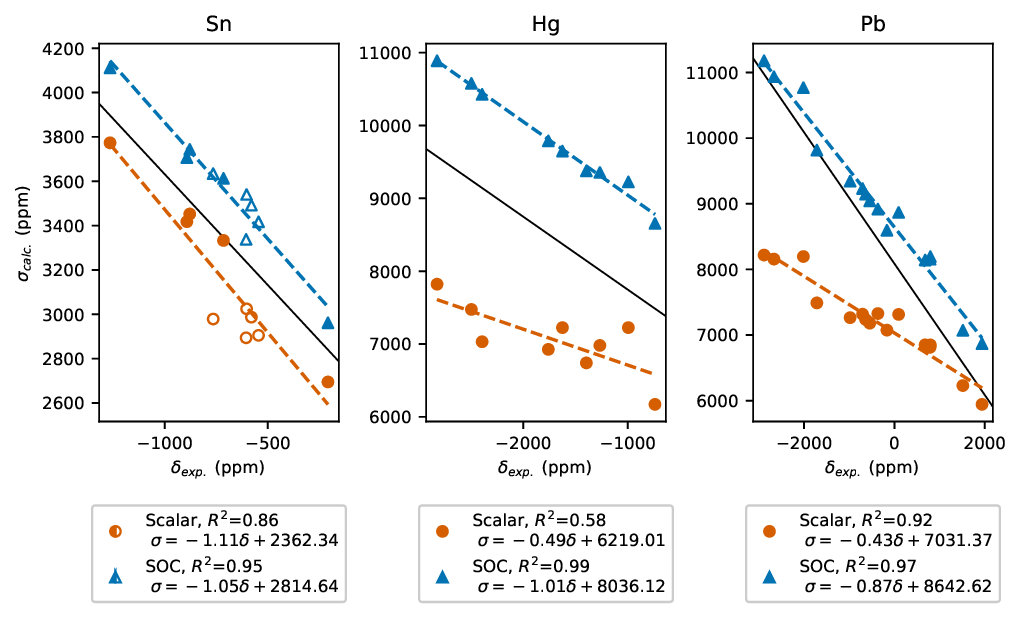}
    \caption{Selection of Alkan calculated isotropic chemical shielding correlated to experimental isotropic chemical shift for Sn, Hg and Pb compounds. Closed symbols represent M(II), open symbols M(IV), fits represented by dashed lines. Solid lines depict the ideal slope of ``$-1$''.}
    \label{SI:fig:CustomAlkanISO}
\end{figure}

Data for Sn at the PBE/ZORA/SC and PBE/ZORA/SO level are from Ref.~\onlinecite{SIAlkan-Sn}, Tables~1 and 2 and S1 (PBE/$10^{-4}$).
Data for Hg at the BP86/ZORA/SC and BP86/ZORA/SO level are from Ref.~\onlinecite{SIAlkan-Hg}, Tables~S4 (ZORA/Scalar) and S1 (Large Cluster) respectively.
Data for Pb at the BP86/ZORA/SC and BP86/ZORA/SO level are from Ref.~\onlinecite{SIAlkan-Pb}, Tables~S4 and S3 (3-VMTO/BV) respectively.
Experimental data in the fits in Figs.~\ref{SI:fig:CustomAlkan}, \ref{SI:fig:CustomAlkanSpan} and \ref{SI:fig:CustomAlkanISO}
below as in Tables~\ref{SI:tab:VASP:Sn}, \ref{SI:tab:VASP:Hg} and \ref{SI:tab:VASP:Pb}.

\FloatBarrier 

\FloatBarrier 

\FloatBarrier

\FloatBarrier 
\clearpage
\subsection{VASP Data}
\subsubsection{Magnetic Susceptibility}

\begin{table}[h!]
\caption{Comparison of VASP calculated $vGv$-susceptibilities $\chi_m$ in cgs units to experimental references.\cite{SISusceptCRC105}}
$\sigma(G=0)$ denotes the corresponding contribution to the isotropic shielding.
\begin{tabular}{@{}rddcdd@{}}
\hline
\hline
\multirow{2}{*}{\begin{tabular}[c]{@{}l@{}}Compound\end{tabular}} & \multicolumn{2}{c}{experiment, Ref.~\onlinecite{SISusceptCRC105}} & & \multicolumn{2}{c}{VASP - $vGv$ (isotropic)} \\
 \cline{2-3} \cline{5-6} \T
 & \multicolumn{1}{c}{$\chi_\text{m}$ [$10^{-6}\text{cm}^3\text{mol}^{-1}$]}
 & \multicolumn{1}{c}{$\sigma(G=0)$ [ppm]} &
 & \multicolumn{1}{c}{$\chi_\text{m}$ [$10^{-6}\text{cm}^3\text{mol}^{-1}$]}
 & \multicolumn{1}{c}{$\sigma(G=0)$ [ppm]} \\
\hline
SnO                         & -19   &  7.63 & &  -15.14 &  6.08 \\
$\text{SnO}_2$              & -41   & 16.62 & &  -30.65 & 12.43 \\
\hline                                                          
$\text{Hg(SCN)}_2$          & -96.5 & 10.80 & &  -87.87 &  9.83 \\
$\text{Hg(CN)}_2$           & -67   &  9.86 & &  -60.37 &  8.89 \\
$\text{HgF}_2$              & -57.3 & 20.13 & &  -39.31 & 13.81 \\
$\text{HgCl}_2$             & -82   & 15.57 & &  -68.49 & 13.01 \\
$\text{HgBr}_2$             & -94.2 & 14.21 & &  -82.96 & 12.52 \\
$\text{Hg}_2\text{Cl}_2$    & -120  & 17.32 & &  -99.54 & 14.36 \\
\hline                                                          
$\beta-\text{PbO}$          & -42   & 15.47 & &  -26.66 &  9.82 \\
$\text{PbF}_2$              & -58.1 & 17.70 & &  -37.81 & 11.52 \\
$\text{PbCl}_2$             & -73.8 & 14.89 & &  -61.37 & 12.38 \\
$\text{PbBr}_2$             & -90.6 & 14.51 & &  -78.85 & 12.62 \\
$\text{Pb}_3\text{(PO4)}_2$ & -182  & 15.37 & & -144.15 & 12.17 \\
\hline
\hline
\end{tabular}\\
\end{table}

\FloatBarrier 
\subsubsection{Sn}
\begin{table}[h!]
\caption{Summary of VASP data for Sn systems. PBE/ZORA/SC: scalar ZORA VASP results, PBE/ZORA/SO: SOC ZORA VASP results excluding the ZORA $K$-factor in the one-center currents, PBE/ZORA/SO+K: SOC ZORA VASP results including the ZORA $K$-factor in the one-center currents. References refer to experimental measurements.}
\label{SI:tab:VASP:Sn}
\begin{tabular}{@{}lrrrrrrrrrrrrrrr@{}}
\hline
\hline
\multirow{2}{*}{\begin{tabular}[c]{@{}l@{}}Compound\end{tabular}} & \multicolumn{3}{c}{Exp.} & \multicolumn{3}{c}{PBE/ZORA/SC} & \multicolumn{3}{l}{PBE/ZORA/SO} & \multicolumn{3}{l}{PBE/ZORA/SO+K} \\
 & $\delta_{11}$ & $\delta_{22}$ & $\delta_{33}$ & $\sigma_{11}$ & $\sigma_{22}$ & $\sigma_{33}$ & $\sigma_{11}$ & $\sigma_{22}$ & $\sigma_{33}$ & $\sigma_{11}$ & $\sigma_{22}$ & $\sigma_{33}$ \\
\hline
SnO\footnotemark[1] & 121 & 121 & -867 & 2804.1 & 2804.1 & 3431.6 & 2293.4 & 2293.4 & 3186.0 & 2454.6 & 2454.6 & 3297.1 \\
$\text{SnC}_2\text{O}_4$\footnotemark[2] & -523 & -639 & -1474 & 3387.2 & 3432.6 & 3893.4 & 2896.2 & 2968.0 & 3745.4 & 3023.0 & 3090.5 & 3831.1 \\
$\text{SnSO}_4$\footnotemark[2] & -1047 & -1070 & -1679 & 3786.8 & 3863.8 & 4180.2 & 3406.9 & 3477.9 & 4050.3 & 3505.7 & 3571.8 & 4118.7 \\
$\text{BaSnF}_4$\footnotemark[2] & -596 & -596 & -1486 & 3389.9 & 3389.9 & 3980.7 & 2928.1 & 2928.1 & 3871.2 & 3053.2 & 3053.2 & 3949.0 \\
$\text{SnHPO}_3$\footnotemark[2] & -290 & -420 & -1435 & 3247.6 & 3366.4 & 3920.2 & 2729.3 & 2883.2 & 3790.5 & 2864.1 & 3009.6 & 3873.3 \\
\hline
$\text{SnO}_2$\footnotemark[3] & -550 & -573 & -686 & 3132.9 & 3138.6 & 3255.7 & 2965.1 & 2982.3 & 3095.7 & 3077.8 & 3095.8 & 3206.7 \\
$\text{Ca}_2\text{SnO}_4$\footnotemark[4] & -459 & -512 & -664 & 3043.6 & 3071.7 & 3211.7 & 2902.5 & 2920.4 & 3073.3 & 3022.0 & 3042.1 & 3187.1 \\
$\text{SnS}_2$\footnotemark[5] & -730 & -730 & -835 & 3172.8 & 3172.9 & 3277.2 & 3182.5 & 3182.5 & 3320.9 & 3296.3 & 3296.3 & 3426.3 \\
$\text{Pb}_2\text{SnO}_4$\footnotemark[6] & -558 & -566 & -692 & 3020.2 & 3088.4 & 3099.8 & 2846.4 & 2883.1 & 2891.2 & 2969.5 & 3004.9 & 3014.1 \\
$\text{Sr}_2\text{SnO}_4$\footnotemark[4] & -510 & -548 & -681 & 3122.7 & 3126.6 & 3275.7 & 2954.5 & 2955.8 & 3133.5 & 3069.3 & 3070.6 & 3235.8 \\
\hline
\hline
\end{tabular}\\
\footnotemark[1]Ref.~\onlinecite{MacGregor2011}, \footnotemark[2]Ref.~\onlinecite{Amornsakchai2004}, \footnotemark[3]Ref.~\onlinecite{Cossement1992}, \footnotemark[4]Ref.~\onlinecite{Clayden1989}, \footnotemark[5]Ref.~\onlinecite{Mundus1996}, \footnotemark[6]Ref.~\onlinecite{Catalano2014}
\end{table}

\FloatBarrier 
\clearpage
\subsubsection{Hg}
\begin{table}[h!]
\caption{Summary of VASP data for Hg systems. PBE/ZORA/SC: scalar ZORA VASP results, PBE/ZORA/SO: SOC ZORA VASP results excluding the ZORA $K$-factor in the one-center currents, PBE/ZORA/SO+K: SOC ZORA VASP results including the ZORA $K$-factor in the one-center currents. References refer to experimental measurements.}
\label{SI:tab:VASP:Hg}
\begin{tabular}{@{}lrrrrrrrrrrrr@{}}
\hline
\hline
\multirow{2}{*}{\begin{tabular}[c]{@{}l@{}}Compound\end{tabular}} & \multicolumn{3}{c}{Exp.} & \multicolumn{3}{c}{PBE/ZORA/SC} & \multicolumn{3}{l}{PBE/ZORA/SO} & \multicolumn{3}{l}{PBE/ZORA/SO+K} \\
 & $\delta_{11}$ & $\delta_{22}$ & $\delta_{33}$ & $\sigma_{11}$ & $\sigma_{22}$ & $\sigma_{33}$ & $\sigma_{11}$ & $\sigma_{22}$ & $\sigma_{33}$ & $\sigma_{11}$ & $\sigma_{22}$ & $\sigma_{33}$ \\
\hline
$\text{Hg}(\text{SCN})_2$\footnotemark[1] & -81 & -328 & -3390 & 7079.4 & 7133.2 & 9565.7 & 5355.5 & 5937.7 & 9320.7 & 5990.1 & 6490.1 & 9478.4 \\
$\text{Hg}(\text{CN})_2$\footnotemark[3] & -33 & -381 & -3773 & 6624.8 & 6674.6 & 9851.8 & 5554.0 & 5665.4 & 9692.0 & 6198.3 & 6323.6 & 9811.2 \\
$\text{Hg}(\text{SeCN})_2$\footnotemark[2] & -503 & -1337 & -3440 & 7032.1 & 7119.1 & 9546.4 & 6014.4 & 6972.2 & 9292.6 & 6624.1 & 7728.2 & 9463.1 \\
$\text{Hg}(\text{CO}_2\text{CH}_3)_2$\footnotemark[1] & -1859 & -1947 & -3685 & 7677.5 & 7905.5 & 9743.1 & 7737.5 & 7897.6 & 9426.9 & 8182.4 & 8318.9 & 9566.5 \\
$\text{HgF}_2$\footnotemark[4] & -2826 & -2826 & -2826 & 8824.9 & 8824.9 & 8824.9 & 8786.4 & 8786.4 & 8786.4 & 9058.1 & 9058.1 & 9058.1 \\
$\text{HgCl}_2$\footnotemark[1] & -282 & -573 & -4019 & 7125.9 & 7135.5 & 10198.8 & 5788.2 & 5821.8 & 10059.7 & 6439.3 & 6469.9 & 10132.4 \\
$\text{HgBr}_2$\footnotemark[4] & -1945 & -1945 & -3293 & 7249.6 & 7254.5 & 9388.7 & 7673.3 & 7687.5 & 9134.3 & 8321.4 & 8333.5 & 9336.6 \\
$\text{Hg}_2\text{Cl}_2$\footnotemark[4] & 236 & 236 & -3452 & 7505.4 & 7505.4 & 9763.7 & 5589.8 & 5589.8 & 9511.7 & 6226.1 & 6226.1 & 9658.3 \\
$\text{K}[\text{Hg}(\text{SCN})_3]$\footnotemark[2] & 49 & -323 & -1941 & 6574.1 & 6689.3 & 8005.9 & 5403.6 & 5612.6 & 7421.8 & 6064.9 & 6200.1 & 7853.9\\
\hline
\hline
\end{tabular}\\
\footnotemark[1]Ref.~\onlinecite{Bowmaker1999}, \footnotemark[2]Ref.~\onlinecite{Bowmaker1998}, \footnotemark[3]Ref.~\onlinecite{Bowmaker1997}, \footnotemark[4]Ref.~\onlinecite{Taylor2011}
\end{table}

\FloatBarrier 
\subsubsection{Pb}
\begin{table}[h!]
\caption{Summary of VASP data for Pb systems. PBE/ZORA/SC: scalar ZORA VASP results, PBE/ZORA/SO: SOC ZORA VASP results excluding the ZORA $K$-factor in the one-center currents, PBE/ZORA/SO+K: SOC ZORA VASP results including the ZORA $K$-factor in the one-center currents. References refer to experimental measurements.}
\label{SI:tab:VASP:Pb}
\begin{tabular}{@{}lrrrrrrrrrrrr@{}}
\hline
\hline
\multirow{2}{*}{\begin{tabular}[c]{@{}l@{}}Compound\end{tabular}} & \multicolumn{3}{c}{Exp.} & \multicolumn{3}{c}{PBE/ZORA/SC} & \multicolumn{3}{l}{PBE/ZORA/SO} & \multicolumn{3}{l}{PBE/ZORA/SO+K} \\
 & $\delta_{11}$ & $\delta_{22}$ & $\delta_{33}$ & $\sigma_{11}$ & $\sigma_{22}$ & $\sigma_{33}$ & $\sigma_{11}$ & $\sigma_{22}$ & $\sigma_{33}$ & $\sigma_{11}$ & $\sigma_{22}$ & $\sigma_{33}$ \\
\hline
$\alpha-\text{PbO}$\footnotemark[1] & 2977.0 & 2977.0 & -137.0 & 6524.7 & 6524.7 & 7462.9 & 2654.9 & 2654.9 & 5238.1 & 3807.3 & 3807.3 & 6044.4 \\
$\beta-\text{PbO}$\footnotemark[1] & 2944.7 & 2572.6 & -972.3 & 6513.7 & 6772.8 & 7673.9 & 2863.8 & 2907.0 & 6098.7 & 4026.9 & 4059.1 & 6851.4 \\
$\text{Pb}_3\text{O}_4$\footnotemark[1] & 1968.5 & 1496.0 & -1079.5 & 6873.2 & 7296.0 & 7855.7 & 3635.2 & 4056.9 & 6151.9 & 4732.6 & 5029.7 & 6888.6 \\
$\text{Pb}_2\text{SnO}_4$\footnotemark[2] & 1810.0 & 1565.0 & -1335.0 & 6843.0 & 7208.4 & 7808.9 & 3768.1 & 3935.9 & 6283.6 & 4849.7 & 4911.2 & 7002.1 \\
 & 1903.0 & 1828.0 & -1365.0 & 6943.6 & 7291.9 & 7909.1 & 3642.3 & 4034.7 & 6347.4 & 4726.4 & 5008.4 & 7063.6\\
$\text{PbF}_2$\footnotemark[1] & -2484.6 & -2588.7 & -2927.6 & 8803.1 & 8823.3 & 8988.5 & 7167.7 & 7211.5 & 7725.0 & 7764.3 & 7800.1 & 8258.8 \\
$\text{PbCl}_2$\footnotemark[1] & -1507.3 & -1603.3 & -2040.3 & 8221.6 & 8292.1 & 8477.9 & 6473.4 & 6600.3 & 7070.7 & 7216.5 & 7333.2 & 7738.9 \\
$\text{PbBr}_2$\footnotemark[3] & -698.9 & -845.1 & -1398.1 & 7867.3 & 7950.2 & 8178.1 & 5744.7 & 5936.5 & 6529.7 & 6598.2 & 6781.1 & 7294.6 \\
PbClOH\footnotemark[3] & 244.1 & -264.7 & -2096.7 & 7826.3 & 8095.2 & 8420.3 & 5051.5 & 5500.6 & 6948.2 & 5902.4 & 6279.4 & 7590.1 \\
PbBrOH\footnotemark[3] & 190.3 & -196.1 & -1910.6 & 7791.3 & 7973.6 & 8321.5 & 5071.0 & 5379.0 & 6725.9 & 5918.8 & 6192.3 & 7402.8 \\
PbIOH\footnotemark[3] & 79.3 & -70.6 & -1647.0 & 7795.5 & 7846.6 & 8297.5 & 5160.1 & 5195.7 & 6580.2 & 5991.7 & 6057.5 & 7290.1 \\
$\text{PbSiO}_3$\footnotemark[1] & 1088.7 & 583.6 & -2170.3 & 7719.1 & 7735.3 & 8461.1 & 4208.0 & 4597.0 & 7100.8 & 5110.7 & 5515.9 & 7726.7 \\
 & 1215.4 & 726.1 & -1662.6 & 7579.1 & 7695.3 & 8312.8 & 3919.0 & 4435.5 & 6749.8 & 4808.2 & 5334.7 & 7374.5 \\
 & 838.0 & 287.0 & -2223.0 & 7886.6 & 7984.2 & 8439.4 & 4538.0 & 5086.9 & 6980.3 & 5377.1 & 5915.9 & 7609.0 \\
$\text{Pb}_3(\text{PO}_4)_2$\footnotemark[1] & -2758.6 & -2930.8 & -2968.6 & 8915.3 & 9033.1 & 9066.0 & 7662.7 & 7881.7 & 7958.7 & 8236.0 & 8411.4 & 8487.5 \\
 & -1314.3 & -1635.4 & -3098.3 & 8620.3 & 8823.5 & 9460.8 & 6664.5 & 7197.8 & 8366.7 & 7315.1 & 7761.0 & 8781.7 \\
\hline
\hline
\end{tabular}\\
\footnotemark[1]Ref.~\onlinecite{Fayon1997}, \footnotemark[2]Ref.~\onlinecite{Catalano2014}, \footnotemark[3]Ref.~\onlinecite{Dybowski1998}
\end{table}

\FloatBarrier
\clearpage
\subsubsection{Fits}
\begin{figure}[h!]
    \centering
    \includegraphics[width=\linewidth]{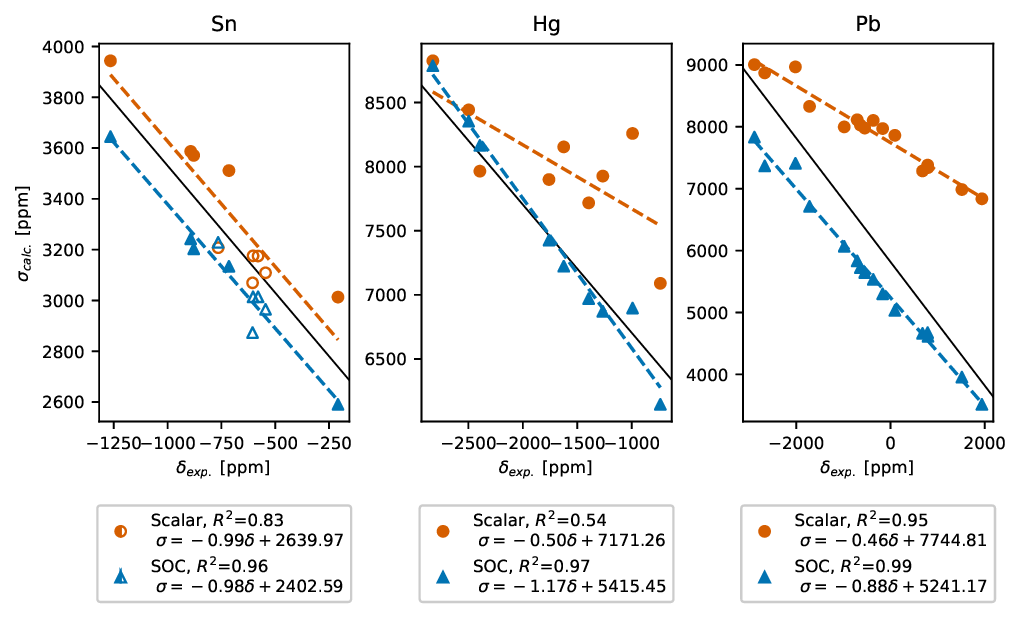}
    \caption{Calculated isotropic chemical shielding correlated to experimental isotropic chemical shift for Sn, Hg and Pb compounds using VASP. SOC series exclude ZORA $K$-factor in the one-center currents. Closed symbols represent M(II), open symbols M(IV), fits represented by dashed lines. Solid lines depict the ideal slope of ``$-1$''.}
\end{figure}

\begin{figure}[h!]
    \centering
    \includegraphics[width=\linewidth]{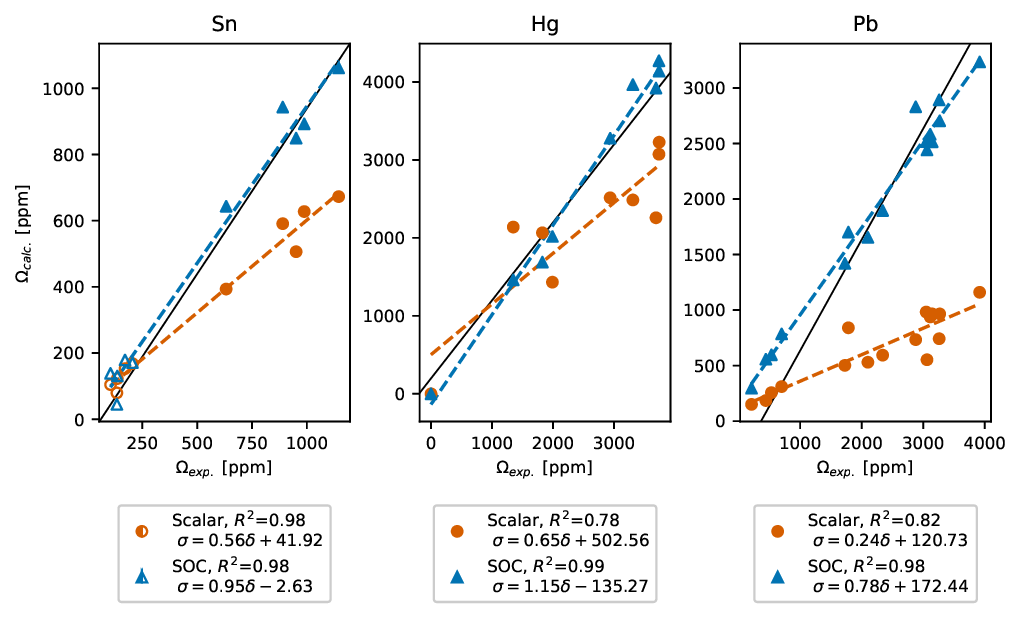}
    \caption{Calculated span correlated to experimental span for Sn, Hg and Pb compounds using VASP. SOC series exclude ZORA $K$-factor in the one-center currents. Closed symbols represent M(II), open symbols M(IV), fits represented by dashed lines. Solid lines depict the ideal slope of ``$-1$''.}
\end{figure}

\begin{figure}[h!]
    \centering
    \includegraphics[width=\linewidth]{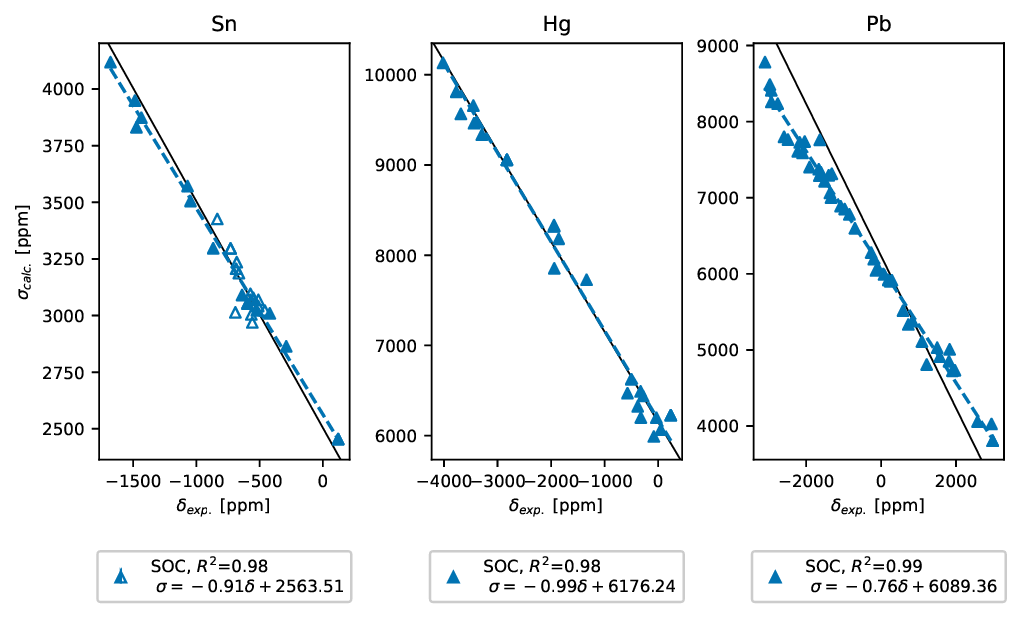}
    \caption{Calculated principal components of chemical shielding correlated to experimental chemical shift principal components for Sn, Hg and Pb compounds using VASP. SOC series include ZORA $K$-factor in the one-center currents. Closed symbols represent M(II), open symbols M(IV), fits represented by dashed lines. Solid lines depict the ideal slope of ``$-1$''.}
\end{figure}

\begin{figure}[h!]
    \centering
    \includegraphics[width=\linewidth]{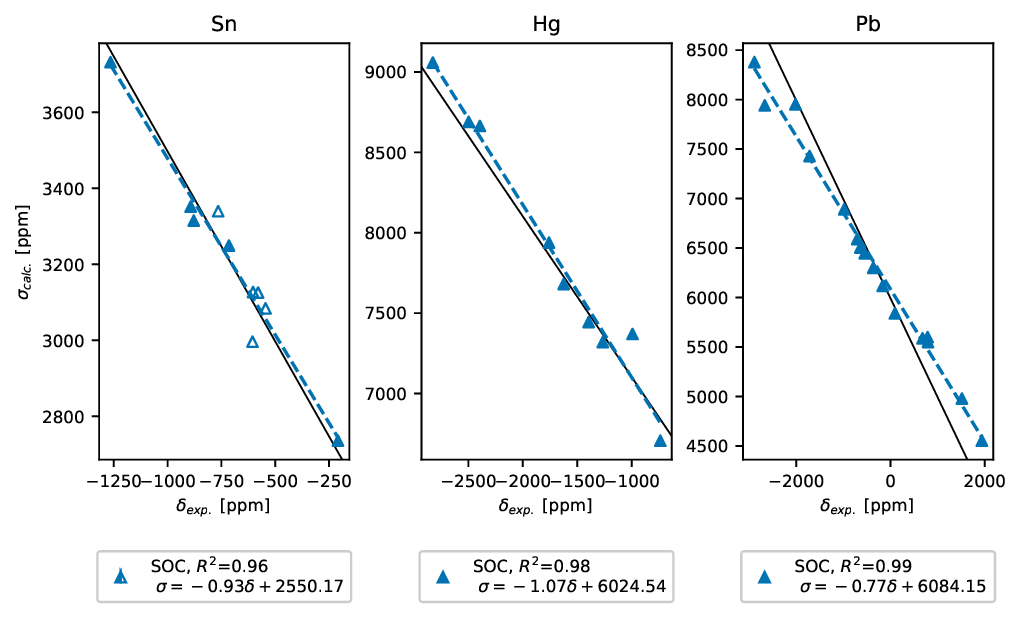}
    \caption{Calculated isotropic chemical shielding correlated to experimental isotropic chemical shift for Sn, Hg and Pb compounds using VASP. SOC series include ZORA $K$-factor in the one-center currents. Closed symbols represent M(II), open symbols M(IV), fits represented by dashed lines. Solid lines depict the ideal slope of ``$-1$''.}
\end{figure}

\begin{figure}[h!]
    \centering
    \includegraphics[width=\linewidth]{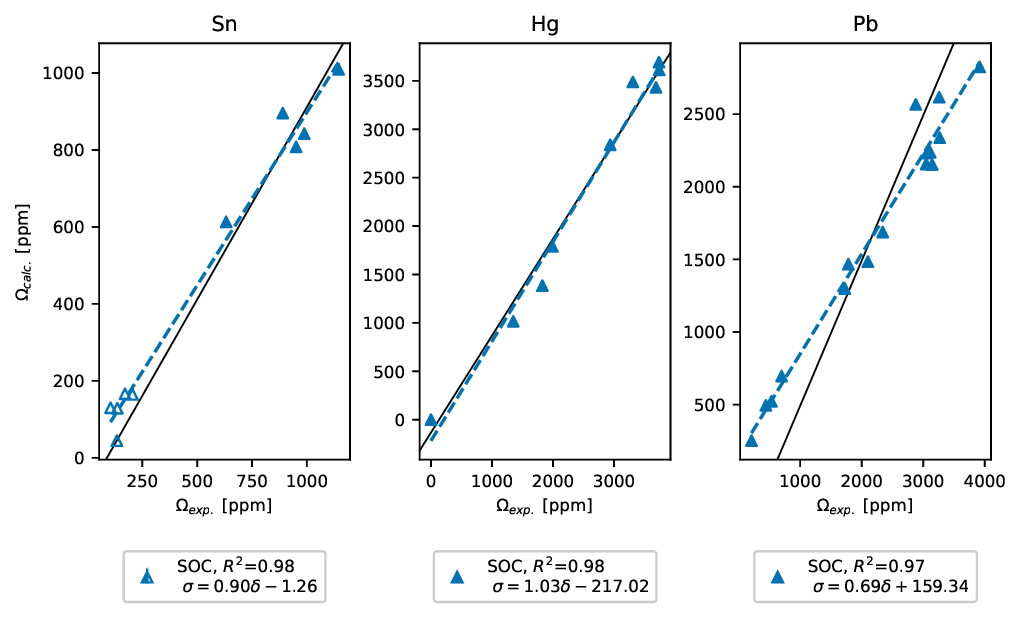}
    \caption{Calculated span correlated to experimental span for Sn, Hg and Pb compounds using VASP. SOC series include ZORA $K$-factor in the one-center currents. Closed symbols represent M(II), open symbols M(IV), fits represented by dashed lines. Solid lines depict the ideal slope of ``$-1$''.}
\end{figure}

\FloatBarrier